\documentclass[%
 aip,
 amsmath,amssymb,
 reprint,%
prb,onecolumn,superscriptaddress]{revtex4-1}

\usepackage{hyperref}
\usepackage{comment}
\usepackage{physics}
\usepackage[normalem]{ulem}
\usepackage{cancel}
\usepackage{xcolor}
\usepackage{graphicx}
\usepackage{dcolumn}
\usepackage{bm}

\usepackage[utf8]{inputenc}
\usepackage[T1]{fontenc}
\usepackage{mathptmx}
\usepackage{etoolbox}
\usepackage{placeins}

\newcommand*{\citen}[1]{%
  \begingroup
    \romannumeral-`\x 
    \setcitestyle{numbers}%
    \citep{#1}%
  \endgroup   
}

\makeatletter
\def\@email#1#2{%
 \endgroup
 \patchcmd{\titleblock@produce}
  {\frontmatter@RRAPformat}
  {\frontmatter@RRAPformat{\produce@RRAP{*#1\href{mailto:#2}{#2}}}\frontmatter@RRAPformat}
  {}{}
}%
\makeatother
\begin{document}

\preprint{AIP/123-QED}

\title[Thermal noise Mn$_3$Sn]{Influence of thermal noise on the field-driven dynamics of the non-collinear antiferromagnet Mn$_3$Sn}
\author{S. Qian}
\affiliation{
Holonyak Micro and Nanotechnology Laboratory, University of Illinois, Urbana-Champaign, IL, USA
}%
\author{A. Shukla}
\affiliation{
Holonyak Micro and Nanotechnology Laboratory, University of Illinois, Urbana-Champaign, IL, USA
}%
\author{S. Rakheja}%
 \email{rakheja@illinois.edu}
\affiliation{
Holonyak Micro and Nanotechnology Laboratory, University of Illinois, Urbana-Champaign, IL, USA
}%


\date{\today}

\begin{abstract}
$\mathrm{Mn_3Sn}(0\overline{1}\overline{1}0)[0001]$ experiences a tensile strain when grown epitaxially on $\mathrm{MgO}(110)[001]$, and thus the energy landscape changes from six-fold symmetry to two-fold symmetry. External magnetic field further breaks the symmetry and the 
{resulting energy landscape is sensitive to}
the field orientation relative to the easy axis. {In the presence of thermal noise,}
the relaxation of the magnetic octupole moment 
in {a strained Mn$_3$Sn film}
is composed of four distinct escape processes involving the two saddle points and two equilibrium states in the energy landscape. Here, we apply harmonic transition-state theory to derive analytical expressions for the inter-well escape time and octupole moment relaxation time, both influenced by an external symmetry-breaking magnetic field and finite thermal noise in the intermediate-to-high damping regime.
The analytical predictions are in strong agreement with comprehensive numerical simulations based on coupled LLG equations.
The results presented here are crucial toward realizing Mn$_3$Sn's applications in  random number generation and probabilistic computing.
\end{abstract}

\maketitle

Non-collinear chiral antiferromagnets (AFMs) of the form Mn$_3$X (X = Sn, Ge, Ir, and Pt)~\cite{shukla2025spintronic, rimmler2025non} 
have gained significant attention owing to their large anomalous Hall effect (AHE),~\cite{nakatsuji2015large,nayak2016large,chen2014anomalous} spin Hall effect (SHE),~\cite{zhang2016giant} anomalous Nernst effect (ANE),~\cite{ikhlas2017large, hong2020large} magneto-optical Kerr effect (MOKE),~\cite{higo2018large} and finite tunneling magnetoresistance (TMR).~\cite{dong2022tunneling, chen2023octupole, qin2023room, wang2024mn3sn} The anomalous Hall conductivity (AHC) of Mn$_3$Sn, a prototypical non-collinear chiral AFM, can range from 30 to 40\, $\Omega^{-1}\cdot$cm$^{-1}$ at room temperature, which is attributed to the breaking of time-reversal symmetry (TRS).~\cite{liu2023anomalous, yano2024giant}
In its bulk form and below its N\'{e}el temperature ($T_\mathrm{N}$) of approximately 420\,K, Mn$_3$Sn exhibits a six-fold degenerate energy landscape.~\cite{higo2018large, sung2018magnetic}
However, in an epitaxially grown Mn$_3$Sn/MgO(110) [001] bilayer structure, the energy landscape is found to possess two-fold symmetry with the emergence of a perpendicular magnetic anisotropy (PMA) owing to an epitaxial tensile strain.~\cite{higo2022perpendicular, yoon2023handedness} 

Current-driven dynamics, including deterministic switching and chiral oscillations, have been experimentally explored in polycrystalline as well as epitaxial films of thicknesses ranging from sub-10\,nm to 100\,nm.~\cite{tsai2020electrical, takeuchi2021chiral, yan2022quantum, pal2022setting, krishnaswamy2022time, higo2022perpendicular, yoon2023handedness, xu2023robust, yoo2024thermal, zheng2025all, lee2025spin,deng2023all,liu2023topological} 
The AHE signal in the case of polycrystalline films, measured in a typical Hall bar setup, is considered to be an average of the final magnetic states in the various composing grains.~\cite{ikeda2018anomalous,xie2022magnetization} 
While in some cases pure spin-orbit torque (SOT) is considered to be the fundamental mechanism behind the observed dynamics,~\cite{tsai2020electrical, takeuchi2021chiral, higo2022perpendicular, yoon2023handedness,deng2023all} others have attributed the observed changes in the AHE signal to a combination of Joule heating and SOT (\emph{i.e.}, demagnetization of the polycrystalline Mn$_3$Sn film as the temperature rises above $T_\mathrm{N}$ due to Joule heating, followed by SOT-driven re-magnetization as the film cools below $T_\mathrm{N}$ during the 
slow decrease in current.~\cite{pal2022setting, krishnaswamy2022time, yoo2024thermal} 
The SOT-driven oscillation~\cite{zhao2021terahertz, shukla2022spin, lund2023voltage} and switching~\cite{shukla2023order, xu2024deterministic, shukla2024impact, zhao2025investigation,deng2023all} dynamics in both six-fold and two-fold energy degenerate Mn$_3$Sn crystals have also been explored theoretically. 
In particular, these works have extended the ferromagnetic theory to shed light on the threshold current requirement for initiating dynamics as well as the dependence of the various threshold currents on the material properties. They have also explored the possible dependence of the oscillation frequency and the switching time on the input stimuli and the material parameters.
The manipulation of the order parameter in Mn$_3$Sn via SOT, together with a finite TMR in all-Mn$_3$Sn junctions, can enable the development of AFM-based spintronic devices that are fully compatible with electronic circuitry.

Despite significant progress in material synthesis and device functionality of $\mathrm{Mn_3Sn}$, 
studies investigating the impact of thermal noise on the dynamics and stability of the octupole moment in scaled Mn$_3$Sn bits remain limited. Thermal noise is known to impose a lower bound on the bit error rate in ferromagnetic memories,~\cite{coffey2012thermal, kaiser2019subnanosecond} yet it can also be harnessed to generate random numbers~\cite{shukla2023true,chun2015high} in hardware and potentially enable the paradigm of probabilistic computing.~\cite{shao2023probabilistic}
The thermal stability of Mn$_3$Sn nanodots of diameters ranging from 175 to 1000\,nm was experimentally investigated in Ref.~[\citen{sato2023thermal}]. The energy barrier for switching between the equilibrium states was extracted using the N\'eel-Arrhenius activation model, and it was found that the energy barrier decreased with the nanodot size below the nucleation diameter ($\sim 300$\, nm for the sample). 
%
Kobayashi \emph{et al.}~\cite{kobayashi2023pulse} employed pulsed SOT measurements to extract the energy barrier by analyzing the relationship between the critical switching voltage and pulse duration. In Kobayashi's sample, a robust AHE signal was measured in sub-100~nm grains, underscoring the promise of Mn$_3$Sn for dense non-volatile memory applications.

Konakanchi~\emph{et al.}~\cite{konakanchi2025electrically} presented a theoretical analysis of the relaxation time of the octupole order parameter, and its electrical tunability, in chiral AFMs. The relaxation of the octupole moment was found to depend either on the precessional dephasing when the energy barrier of the octupole was much smaller than the thermal energy, or given by the thermally activated escape over the barrier in the high-barrier regime.  
\begin{figure}[b!]
    \centering    
    \includegraphics[width=0.6\linewidth]{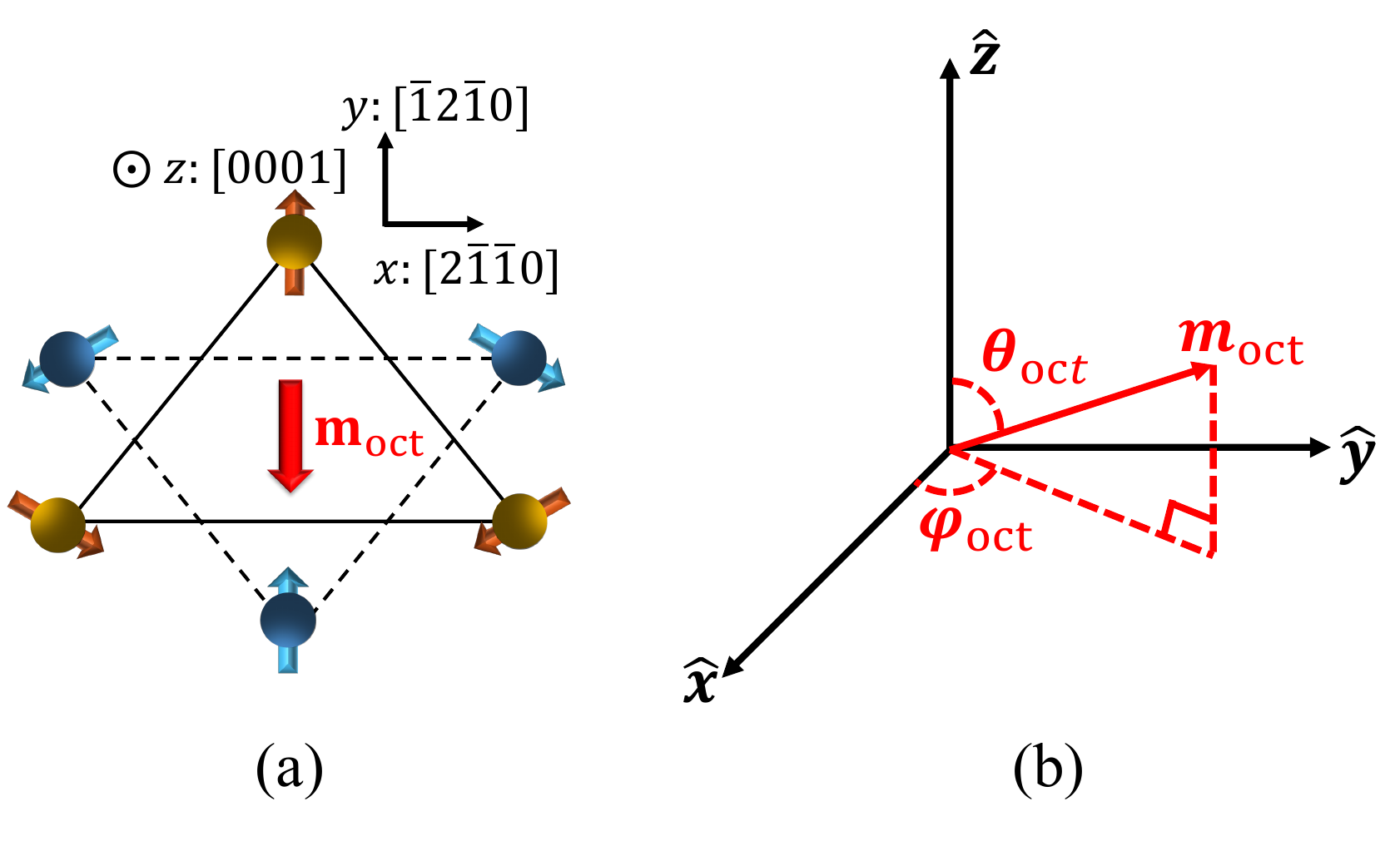}
    \vspace{-12pt}
    \caption{(a) 
    {Unit Cell of Mn$_3$Sn with two Kagome planes. Mn spins in different planes are marked with different colors.} (b) Octupole moment and relevant angles in Cartesian coordinate system.}
    \label{fig:Structure}
    \vspace{-12pt}
\end{figure}

\begin{figure*}
    \centering
    \includegraphics[width=\columnwidth]{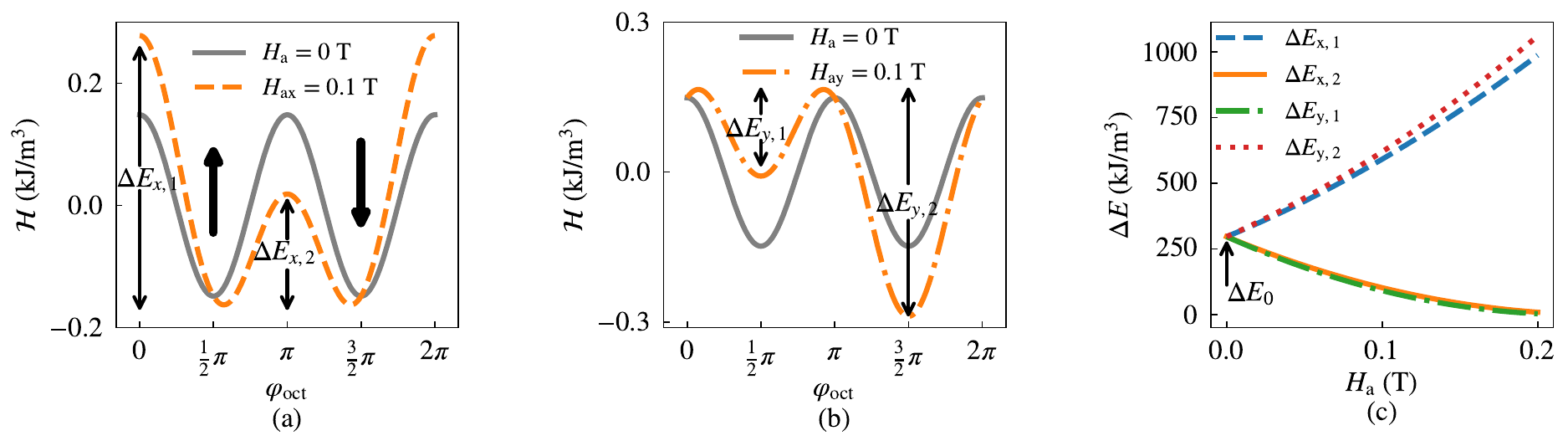}
   \vspace{-10pt}
    \caption{Hamiltonian of $\mathrm{Mn_3Sn}$ as a function of $\varphi_\mathrm{oct}$ with $H_\mathrm{a} = 0~\mathrm{T}$ and (a) $H_\mathrm{ax} = 0.1~\mathrm{T}$ and (b) $H_\mathrm{ay} = 0.1~\mathrm{T}$. Two equilibrium states of octupole moment are referred to as up state and down state in (a). (c) Splitting of the energy barriers as a function of the external magnetic field magnitude. $\Delta E_0$ denotes the energy barrier in absence of magnetic field.}
    \label{fig:Energy}
    \vspace{-12pt}
\end{figure*}
%
Konakanchi's analysis indicates that the markedly shorter relaxation times of chiral AFMs—on the order of picoseconds—are driven by strong exchange interactions. However, their model neglects any external magnetic fields, which are 
typically incorporated in experimental studies and are essential for achieving more diverse manipulations of the order parameter.

In this paper, we investigate the impact of thermal noise on the stability of a strained 
$\mathrm{Mn_3Sn}$ bit 
in the presence of an external magnetic field. We present a comprehensive analysis of the mechanisms underlying the free-energy symmetry breaking, examining  the dependence of free energy on both the magnitude and orientation of the applied field, as well as the associated fluctuation modes. The interwell escape time of the magnetic octupole moment is theoretically evaluated as a function of the external field and temperature, with results corroborated by numerical simulations in the low-energy barrier regime ($\Delta E_0 \leq 
{6}k_\mathrm{B}T$). We further extend our theoretical framework to encompass the relaxation dynamics of octupole moment and discuss prospective applications in random number generation and probabilistic computing.



Mn$_3$Sn crystallizes into a hexagonal Kagome $D0_{19}$ lattice below its N\'eel temperature, exhibiting a slight excess of Mn atoms.~\cite{shukla2025spintronic} Sn atoms occupy the centers of the hexagons, whereas Mn atoms are located at the vertices. 
{Each unit cell of Mn$_3$Sn includes two Kagome planes with three Mn spins per plane. 
The magnetic moments of Mn atoms, forming an inverse triangular configuration, are 
slightly canted toward the in-plane easy axis, giving rise to a non-zero net magnetization as shown in Fig.~\ref{fig:Structure}(a). The magnetic ground state typically exhibits a six-fold degeneracy;~\cite{shukla2023order} however, in the presence of an in-plane uniaxial strain,  the symmetry is reduced to two-fold.~\cite{higo2022perpendicular, yoon2023handedness} It is worth mentioning that intra-layer interactions are dominant as validated in Sec.~\ref{sec:six spin} of the Supplementary Material, so the following investigation focuses on one Kagome plane.}


The Hamiltonian of a single-domain $\mathrm{Mn_3Sn}$ is~\cite{tomiyoshi1982magnetic, park2018magnetic}
\begin{equation}
    \begin{aligned}
        \mathcal{H}(\vb{m}) &= J_\mathrm{E}((1+\delta_\mathrm{E})\vb{m}_1 \cdot \vb{m}_2+\vb{m}_2\cdot \vb{m}_3+\vb{m}_3\cdot \vb{m}_1)\\
        &+D_\mathrm{M} \vb{z}\cdot (\vb{m}_1\times \vb{m}_2+\vb{m}_2\times \vb{m}_3+\vb{m}_3\times \vb{m}_1)\\
        &-\sum_{i=1}^{3}{\left[K_\mathrm{u}(\vb{m}_i\cdot \vb{u}_i)^2+M_\mathrm{s} \vb{H}_\mathrm{a}\cdot \vb{m}_i)\right]},
    \end{aligned}
    \label{eq:energy m}
\end{equation}
where $\vb{m}_1$, $\vb{m}_2$, and $\vb{m}_3$ are the magnetic moments of three sublattices, respectively. The parameters $M_\mathrm{s}$, $J_\mathrm{E}$, $D_\mathrm{M}$ and $K_\mathrm{u}$ represent the saturation magnetization of each sublattice, the symmetric exchange interaction constant, the asymmetric Dzyaloshinskii-Moriya Interaction (DMI) constant and the single-ion uniaxial magnetocrystalline anisotropy constant, respectively. In the Cartesian coordinates, $\vb{u}_1 = -1/2 \vb{x} +\sqrt{3}/2 \vb{y}$, $\vb{u}_2 = -1/2 \vb{x} -\sqrt{3}/2 \vb{y}$ and $\vb{u}_3 = \vb{x}$ are the uniaxial easy axes of each sublattice.  A positive (negative) $\delta_E$ corresponds to compressive (tensile) strain, enhancing (weakening) the exchange interaction among the corresponding sublattices. Unless otherwise noted, the values of material parameters used in our simulations are those listed in Table~\ref{tab:parameters}.~\cite{yoon2023handedness}



Using perturbation techniques, the Hamiltonian $\mathcal{H}_\mathrm{oct}$ of the single-domain Mn$_3$Sn can be expressed in terms of $\varphi_\mathrm{oct}$ and $\theta_\mathrm{oct}$ (see Fig.~\ref{fig:Structure}(b)), which represent the azimuthal and polar angles of the octupole moment, respectively, as~\cite{yoon2023handedness, he2024magnetic, shukla2024impact}
\begin{equation}
    \begin{aligned}
        &\mathcal{H}(\theta_\mathrm{oct}, \varphi_\mathrm{oct}) = \frac{3}{4}M_\mathrm{s} H_\mathrm{K} \cos(2\varphi_\mathrm{oct}) + \frac{3}{2}M_\mathrm{s} H_\mathrm{J} \cos^2 \theta_\mathrm{oct}\\
        &-\frac{M_\mathrm{s} H_\mathrm{a} \cos{(\theta_\mathrm{H})}}{J_\mathrm{E}+\sqrt{3}D_\mathrm{M}}\Big[ K_\mathrm{u} \cos{(\varphi_{\mathrm{oct}}-\varphi_{\mathrm{H}})}+J_\mathrm{E}\delta_E\cos{(\varphi_{\mathrm{oct}}+\varphi_{\mathrm{H}})}\Big].
    \end{aligned}
    \label{eq:energy phi}
\end{equation}
Here, we used $\vb{m}_\mathrm{oct}  = \frac{1}{3} \mathcal{M}_{zx} \qty[ R\left(\frac{2\pi}{3}\right) \vb{m}_1 + R\left(-\frac{2\pi}{3}\right) \vb{m}_2 + \vb{m}_3 ]$ while $H_\mathrm{J} = \frac{(3J_\mathrm{E}+\sqrt{3}D_\mathrm{M})}{M_s}$ and $H_\mathrm{K} = -\frac{(4K_\mathrm{u} J_\mathrm{E}\delta_\mathrm{E})}{3M_s(J_\mathrm{E}+\sqrt{3}D_\mathrm{M})}$ are the strength of exchange field and uniaxial anisotropy field, respectively. $\theta_\mathrm{H}$ and $\varphi_\mathrm{H}$ are the polar and the azimuthal angle of the applied external field $\vb{H}_\mathrm{a}$, respectively. {In addition, $\mathcal{M}_zx$ and $R$ represent mirror operation against $zx$ plane and anti-clockwise rotation operation against $z$ axis, respectively.}

\begin{table}[h!]
\vspace{-15pt}
    \caption{Material parameters of strained Mn$_3$Sn thin film.~\cite{yoon2023handedness}}
    \begin{tabular}{|p{2cm}|p{4cm}|p{2cm}|}
        \hline
       \textbf{Parameter}                & \textbf{Definition}                         & \textbf{Value}              \\ \hline
        $\alpha$                  & Damping coefficient                & 0.003              \\ \hline
        $J_\mathrm{E}~(\mathrm{J/m^3})$                     & Exchange constant                  & $10^8$             \\ \hline
        $D_\mathrm{M}~(\mathrm{J/m^3})$    & DMI constant                       & $10^7 $            \\ \hline
        $K_\mathrm{u}~(\mathrm{J/m^3})$ & Uniaxial anisotropy constant       & $6.7 \times 10^5$  \\ \hline
        $M_\mathrm{s}~(\mathrm{A/m})$      & Saturation magnetization           & $4.46 \times 10^5$ \\ \hline
        $\delta_\mathrm{E}$              & Strain parameter                   & $-2.6\times10^{-4}$   \\ \hline
        $d_\mathrm{z}$~(nm)                    & Thickness                          & 8.8 \\ \hline
    
    \end{tabular}
    \label{tab:parameters}
\end{table}


From Eq.~(\ref{eq:energy phi}), strong exchange interactions constrain $\vb{m}_\mathrm{oct}$ to lie within the Kagome plane, resulting in two degenerate energy maxima at $\varphi_\mathrm{oct} = 0$ and $\pi$, and two minima at $\varphi_\mathrm{oct} = \pi/2$ and $3\pi/2$, in the absence of an external field. The equilibrium states near $\varphi_\mathrm{oct} = \pi/2$ and $3\pi/2$ are hereafter referred to as the up and down states, respectively. 
An applied external field breaks the symmetry of the local extrema. As shown in Fig.~\ref{fig:Energy}(a), when $\vb{H}_\mathrm{a}$ is applied along the $-x$ direction (\emph{i.e.}, $\varphi_\mathrm{H} = \pi$), the degeneracy of the energy maxima is lifted. For a film of volume $\mathcal{V}$, the 
energy barriers, $\Delta E_{\mathrm{x},1}$ and $\Delta E_{\mathrm{x},2}$, evaluated at $\varphi_\mathrm{oct} = 0$ and $\pi$, respectively, are given as
\begin{equation}
    \Delta E_\mathrm{x, 1(2)} = \mathcal{V} \Big[ \frac{3}{2}M_\mathrm{s} H_\mathrm{K} \pm 3M_\mathrm{s} H_\mathrm{x,eff} +\frac{3}{2} M_\mathrm{s} \frac{H_\mathrm{x,eff}^2}{H_\mathrm{K}} \Big],
    \label{eq:barrier x}
\end{equation}
where $H_\mathrm{x,eff} = H_\mathrm{ax}\frac{(K_\mathrm{u}+J_\mathrm{E}\delta_\mathrm{E})}{3(J_\mathrm{E}+\sqrt{3}D_\mathrm{M})}$ is the normalized external field applied along the $-x$ direction {with $H_\mathrm{ax}$ being the magnitude of the external field applied along $-x$ axis}. In this configuration, the up and down states remain energetically equivalent; however, thermal fluctuations preferentially promote transitions across $\Delta E_\mathrm{x,2}$ due to its lower energy barrier. {Furthermore, the two-state system reduces to one stable state at $\varphi_\mathrm{oct} = \pi$ as the magnitude of $H_\mathrm{ax}$ increases up to a certain magnitude, which for the parameters chosen here is 0.25\,T.}

In contrast, when the external field is applied along the $-y$ direction, it primarily perturbs the energy minima, as shown in Fig.~\ref{fig:Energy}(b). The energy maxima remain degenerate, resulting in symmetric energy barriers on either side of each state. The analytical expressions for the energy barriers $\Delta E_\mathrm{y,1}$ and $\Delta E_\mathrm{y,2}$, corresponding to the up and down states, respectively, are
\begin{equation}
    \Delta E_\mathrm{y, 1(2)} = \mathcal{V} \Big[ \frac{3}{2}M_\mathrm{s}H_\mathrm{K} \mp 3M_\mathrm{s}H_\mathrm{y,eff} +\frac{3}{2}M_\mathrm{s} \frac{H_\mathrm{y,eff}^2}{H_\mathrm{K}} \Big],
    \label{eq:barrier y}
\end{equation}
where $H_\mathrm{y,eff} = H_\mathrm{ay}\frac{(K_\mathrm{u}-J_\mathrm{E}\delta_\mathrm{E})}{3(J_\mathrm{E}+\sqrt{3}D_\mathrm{M})}$ denotes the normalized external field applied along the $-y$ direction {with $H_\mathrm{ay}$ being the magnitude of the external field applied along $-y$ axis}. {When $H_\mathrm{ay}\geq 0.23~\mathrm{T}$, $\Delta E_\mathrm{y,1}$ vanishes and the two-state system reduces to one stable state at $\varphi_\mathrm{oct} = 3\pi/2$.} For comparison, Fig.~\ref{fig:Energy}(c) quantitatively illustrates the quadratic dependence of the energy barrier density on the magnitude of the external field. Additionally, we define $\Delta E_0$ as the energy barrier height in the absence of an external field, which serves as a reference in subsequent analyses.

Based on the harmonic transition-state theory (HTST), {a a widely accepted framework for analyzing thermal behavior in ferromagnets}~\cite{coffey2012thermal,morse1946methods} {and collinear antiferromagnets},~\cite{rozsa2019reduced}
the escape time with intermediate-to-high (IHD) damping is formulated per Eq.~(\ref{eq:ihd}). In this equation,
%
%
$\varepsilon_{j,\mathrm{min(sp)}}$ denotes the $j$-th eigenvalue of the Hessian matrix in the harmonic approximation near the energy minimum (saddle point), and $\lambda_+$ is the positive eigenvalue of the linearized Landau-Lifshitz-Gilbert (LLG) equation at the saddle point. $P_\mathrm{min(sp)}$ represents the number of Goldstone modes at the minimum (saddle point), and $V_\mathrm{min(sp)}$ denotes the corresponding phase-space volume. 
Additionally, $A(\delta E/k_\mathrm{B}T)$ is the depopulation factor that accounts for the coupling strength to the thermal bath, where $\delta E$ is the energy dissipated along an iso-energy contour encompassing the energy maximum. 
See Eq. (\ref{eq:Ax}) for $A(x)$. 

In the theoretical analysis of escape time, 
we employ the angular variables $\theta_\mathrm{oct}$ and $\varphi_\mathrm{oct}$ to describe the dynamics of the octupole moment $\vb{m}_\mathrm{oct}$ within a perturbative framework.~\cite{he2024magnetic} 
The escape time due to thermal fluctuations over a single saddle point, from the up state to the down state under an external field $H_\mathrm{ax}$ or $H_\mathrm{ay}$ is given in Eq.~(\ref{eq:tau_esc}), where $k = 1$ and $k = 2$ correspond to the magnetic octupole crossing the saddle point near $\varphi_\mathrm{oct} = 0$ and $\pi$, respectively.
The parameters $\beta_x$ and $\beta_y$ in Eq. (\ref{eq:tau_esc}) are given as $\left((-1)^k \frac{H_\mathrm{x,eff}}{H_\mathrm{K}} - 1\right)$ and $\left(3 \frac{H_\mathrm{y,eff}^2}{H_\mathrm{K}^2} - 1\right)$, respectively. 
{The derivation of Eqs.~(\ref{eq:ihd})-(\ref{eq:tau_esc}) is provided in Sec.~\ref{sec:esc derive} of the Supplementary Material.}
Our calculations suggest that for the magnetic fields and energy barriers considered in this work, $\qty[A(\delta E / k_\mathrm{B}T)]^{-1}$ can be approximated to unity with less than
{1}\% error {as shown in Sec.~\ref{sec:depopulation} of the Supplementary Material.}

\begin{figure*}
    \begin{widetext}
    \begin{equation} 
    \tau_\mathrm{esc} = \left[A\left(\frac{\delta E}{k_\mathrm{B}T}\right)\right]^{-1}\frac{2\pi}{\lambda_+}\frac{V_\mathrm{min}}{V_\mathrm{sp}}(2\pi k_\mathrm{B}T)^{\frac{P_\mathrm{sp}-P_\mathrm{min}}{2}} \sqrt{\frac{\Pi'_j|\varepsilon_{j,\mathrm{,sp}}|}{\Pi'_j\varepsilon_{j,\mathrm{min}}}} e^{\frac{\Delta E}{k_\mathrm{B}T}}.
    \label{eq:ihd}
    \end{equation}

    \begin{equation}
    A(x) = \exp\left(\frac{1}{2\pi}\int_{-\infty}^{\infty} \ln\left(1 - e^{-x(0.25 + y^2)}\right) \frac{1}{0.25 + y^2} \, dy\right).
    \label{eq:Ax}
    \end{equation}
    
    \begin{subequations} \label{eq:tau_esc}
        \begin{flalign}
            &\tau_\mathrm{escx,k} = \qty[A\qty(\frac{\delta E}{k_\mathrm{B}T})]^{-1} \frac{4\pi(1+\alpha^2) \exp(\Delta E_\mathrm{x,k}/k_\mathrm{B}T)}{\gamma \sqrt{|H_\mathrm{x,eff} - (-1)^{k+1} H_\mathrm{K}|}}  \times \frac{\sqrt{H_\mathrm{K}}}{-\alpha \left[ H_J+\beta_x H_K  \right]+\sqrt{\alpha^2\left( H_\mathrm{J}-\beta_x H_\mathrm{K} \right)^2-4\beta_xH_\mathrm{J} H_\mathrm{K}}}. \\
            &\tau_\mathrm{escy,k} = \qty[A\qty(\frac{\delta E}{k_\mathrm{B}T})]^{-1} \frac{4\pi(1+\alpha^2) \exp(\Delta E_\mathrm{y,1}/k_\mathrm{B}T)}{\gamma H_K} \times \frac{\sqrt{| H_\mathrm{y,eff}^2-H_K^2|}}{-\alpha \left[ H_J+\beta_y H_K  \right]+\sqrt{\alpha^2\left( H_J-\beta_y H_K \right)^2-4\beta_yH_J H_K}}.
        \end{flalign}
    \end{subequations}

    \begin{equation}
        \frac{1}{\tau_{\mathrm{escx(y),}\uparrow }} = \frac{1}{\tau_\mathrm{escx(y),1}} + \frac{1}{\tau_\mathrm{escx(y),2}}.
        \label{eq:tau up}
    \end{equation}
\refstepcounter{equation}  
\setcounter{parentequation}{\value{equation}}  
\addtocounter{equation}{-1}  

    \begin{subequations}
        \noindent
        \begin{minipage}{0.45\linewidth}
        \begin{equation}
        \frac{dn_{\uparrow}}{dt} = -\frac{n_{\uparrow}}{\tau_{\mathrm{esc,}\uparrow }} +\frac{n_{\downarrow}}{\tau_{\mathrm{esc,}\downarrow}}. \tag{9a}
        \end{equation}
        \end{minipage}
        \hfill
        \begin{minipage}{0.45\linewidth}
        \begin{equation}
        \frac{dn_{\downarrow}}{dt} = \frac{n_{\uparrow}}{\tau_{\mathrm{esc,}\uparrow}} -\frac{n_{\downarrow}}{\tau_{\mathrm{esc,}\downarrow}}. \tag{9b}
        \end{equation}
        \end{minipage}
    \label{eq:n diff}
    \vspace{10pt}
    \end{subequations}

    \begin{equation}
        \begin{aligned}
            m_\mathrm{oct,y}(t) &= \frac{\tau_{\mathrm{esc,}\uparrow }-\tau_{\mathrm{esc,}\downarrow}}{\tau_{\mathrm{esc,}\uparrow } + \tau_{\mathrm{esc,}\downarrow}}+ 2 \left[ m_\mathrm{oct,y}(0)- \frac{\tau_{\mathrm{esc,}\uparrow}}{\tau_{\mathrm{esc,}\uparrow} + \tau_{\mathrm{esc,}\downarrow}} \right ] \exp\qty(-\frac{\tau_{\mathrm{esc,}\uparrow}+\tau_{\mathrm{esc,}\downarrow}}{\tau_{\mathrm{esc,}\uparrow} \tau_{\mathrm{esc,}\downarrow}})t.
        \end{aligned}
        \label{eq:moct tempo}
    \end{equation}

    \end{widetext}
\end{figure*}


As indicated by Eq.~(\ref{eq:tau_esc}), under an external field applied along the $y$-axis, the escape times of the magnetic octupole in the up state over the two saddle points to the down state are identical due to the preserved symmetry of the energy maxima. In contrast, an external field applied along the $x$-axis introduces asymmetry between the energy maxima, resulting in different escape times for the magnetic octupole over the two barriers. 
For the magnetic octupole in the down state, the escape time over the two barriers to the up state can be evaluated using Eq.~\ref{eq:tau_esc}(a), when the magnetic field is applied along the $x$-axis. However, for the magnetic field along the $y$-axis, $\Delta E_\mathrm{y, 1}$ in Eq.~\ref{eq:tau_esc}(b) should be replaced by $\Delta E_\mathrm{y, 2}$ for an accurate definition of the escape time. 
Assuming the transitions from the up (down) state to the down (up) state over the two barriers to be independent, the total escape time satisfies the relationship in Eq.~(\ref{eq:tau up}).




\begin{figure*}
\vspace{-10pt}
    \centering
    \includegraphics[width=\columnwidth]{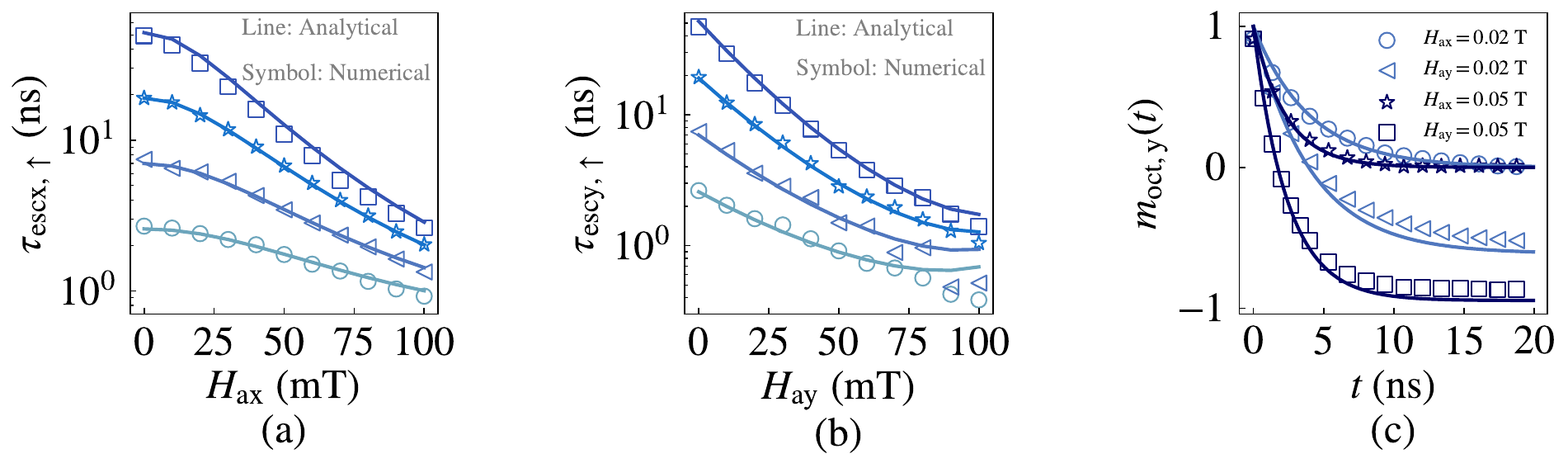}
    \vspace{-10pt}
    \caption{The schematic of escape time from up state to down state while numerical extracted data is in markers and analytical predictions are plotted with lines. $\Delta E_0$ ranges from ${3}k_\mathrm{B}T$ to ${6}k_\mathrm{B}T$ and field is applied along (a) $x$-axis and (b) $y$-axis. (c) The time evolution of octupole moment with external field and $\Delta E_0 = 4k_\mathrm{B}T$.}
    \label{fig:Tau}
    \vspace{-5pt}
\end{figure*}

To validate Eqs.~(\ref{eq:tau_esc}) and~(\ref{eq:tau up}), the dynamics of the magnetic octupole is numerically explored via three coupled stochastic LLG equations,~\cite{mayergoyz2009nonlinear, go2022noncollinear} for $\Delta E_0$ ranging from $k_\mathrm{B}T$ to $5k_\mathrm{B}T$ and $H_\mathrm{a} \leq 0.1~\mathrm{T}$. 
For each sublattice, $i$~(= 1,2,3), the LLG equation is given as $\pdv{\vb{m}_i}{t} = -\gamma (\vb{m}_i \cross \vb{H}_i^{\mathrm{eff}}) +\alpha \qty(\vb{m}_i \times \pdv{\vb{m}_i}{t})$,
where $\vb{H}_i^{\mathrm{eff}} = -\frac{1}{M_\mathrm{s}}\frac{\partial \mathcal{H}}{\partial \vb{m}_i} + \vb{H}_{\mathrm{thermal},i}$ is the effective magnetic field and includes the thermal noise field, $\vb{H}_{\mathrm{thermal},i} = \sqrt{\frac{2\alpha k_\mathrm{B} T}{\gamma \mu_0 M_\mathrm{s}^2 \mathcal{V} \Delta t}} \boldsymbol{\xi}^i_t$. 
In this work, we assume $\vb{H}_{\mathrm{thermal},i}$ to be Gaussian with zero mean as well as uncorrelated in space, time, and between sublattices, such that 
$\boldsymbol{\xi}^i_t \sim \mathcal{N}(0,1)$.~\cite{go2022noncollinear, ament2016solving}
Here, $\gamma$ and $\mu_0$ denote the gyromagnetic ratio and vacuum permeability, respectively, while $\Delta t$ is the time step used in the simulation. Other parameters are defined in Table~\ref{tab:parameters}. {The coupled LLG equations are solved using the finite difference method implemented within a self-developed CUDA framework, which has been benchmarked against published work with~\cite{konakanchi2025electrically} and without~\cite{shukla2024impact} thermal noise. Additional details of the numerical LLG framework, its validation, and numerical extraction of escape time are provided in Sec.~\ref{sec:numerical} of the Supplementary Material.}


In Figs.~\ref{fig:Tau}(a) and~\ref{fig:Tau}(b), the numerically computed escape time from up state to down state as a function of $H_\mathrm{ax}$ and $H_\mathrm{ay}$, for five different $\Delta E_0$, shown using symbols, is compared to the analytic predictions of 
Eq.~(\ref{eq:tau_esc}) and~(\ref{eq:tau up}), represented by solid lines.
We find that, for both directions of the applied field, the analytical predictions agree very well with the numerical data for $\Delta E_0 \geq 
{3}k_\mathrm{B}T$, though a small discrepancy emerges at higher fields ($H_\mathrm{a} \gtrsim 75~\mathrm{mT}$) and increases with both $H_\mathrm{ax}$ and $H_\mathrm{ay}$. 
{This discrepancy arises from a reduction in the effective barrier heights ($\Delta E_\mathrm{x,1}$ and $\Delta E_\mathrm{y,1}$) below the valid limits of HTST. In addition, increasing the field may further amplify the deviation of Eq.~(2) from Eq.~(1), ultimately leading to inaccuracies in $\tau_\mathrm{esc}$. A detailed discussion of this issue is provided in Sec.~\ref{sec:deviation} of the Supplementary Material.
}

For the two directions of magnetic fields, the escape time appears to be more sensitive to $H_\mathrm{ay}$ in the low-field regime.
This is because an increase in $H_\mathrm{ay}$ decreases $\Delta E_\mathrm{y, 1}$, favoring the octupole transitions from the up state to the down state over both the saddle points equally, thereby speeding up the process. On the other hand, when $H_\mathrm{ax}$ increases above zero, thermal field-driven transitions from the up state to the down state are comparatively slower as only $\Delta E_\mathrm{x, 2}$ decreases but  $\Delta E_\mathrm{x, 1}$ increases.  
{The sensitivity of escape time to various magnetic parameters is quantified in Sec.~\ref{sec:sensitivity} of the Supplementary Material.}
Although not shown here, we find $\tau_\mathrm{escx, \downarrow}$ to be same as the results shown in Fig.~\ref{fig:Tau}(a), but $\tau_\mathrm{escy, \downarrow}$ increases exponentially with $H_\mathrm{ay}$ due to increase in $\Delta E_\mathrm{y, 2}$ {as shown in Sec.~\ref{sec:damping} of the Supplementary Material.} 

The thermal relaxation in Mn$_3$Sn arises from the collective dynamics of multiple magnetic moments transitioning back and forth between two energy wells. The temporal evolution of the spin population is governed by the rate equations~\cite{brown1963thermal,brown1979thermal} specified in Eq.~(\ref{eq:n diff}),
where $n_{\uparrow}$ and $n_{\downarrow}$ denote the number of spins in the up and down states, respectively. Given the constraint $n_{\uparrow} + n_{\downarrow} = N_\mathrm{total}$ and the definition of the octupole moment's $y$ component as $m_\mathrm{oct,y} = \frac{(n_{\uparrow} - n_{\downarrow})}{N_\mathrm{total}}$, the time evolution of $m_\mathrm{oct,y}$ is given in Eq.~(\ref{eq:moct tempo}),
where $m_\mathrm{oct,y}(0)$ denotes the initial octupole moment. From this expression, the thermal equilibrium state $m_\mathrm{oct,y}(\infty)$ is given by 
$m_\mathrm{oct,y}(\infty) = \frac{\tau_{\mathrm{esc,}\uparrow }-\tau_{\mathrm{esc,}\downarrow}}{\tau_{\mathrm{esc,}\uparrow } + \tau_{\mathrm{esc,}\downarrow}}.$
In particular, when the external field is applied along the $x$-axis, the escape times become equal, $\tau_{\mathrm{esc,}\uparrow} = \tau_{\mathrm{esc,}\downarrow}$, leading to a vanishing steady-state magnetization, \emph{i.e.}, $m_\mathrm{oct,y}(\infty) = 0$. On the other hand, as the external field along the $y$-axis increases, thermal transitions in one direction become more favorable than the reverse process, leading to a well defined steady-state magnetization, \emph{i.e.}, $m_\mathrm{oct,y}(\infty) \to \pm1$.   
The relaxation time associated with this two-state system is 
$ \tau_\mathrm{relax} = \frac{\tau_{\mathrm{esc,}\uparrow} \, \tau_{\mathrm{esc,}\downarrow}}{\tau_{\mathrm{esc,}\uparrow} + \tau_{\mathrm{esc,}\downarrow}}$,
which characterizes the timescale over which $m_\mathrm{oct,y}$ decays to $1/e$ of its initial deviation from equilibrium. The temporal evolution of $m_\mathrm{oct,y}$ for $\Delta E_0 = 4k_\mathrm{B}T$, as shown in Fig.~\ref{fig:Tau}(c), illustrates the dependence of the relaxation time and steady-state magnetization on the magnitude and direction of the external field.

\begin{figure}
    \centering
    \includegraphics[width=0.8\columnwidth]{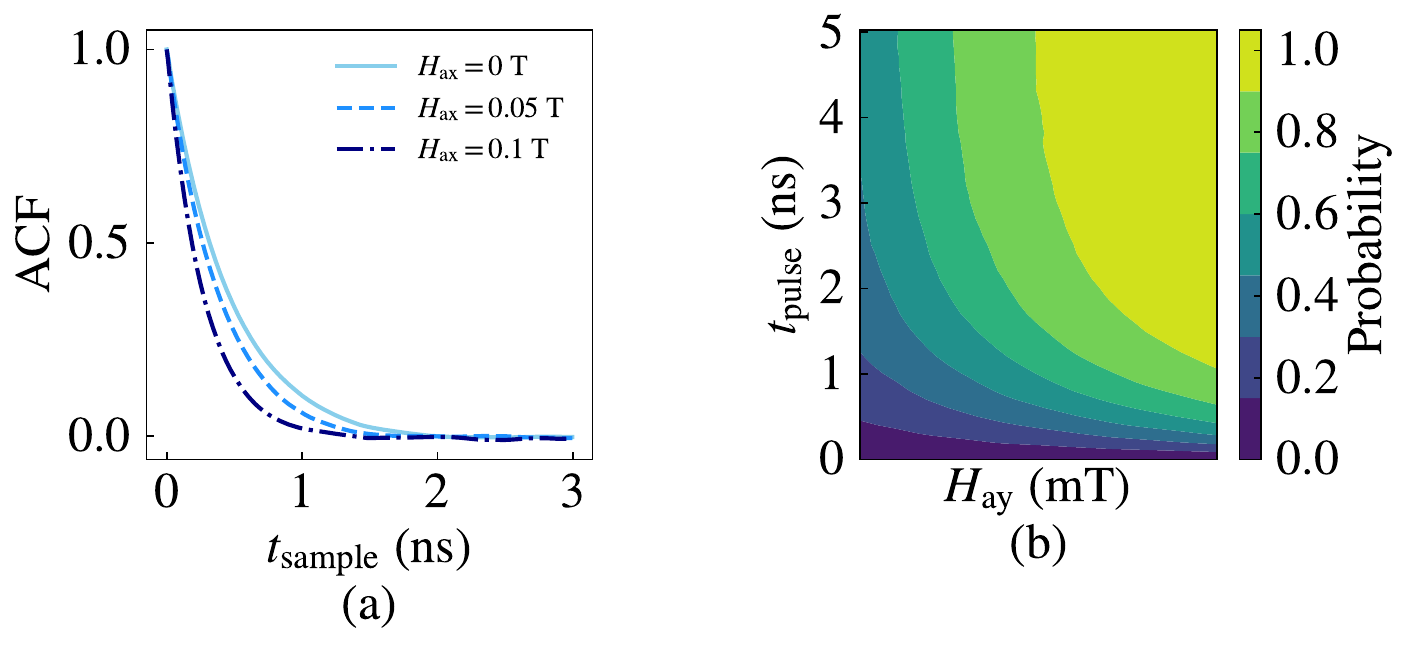}
    \vspace{-10pt}
    \caption{(a) The ACF as a function of sample interval with $x$-axis external field. (b) The probability of finding the octupole moment in the down state at steady state  with a sweep of $H_\mathrm{ay}$ and pulse width.}
    \label{fig:application}
\end{figure}

The distinct thermal fluctuation mechanisms provide valuable insights into the prospective applications of $\mathrm{Mn_3Sn}$. When an external field is applied along the $x$-axis, the two equilibrium states remain symmetric, resulting in equal occupation probabilities. This inherent symmetry renders the system a promising candidate for random number generation. As shown in Fig.~\ref{fig:application}(a), for a single $\mathrm{Mn_3Sn}$-based magnetic tunnel junction, the autocorrelation function (ACF) of the octupole moment decays to zero with a 2-ns sampling interval. Moreover, increasing the magnitude of the applied field effectively reduces the required sampling interval, whereas the typical sampling interval for ferromagnetic random number generators is on the order of $\mu s$ or even higher.~\cite{wang2022single,chen2022magnetic} In contrast, when the external field is applied along the $y$-axis, the symmetry between the two equilibrium states is broken, enabling probabilistic control of the octupole moment $\vb{m}_\mathrm{oct}$. Figure~\ref{fig:application}(b) presents the temporal evolution of the probability of occupying the down state under various field magnitudes, thereby confirming the effectiveness of $H_\mathrm{ay}$ in modulating the switching probability.

In conclusion, we comprehensively demonstrated the external field-assisted thermal behaviors of $\mathrm{Mn_3Sn}$ {for energy barriers in the range of $(3-6)k_\mathrm{B}T$}. The energy landscape symmetry breakdown is dependent on the field orientation, in which case the theoretical prediction of escape time varies and contributes to two different relaxation modes. The two modes provide prospective applications in random number generation and probabilistic computing while the frequency can reach above GHz. This work focuses on low barrier magnets~{$\left((3-6)k_\mathrm{B}T\right)$} and low external field~($\leq 0.1~\mathrm{T}$), which is the effective regime of perturbation theory. However, current experiments are reported for high barrier regime~($\geq 40k_\mathrm{B}T$) and the switching field is up to $0.5~\mathrm{T}$ so our future work will explore the thermal behaviors of $\mathrm{Mn_3Sn}$ within this regime. 
Our work highlights the relevance of thermal noise in establishing the physical understanding and operating principles of scaled $\mathrm{Mn_3Sn}$ devices and enable applications such as random number generation and probabilistic computing. 

\vspace{-10pt}
\section*{Supplementary Material}
\vspace{-10pt}
See the supplementary material for a derivation of the escape time of the octupole, numerical LLG simulation, and regimes of model validity.

\vspace{-10pt}
\section*{Acknowledgment}
\vspace{-10pt}
This research was supported by the NSF through the University of Illinois at Urbana-Champaign Materials Research Science, Engineering Center DMR-1720633. 

\vspace{-10pt}
\section*{Author Declarations}
\vspace{-10pt}
\subsection*{Conflict of Interest}
\vspace{-10pt}
The authors have no conflicts to disclose.
\vspace{-10pt}
\subsection*{Author Contributions}
\vspace{-10pt}
\textbf{Siyuan Qian}: Conceptualization (equal); Data curation (lead);
Formal analysis (lead); Investigation (lead); Methodology (lead);
Software (lead); Validation (lead); Writing – original draft (lead);
Writing – review \& editing (supporting).
\textbf{Ankit Shukla}: Data curation
(supporting); Formal analysis (supporting); Software (supporting), Writing – review \& editing (lead).
\textbf{Shaloo Rakheja}: Conceptualization (lead); Funding acquisition
(lead); Project administration (lead); Resources (lead); Supervision
(lead); Writing – original draft (supporting); Writing – review \&
editing (supporting).

\section*{Data Availability}
\vspace{-10pt}
The data that support the findings of this study are available
from the corresponding author upon reasonable request.


\newpage

\section{Supplementary Material}

{\subsection{Analytical derivation of escape time}
\label{sec:esc derive}
}

{
Following steps must be implemented to derive Eq.~(5) in the main text.:
\begin{itemize}
    \item Linearizing the equations describing the motion of $\vb{m}_\mathrm{oct}$ to get the only positive eigenvalue $\lambda_+$ at the saddle point.
    \item Evaluating the Hessian matrix to obtain the eigenvalues $\varepsilon_{\mathrm{j,min}}$ at the energy minimum and $\varepsilon_{\mathrm{j,sp}}$ at the saddle point. The number of eigenvalues equal to zero determines the number of Goldstone states and the associated parameters such as $V_\mathrm{min}$, $P_\mathrm{min}$, $V_\mathrm{sp}$, and $P_\mathrm{sp}$.
    \item Calculating the depopulation factor $A(\frac{\delta E}{k_\mathrm{B}T})$ through numerical analysis. 
\end{itemize}
}

{
\subsubsection{The Landau-Lifshitz-Gilbert (LLG) equation}
In preparation for the linearization of the LLG equation describing the motion of $\vb{m}_\mathrm{oct}$, we first derive the equations directly in terms of $\vb{m}_\mathrm{oct} = (\sin \theta_\mathrm{oct}\cos \varphi_\mathrm{oct}, \sin \theta_\mathrm{oct} \sin \varphi_\mathrm{oct}, \cos \theta_\mathrm{oct})$ where $\varphi_\mathrm{oct}$ and $\theta_\mathrm{oct}$ are the azimuthal and polar angles of the octupole moment as depicted in Fig.~1(b) in the main text; therefore the equations of motion are explicitly dependent on 
$\theta_\mathrm{oct}$ and $\varphi_\mathrm{oct}$.

The LLG equation for a magnetic moment vector $(\hat{r},\hat{\theta},\hat{\varphi})$ in the spherical coordinates is given as
\begin{subequations}
    \begin{flalign}
        \dot{\theta} &= \frac{\gamma}{1+\alpha^2} \left( H_{\varphi} +\alpha H_{\theta} \right), \\
        \dot{\varphi} &= \frac{\gamma}{1+\alpha^2} \left( -H_{\theta} + \alpha H_{\varphi} \right) \csc \theta,
    \end{flalign}
    \label{eq:LLG spherical}
\end{subequations}
where $|\hat{r}|\equiv1$ and the effective magnetic field in spherical coordinate is obtained from the Hamiltonian, $\mathcal{H}$, as 
\begin{subequations}
    \begin{flalign}
        H_\varphi &= -\frac{1}{M_\mathrm{s}\sin \theta}\frac{\partial \mathcal{H}}{\partial \varphi}, \\
        H_\theta &=  -\frac{1}{M_\mathrm{s}}\frac{\partial \mathcal{H}}{\partial \theta}. 
    \end{flalign}
\end{subequations}
In these equations, $\gamma$, $\alpha$, and $M_\mathrm{s}$ represent the gyromagnetic ratio, damping coefficient, and saturation magnetization, respectively, of the magnetic material.

In the case of 
$\mathrm{Mn_3Sn}$, we model the dynamics of the magnetic octupole moment, which is defined as 
\begin{equation}
    \vb{m}_\mathrm{oct} = \frac{1}{3} \mathcal{M}_{zx}\left[R\left(\frac{2\pi}{3}\right) \vb{m}_1 + R\left(-\frac{2\pi}{3}\right) \vb{m}_2 + \vb{m}_3\right],
\end{equation}
where $\mathcal{M}_{zx}$ is the mirror operation against $zx$ plane and $\mathcal{R}$ is the rotation operation in $xy$ plane, so $\varphi$ in Eq.~(\ref{eq:LLG spherical}) is replace by $-\varphi$. Owing to the strong exchange interaction between the sublattice vectors, $\vb{m}_i$ (with $i=1, 2, 3$), $\mathrm{\theta}_\mathrm{oct}$ is restricted around $\pi/2$ and, therefore, the LLG equation and the effective field for $\vb{m}_\mathrm{oct}$ is
\begin{subequations}
    \begin{flalign}
        \dot{\theta}_\mathrm{oct} &= \frac{\gamma}{1+\alpha^2} \left( H_{\varphi}(-\varphi_\mathrm{oct}) +\alpha H_{\theta} \right), \\
        \dot{\varphi}_\mathrm{oct} &= \frac{\gamma}{1+\alpha^2} \left( -H_{\theta} + \alpha H_{\varphi}(-\varphi_\mathrm{oct}) \right).
    \end{flalign}
    \label{eq:LLG moct general}
\end{subequations}
To average the joint contribution of the three sublattices, the effective field is scaled by a factor of $\frac{1}{3}$,~\cite{he2024magnetic, konakanchi2025electrically} leading to
\begin{subequations}
    \begin{flalign}
        H_\varphi &= -\frac{1}{3M_\mathrm{s}\sin \theta_\mathrm{oct}}\frac{\partial \mathcal{H}_\mathrm{oct}}{\partial \varphi_\mathrm{oct}}, \\
        H_\theta &=  -\frac{1}{3M_\mathrm{s} }\frac{\partial \mathcal{H}_\mathrm{oct}}{\partial \theta_\mathrm{oct}}.
    \end{flalign}
\end{subequations}
Further simplification yields
\begin{subequations}
    \begin{flalign}
        H_\varphi &= -\frac{1}{3M_\mathrm{s} \sin \theta_\mathrm{oct}}\frac{\partial \mathcal{H}_\mathrm{oct}}{\partial \varphi_\mathrm{oct}} = \frac{1}{\sin \theta_\mathrm{oct}}\Big[\frac{1}{2}H_\mathrm{K} \sin(2\varphi_\mathrm{oct})-\frac{H_\mathrm{a} \cos{(\theta_\mathrm{H})}}{3(J_\mathrm{E}+\sqrt{3}D_\mathrm{M})}\Big[ K_\mathrm{u} \sin{(\varphi_{\mathrm{oct}}-\varphi_{\mathrm{H}})}+J_\mathrm{E}\delta_E\sin{(\varphi_{\mathrm{oct}}+\varphi_{\mathrm{H}})}\Big]\Big], \\
        H_\theta &=  -\frac{1}{3M_\mathrm{s}}\frac{\partial \mathcal{H}_\mathrm{oct}}{\partial \theta_\mathrm{oct}} = H_\mathrm{J}\cos{\theta_\mathrm{oct}} \sin{\theta_\mathrm{oct}} .
    \end{flalign}
    \label{eq: H detail}
\end{subequations}
}

\newpage
{\subsubsection{Escape time with $H_\mathrm{ay}$}}

{With $\varphi_\mathrm{H} = 3 \pi /2$, the Hamiltonian {from Eq.~(2) in the main text} is
\begin{equation} \label{eq:H1}
    \mathcal{H}_\mathrm{oct} = \frac{3}{4}M_\mathrm{s} H_\mathrm{K} \cos(2\varphi_\mathrm{oct}) + \frac{3}{2}M_\mathrm{s} H_\mathrm{J} \cos^2 \theta_\mathrm{oct} +3 M_\mathrm{s}H_\mathrm{y,eff} \sin \varphi_\mathrm{oct},
\end{equation}
in which case local minimum is at $\varphi_\mathrm{oct} = \pi/2$ and $3\pi/2$, while local maximum is at $\sin \varphi_\mathrm{oct} = \frac{H_\mathrm{y,eff}}{H_\mathrm{K}}$.}

\vspace{10pt}
\begin{itemize}
    \item \textbf{Eigenvalue of linearized LLG}
\end{itemize}



{Substituting Eq.~(\ref{eq:H1}) into Eq.~(\ref{eq:LLG moct general}) and Eq.~(\ref{eq: H detail}), the LLG equation of $\vb{m}_\mathrm{oct}$ is written as}
\begin{subequations}
    \begin{flalign}
        \dot{\theta}_{\mathrm{oct}} &= \frac{\gamma}{1+\alpha^2} \left( H_\mathrm{K}\sin\varphi_\mathrm{oct}\cos\varphi_\mathrm{oct}-H_\mathrm{y,eff}\cos\varphi_\mathrm{oct} +\alpha H_J\cos \theta_{\mathrm{oct}} \sin \theta_{\mathrm{oct}} \right), \\
        \dot{\varphi}_{\mathrm{oct}} &= \frac{\gamma}{{(}1+\alpha^2{)\sin \theta_\mathrm{oct}}} \left( -H_J\cos \theta_{\mathrm{oct}} \sin \theta_{\mathrm{oct}} + \alpha H_\mathrm{K}\sin\varphi_\mathrm{oct}\cos\varphi_\mathrm{oct}-\alpha H_\mathrm{y,eff}\cos \varphi_\mathrm{oct} \right).
    \end{flalign}
\end{subequations}
The linearized LLG equation of $\vb{m}_\mathrm{oct}$ at the saddle point $\theta_\mathrm{oct} = \pi/2$ and $\sin \varphi_\mathrm{oct} = H_\mathrm{y,eff}/H_\mathrm{K}$ is

\begin{align}
\begin{pmatrix}
\delta \dot{\theta}_\mathrm{oct} \\
\delta \dot{\varphi}_\mathrm{oct}
\end{pmatrix}
= \frac{\gamma}{1+\alpha^2}
\begin{pmatrix}
-\alpha H_J &  \beta_y H_K\\
-H_J & -\alpha \beta_y H_K
\end{pmatrix}
\begin{pmatrix}
\delta \dot{\theta}_\mathrm{oct} \\
\delta \dot{\varphi}_\mathrm{oct}
\end{pmatrix},
\end{align}
where $\beta_y = 3\frac{H_\mathrm{y,eff}^2}{H_\mathrm{K}^2}-1$. The positive eigenvalue of the matrix is 
\begin{equation}
    \lambda_+ = \frac{\gamma}{(1+\alpha^2)}\frac{-\alpha \left[ H_\mathrm{J}+\beta_y H_\mathrm{K}  \right]+\sqrt{\alpha^2\left( H_\mathrm{J}-\beta_y H_\mathrm{K}  \right)^2-4\beta_y H_\mathrm{J} H_\mathrm{K}}}{2}.    
\end{equation}

\begin{itemize}
    \item \textbf{Eigenvalue of Hessian matrix}
\end{itemize}

The eigenvalues of the Hessian matrix at the minimum and saddle points (sp) are 
\begin{subequations}
    \begin{flalign}
        \varepsilon_\mathrm{1,min} &= \mathcal{V}\frac{\partial^2 \mathcal{H}_\mathrm{oct}}{\partial \theta_\mathrm{oct}^2}\Big|_{\theta_\mathrm{oct} = \pi/2,\varphi_\mathrm{oct}=\pi/2} = 3M_s H_J \mathcal{V}, \\
        \varepsilon_\mathrm{2,min} &= \mathcal{V}\frac{\partial^2 \mathcal{H}_\mathrm{oct}}{\partial \theta_\mathrm{oct}^2}\Big|_{\theta_\mathrm{oct} = \pi/2,\varphi_\mathrm{oct}=3\pi/2} = 3M_s H_J \mathcal{V}, \\
        \varepsilon_\mathrm{1,sp} &= \mathcal{V}\frac{\partial^2 \mathcal{H}_\mathrm{oct}}{\partial \theta_\mathrm{oct}^2}\Big|_{\theta_\mathrm{oct} = \pi/2,\varphi_\mathrm{oct}=\sin^{-1}\qty(\frac{H_\mathrm{y,eff}}{H_K})} = 3M_s H_J \mathcal{V}, \\
        \varepsilon_\mathrm{2,sp} &= \mathcal{V}\frac{\partial^2 \mathcal{H}_\mathrm{oct}}{\partial \theta_\mathrm{oct}^2}\Big|_{\theta_\mathrm{oct} = \pi/2,\varphi_\mathrm{oct}=\pi-\sin^{-1}\qty(\frac{H_\mathrm{y,eff}}{H_K})} = 3 M_s H_J \mathcal{V}, \\
        \varepsilon_\mathrm{3,min} &= \mathcal{V}\frac{\partial^2 \mathcal{H}_\mathrm{oct}}{\partial \varphi_\mathrm{oct}^2}\Big|_{\theta_\mathrm{oct} = \pi/2,\varphi_\mathrm{oct}=\pi/2}= 3M_s \mathcal{V}(H_K - H_\mathrm{y,eff}), \\
        \varepsilon_\mathrm{4,min} &= \mathcal{V}\frac{\partial^2 \mathcal{H}_\mathrm{oct}}{\partial\varphi_\mathrm{oct}^2}\Big|_{\theta_\mathrm{oct} = \pi/2,\varphi_\mathrm{oct}=3\pi/2}= 3M_s\mathcal{V}(H_K+H_\mathrm{y,eff}), \\
        \varepsilon_\mathrm{3,sp} &= \mathcal{V}\frac{\partial^2 \mathcal{H}_\mathrm{oct}}{\partial \varphi_\mathrm{oct}^2}\Big|_{\theta_\mathrm{oct} = \pi/2,\varphi_\mathrm{oct}=\sin^{-1}\qty(\frac{H_\mathrm{y,eff}}{H_K})}= 3 M_s \mathcal{V}\qty(\frac{H_\mathrm{y,eff}^2}{H_K} - H_K), \\
        \varepsilon_\mathrm{4,sp} &= \mathcal{V}\frac{\partial^2 \mathcal{H}_\mathrm{oct}}{\partial \varphi_\mathrm{oct}^2}\Big|_{\theta_\mathrm{oct} = \pi/2,\varphi_\mathrm{oct}=\pi-\sin^{-1}\qty(\frac{H_\mathrm{y,eff}}{H_K})}= 3M_s\mathcal{V}\qty(\frac{H_\mathrm{y,eff}^2}{H_K}-H_K),
    \end{flalign}
\end{subequations}
where $\mathcal{V}$ is the sample volume, and, therefore, we have
\begin{equation}
    \sqrt{\frac{\prod_j \left| \varepsilon_{j,\text{sp}} \right|}{\prod_j \varepsilon_{j,\min}}} = \sqrt{\frac{\qty|\qty(\frac{H_\mathrm{y,eff}}{H_K})^2-1|^2}{\qty|1-\qty(\frac{H_\mathrm{y,eff}}{H_K})^2|}} = \sqrt{\qty| \frac{H_\mathrm{y,eff}^2-H_K^2}{H_K^2}|}.
\end{equation}
In this work, we only analyze $H_\mathrm{y,eff}< H_\mathrm{K}$ so there is no zero eigenvalue, which indicates no Goldstone mode~\cite{rozsa2019reduced} in this system. So we have $P_\mathrm{sp} = P_\mathrm{min} = 0$ and $V_\mathrm{min}/V_\mathrm{sp} = 1$. 

Finally, the escape time with external field applied along $-y$ axis can be evaluated as 
\begin{equation}
    \begin{aligned}
        \tau_\mathrm{escy} &= \qty[A\qty(\frac{\delta E}{k_\mathrm{B}T})]^{-1}\frac{2\pi}{\lambda_+} \sqrt{\frac{\Pi'_j|\varepsilon_{j,\mathrm{,sp}}|}{\Pi'_j\varepsilon_{j,\mathrm{min}}}} e^{\frac{\Delta E}{k_\mathrm{B}T}}\\
        & = \qty[A\qty(\frac{\delta E}{k_\mathrm{B}T})]^{-1} \frac{4\pi(1+\alpha^2) \exp(\frac{\Delta E_\mathrm{y,1}}{k_\mathrm{B} T}) \sqrt{\qty| H_\mathrm{y,eff}^2-H_K^2|}}{\gamma H_K \qty(-\alpha \left[ H_J+\beta_y H_K  \right]+\sqrt{\alpha^2\left( H_J-\beta_y H_K \right)^2-4\beta_yH_J H_K})}.
    \end{aligned} 
    \label{eq:H2}
\end{equation}

{\subsubsection{Escape time with $H_\mathrm{ax}$}}

{With $\varphi_\mathrm{H} = \pi$, the Hamiltonian is
\begin{equation}
    \mathcal{H}_\mathrm{oct} = \frac{3}{4}M_\mathrm{s} H_\mathrm{K} \cos(2\varphi_\mathrm{oct}) + \frac{3}{2}M_\mathrm{s} H_\mathrm{J} \cos^2 \theta_\mathrm{oct} +3 M_\mathrm{s}H_\mathrm{x,eff} \cos \varphi_\mathrm{oct}.
\end{equation}}

\begin{itemize}
    \item \textbf{Eigenvalue of linearized LLG}
\end{itemize}

{Substituting Eq.~\ref{eq:H2} into Eq.~(\ref{eq:LLG moct general}) and Eq.~(\ref{eq: H detail}),the LLG equations of $\vb{m}_\mathrm{oct}$ is written as:}
\begin{subequations}
    \begin{flalign}
        \dot{\theta}_\mathrm{oct} &= \frac{\gamma}{1+\alpha^2} ( H_\mathrm{K} \sin\varphi_\mathrm{oct} \cos \varphi_\mathrm{oct} +  H_\mathrm{x,eff}\sin \varphi_\mathrm{oct} + \alpha H_\mathrm{J} \cos \theta_\mathrm{oct} \sin \theta_\mathrm{oct}), \\
        \dot{\varphi}_\mathrm{oct} &=  \frac{\gamma}{{(}1+\alpha^2{)\sin \theta_\mathrm{oct}}} (-H_\mathrm{J}\cos \theta_\mathrm{oct} \sin \theta_\mathrm{oct} + \alpha  H_\mathrm{K} \sin\varphi_\mathrm{oct} \cos \varphi_\mathrm{oct} +  \alpha H_\mathrm{x,eff}\sin \varphi_\mathrm{oct}).
    \end{flalign}
\end{subequations}

{Unlike the case with $H_\mathrm{ay}$, the two saddle points are not degenerate. Specifically, $\cos \varphi_\mathrm{oct} = 1$ and $-1$ correspond to two distinct saddle points at $\varphi_\mathrm{oct} = 0$ and $\pi$, respectively. As a result, the corresponding matrices and their eigenvalues must be evaluated separately for each saddle point.}

\begin{align}
\begin{pmatrix}
\delta \dot{\theta}_\mathrm{oct} \\
\delta \dot{\varphi}_\mathrm{oct}
\end{pmatrix}
= \frac{\gamma}{1+\alpha^2}
\begin{pmatrix}
-\alpha H_J &  \beta_x H_\mathrm{K} \\
-H_J & -\alpha \beta_x H_\mathrm{K}
\end{pmatrix}
\begin{pmatrix}
\delta \dot{\theta}_\mathrm{oct} \\
\delta \dot{\varphi}_\mathrm{oct}
\end{pmatrix},
\end{align}
\begin{equation}
    \lambda_+ = \frac{\gamma}{(1+\alpha^2)}\frac{-\alpha \left[ H_J+\beta_x H_K  \right]+\sqrt{\alpha^2\left( H_J-\beta_x H_K  \right)^2-4\beta_x H_J H_K}}{2}.
\end{equation}

{where $\beta_x$ is $\varphi_\mathrm{oct}$ dependent:
\begin{equation}
    \beta_x = 
    \begin{cases}
    -\frac{H_\mathrm{x,eff}}{H_\mathrm{K}}-1 \ \ \ &\varphi_\mathrm{oct} = 0 \\[10pt]
    \frac{H_\mathrm{x,eff}}{H_\mathrm{K}}-1 \ \ \ \ &\varphi_\mathrm{oct} = \pi
    \end{cases}
\end{equation}
}

\begin{itemize}
    \item \textbf{Eigenvalue of Hessian matrix}
\end{itemize}


{Because the saddle points are not identical, the eigenvalues of Hessian matrix with $H_\mathrm{ax}$ need to be evaluated separately:}

\begin{subequations}
    \begin{flalign}
        \varepsilon_\mathrm{1,min} &= \mathcal{\mathcal{V}}\frac{\partial^2 \mathcal{H}_\mathrm{oct}}{\partial \theta_\mathrm{oct}^2}\Big|_{\theta_\mathrm{oct} = \pi/2,\varphi_\mathrm{oct}=\cos^{-1}\qty(\frac{H_\mathrm{x,eff}}{H_\mathrm{K}})} = 3 M_s H_J \mathcal{V}, \\
        \varepsilon_\mathrm{1,sp} &= \mathcal{V}\frac{\partial^2 \mathcal{H}_\mathrm{oct}}{\partial \theta_\mathrm{oct}^2}\Big|_{\theta_\mathrm{oct} = \pi/2,\varphi_\mathrm{oct}=0{/\pi}} = 3 M_s H_J \mathcal{V}, \\
        \varepsilon_\mathrm{2,min} &= \mathcal{V}\frac{\partial^2 \mathcal{H}_\mathrm{oct}}{\partial \varphi_\mathrm{oct}^2}\Big|_{\theta_\mathrm{oct} = \pi/2,\varphi_\mathrm{oct}=\cos^{-1}\qty(\frac{H_\mathrm{x,eff}}{H_\mathrm{K}})}= 3 M_s \mathcal{V} H_\mathrm{K}\qty(1-\frac{H_\mathrm{x,eff}^2}{H_\mathrm{K}^2}), \\
        {\varepsilon_\mathrm{2,sp}} &{=} 
        \begin{cases}
            {\mathcal{V}\frac{\partial^2 \mathcal{H}_\mathrm{oct}}{\partial \varphi_\mathrm{oct}^2}\Big|_{\theta_\mathrm{oct} = \pi/2,\varphi_\mathrm{oct}=0}= -3 M_s \mathcal{V}(H_\mathrm{K}+H_\mathrm{x,eff})\ \ \ \varphi_\mathrm{oct} = 0,}\\
            {\mathcal{V}\frac{\partial^2 \mathcal{H}_\mathrm{oct}}{\partial \varphi_\mathrm{oct}^2}\Big|_{\theta_\mathrm{oct} = \pi/2,\varphi_\mathrm{oct}=\pi}= -3 M_s \mathcal{V}(H_\mathrm{K}-H_\mathrm{x,eff})\ \ \ \varphi_\mathrm{oct} = \pi,}
        \end{cases}
    \end{flalign}
\end{subequations}
which gives
{
\begin{equation}
    \sqrt{\frac{\prod_j \left| \varepsilon_{j,\text{sp}} \right|}{\prod_j \varepsilon_{j,\min}}} = 
    \begin{cases}
        \sqrt{\qty| \frac{H_\mathrm{K}\qty(H_\mathrm{K}+H_\mathrm{x,eff})}{H_\mathrm{K}^2-H_\mathrm{x,eff}^2} | } = \sqrt{\qty| \frac{H_\mathrm{K}}{H_\mathrm{K}-H_\mathrm{x,eff}}| },\\[20pt]
        \sqrt{\qty| \frac{H_\mathrm{K}(H_\mathrm{K}+H_\mathrm{x,eff})}{H_\mathrm{K}^2-H_\mathrm{x,eff}^2}| } = \sqrt{\qty| \frac{H_\mathrm{K}}{H_\mathrm{K}+H_\mathrm{x,eff}}| }
    \end{cases}
\end{equation}}

\noindent

Therefore, the escape time with $H_\mathrm{ax}$-assisted switching at saddle point $\varphi_\mathrm{oct} = 0$ is
\begin{equation}
    \begin{aligned}
        &\tau_\mathrm{escx,1} = \qty[A\qty(\frac{\delta E}{k_\mathrm{B}T})]^{-1} \frac{4\pi(1+\alpha^2) \exp(\Delta E_\mathrm{x,k}/k_\mathrm{B}T) \sqrt{H_\mathrm{K}}}{\gamma \sqrt{|H_\mathrm{x,eff} - H_\mathrm{K}|} \qty(-\alpha \left[ H_J+\beta_x H_K  \right]+\sqrt{\alpha^2\left( H_J-\beta_x H_K \right)^2-4\beta_xH_J H_K})},
    \end{aligned}
\end{equation}
and the escape time at the saddle point $\varphi_\mathrm{oct} = \pi$ is
\begin{equation}
    \begin{aligned}
        &\tau_\mathrm{escx,2} = \qty[A\qty(\frac{\delta E}{k_\mathrm{B}T})]^{-1} \frac{4\pi(1+\alpha^2) \exp(\Delta E_\mathrm{x,k}/k_\mathrm{B}T) \sqrt{H_\mathrm{K}}}{\gamma \sqrt{|H_\mathrm{x,eff} + H_\mathrm{K}|} \qty(-\alpha \left[ H_J+\beta_x H_K  \right]+\sqrt{\alpha^2\left( H_J-\beta_x H_K \right)^2-4\beta_xH_J H_K})}.
    \end{aligned}
\end{equation}

\newpage
\subsubsection{Analysis of depopulation factor}
\label{sec:depopulation}

To evaluate the depopulation factor, we define $\delta E = \delta \mathcal{F}_\mathrm{oct} \mathcal{V}$ where $\delta \mathcal{F}_\mathrm{oct}$ represents the energy density dissipated over an equal free energy contour containing the energy maximum. $\delta \mathcal{F}_\mathrm{oct}$ as a function of out-of-plane component $m_\mathrm{oct,z}$ and in-plane polar angle $\varphi_\mathrm{oct}$ is given as~\cite{rozsa2019reduced,konakanchi2025electrically}
\begin{equation}
    \delta \mathcal{F}_{\text{oct}} = \int_{\varphi_{\text{oct}}} {-\alpha}\left( \frac{\partial \mathcal{F}_{\text{oct}}}{\partial m_{\mathrm{oct},z}} \right)
     \, d\varphi_{\text{oct}} + \int_{m_{\mathrm{oct},z}} {\alpha}\left( \frac{\partial \mathcal{F}_{\text{oct}}}{\partial \varphi_{\text{oct}}} \right)\, dm_{\mathrm{oct},z}.
    \label{eq:energy dissipation}
\end{equation}

{{In order to calculate the integration numerically, we have to get the $m_\mathrm{oct,z}-\varphi_\mathrm{oct}$ relationship and make the integration $\varphi_\mathrm{oct}$-dependent only.}}

\begin{itemize}
    \item \textbf{Depopulation factor with $H_\mathrm{ay}$}
\end{itemize}


{The energy of the system remains constant in the equal free energy contour, so we express the Hamiltonian at the saddle point$\Big(\theta_\mathrm{oct} = \pi/2,\varphi_\mathrm{oct} = \sin^{-1}(\frac{H_\mathrm{y,eff}}{H_\mathrm{K}}) / \pi-\sin^{-1}(\frac{H_\mathrm{y,eff}}{H_\mathrm{K}}) \Big)$ }:
\begin{equation}
    \begin{aligned}
        \mathcal{H}_\mathrm{oct}(H_\mathrm{ay}) &= \frac{3}{4}M_\mathrm{s} H_\mathrm{K} \cos(2\varphi_\mathrm{oct}) + \frac{3}{2}M_\mathrm{s} H_\mathrm{J} m^2_z +3 M_\mathrm{s}H_\mathrm{y,eff} \sin \varphi_\mathrm{oct}\\
        &= \frac{3}{4} M_\mathrm{s} H_\mathrm{K} -\frac{3}{2}M_\mathrm{s} H_\mathrm{K} \qty(\frac{H_\mathrm{y,eff}}{H_\mathrm{K}})^2 + 3M_\mathrm{s} H_\mathrm{y,eff} \frac{H_\mathrm{y,eff}}{H_\mathrm{K}},
    \end{aligned}
\end{equation}
and, therefore, 
{the $m_{\mathrm{oct},z}-\varphi_\mathrm{oct}$ relationship is given as:}
\begin{equation}
    m_{\mathrm{oct},z} = \sqrt{\frac{H_\mathrm{K}^2 \sin^2\varphi_\mathrm{oct} +H_\mathrm{y,eff}^2 -2H_\mathrm{K} H_\mathrm{y,eff} \sin \varphi_\mathrm{oct}   }{H_\mathrm{J} H_\mathrm{K}}}.
\end{equation}
Consequently, the integration in Eq.~(\ref{eq:energy dissipation}) is only dependent on $\varphi_\mathrm{oct}$, which gives
\begin{equation}
    \begin{aligned}
        \delta \mathcal{F}_\mathrm{oct} &= \int_{\pi/2}^{\pi} -3 \alpha M_\mathrm{s} H_J \sqrt{\frac{H_\mathrm{K}^2 \sin^2\varphi_\mathrm{oct} +H_\mathrm{y,eff}^2 -2H_\mathrm{K} H_\mathrm{y,eff} \sin \varphi_\mathrm{oct}    }{H_\mathrm{J} H_\mathrm{K}}} d \varphi_\mathrm{oct} \\
        &\times\int_{\pi/2}^{\pi} \alpha (-3 M_\mathrm{s} H_\mathrm{K} \sin \varphi_\mathrm{oct} \cos \varphi_\mathrm{oct} + 3 M_\mathrm{s} H_\mathrm{y,eff} \cos \varphi_\mathrm{oct}) \\
        &\times \sqrt{\frac{H_\mathrm{J} H_\mathrm{K}}{H_\mathrm{K}^2 \sin^2\varphi_\mathrm{oct} +H_\mathrm{y,eff}^2 -2H_\mathrm{K} H_\mathrm{y,eff} \sin \varphi_\mathrm{oct}}}\qty(2H_\mathrm{K}^2 \sin \varphi_\mathrm{oct}\cos \varphi_\mathrm{oct} -2 H_\mathrm{K} H_\mathrm{y,eff} \cos \varphi_\mathrm{oct}) d \varphi_\mathrm{oct}.
    \end{aligned}
    \label{eq: deltaE y}
\end{equation}

\begin{figure}[h]
    \centering
    \includegraphics[width=0.6\linewidth]{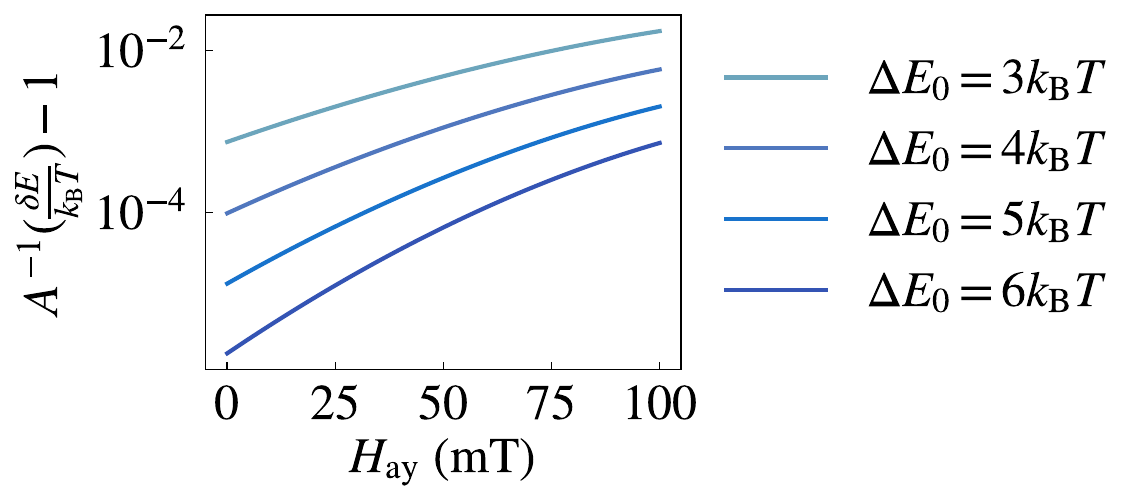}
    \caption{Deviation of the depopulation factor from unity as a function of $H_\mathrm{ay}$}
    \label{fig:Ay}
\end{figure}

Numerically integrating the above equation enables us to compute the inverse of the depopulation factor $\qty[A\qty(\frac{\delta E}{k_\mathrm{B}T})]^{-1}$. For the range of $H_\mathrm{ay}$ and $\mathcal{V}$ considered in this work, $\qty[A\qty(\frac{\delta E}{k_\mathrm{B}T})]^{-1} \approx 1$ with less than $1\%$ error, as presented in Fig.~\ref{fig:Ay} and the error increases with $H_\mathrm{ay}$. 


\begin{itemize}
    \item \textbf{Depopulation factor with $H_\mathrm{ax}$}
\end{itemize}
{Due to the asymmetry of the two saddle points induced by $H_\mathrm{ax}$, their free energies are no longer identical; consequently, the two equal-free-energy contours need to be treated separately.}
In an equal-free-energy contour including the saddle point ($\theta_\mathrm{oct} = \pi/2$, $\varphi_\mathrm{oct} = 0$),
\begin{equation}
    \begin{aligned}
        \mathcal{H}_\mathrm{oct}(H_\mathrm{ax}) &= \frac{3}{4}M_\mathrm{s} H_\mathrm{K} \cos(2\varphi_\mathrm{oct}) + \frac{3}{2}M_\mathrm{s} H_\mathrm{J} m^2_z +3 M_\mathrm{s}H_\mathrm{x,eff} \cos \varphi_\mathrm{oct} = \frac{3}{4} M_\mathrm{s} H_\mathrm{K} + 3M_\mathrm{s} H_\mathrm{x,eff}, 
    \end{aligned}
\end{equation}

\begin{figure}[h]
    \centering
    \includegraphics[width=0.95\linewidth]{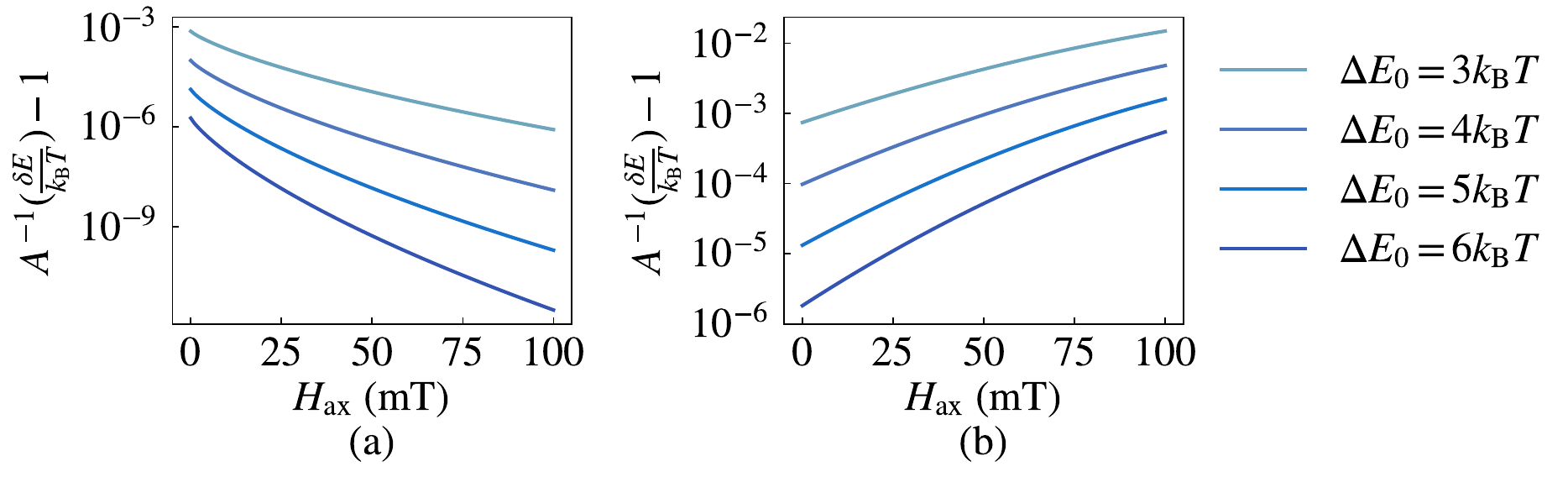}
\caption{Deviation of the depopulation factor from unity with $H_\mathrm{ax}$ for (a) $\varphi_\mathrm{oct} = 0$ and (b) $\varphi_\mathrm{oct} = \pi$.}
    \label{fig:Ax}
\end{figure}

{the $m_{\mathrm{oct},z}-\varphi_\mathrm{oct}$ relationship is given as:}

\begin{equation}
    m_{\mathrm{oct},z} = \sqrt{\frac{H_\mathrm{K} \sin^2 \varphi_\mathrm{oct}+2H_\mathrm{x,eff}(1-\cos \varphi_\mathrm{oct})}{H_\mathrm{J}}}.
\end{equation}
Then Eq.~(\ref{eq:energy dissipation}) can be written as
\begin{equation}
    \begin{aligned}
        \delta \mathcal{F}_\mathrm{oct} &= \Bigg| \int_{\pi/2}^{0} -3 \alpha M_\mathrm{s} H_J \sqrt{\frac{H_\mathrm{K} \sin^2 \varphi_\mathrm{oct}+2H_\mathrm{x,eff}(1-\cos \varphi_\mathrm{oct})}{H_\mathrm{J}}} \mathrm{d} \varphi_\mathrm{oct} \\
        &\times\int_{\pi/2}^{0} \alpha (-3 M_\mathrm{s} H_\mathrm{K} \sin \varphi_\mathrm{oct} \cos \varphi_\mathrm{oct} - 3 M_\mathrm{s} H_\mathrm{x,eff} \sin \varphi_\mathrm{oct} ) \\
        &\times \sqrt{\frac{H_\mathrm{J} }{H_\mathrm{K} \sin^2 \varphi_\mathrm{oct}+2H_\mathrm{x,eff}(1-\cos \varphi_\mathrm{oct})}}(2H_\mathrm{K}\sin \varphi_\mathrm{oct} \cos \varphi_\mathrm{oct} + 2H_\mathrm{x,eff} \sin \varphi_\mathrm{oct}) \mathrm{d} \varphi_\mathrm{oct} \Bigg|.
    \end{aligned}
    \label{eq: deltaE x1}
\end{equation}
Numerically integrating the above equation allows us to compute $\qty[A\qty(\frac{\delta E}{k_\mathrm{B}T})]^{-1}$, which can again be approximated to $1$ with an error of less than $1\%$ for $\Delta E_0 \geq 3k_\mathrm{B}T$, as shown in Fig.~\ref{fig:Ax}(a) and the error exponentially decreases as $\Delta E_0$ increases. 

In then equal-free-energy contour including the saddle point ($\theta_\mathrm{oct} = \pi/2$, $\varphi_\mathrm{oct} = \pi$),
\begin{equation}
    \begin{aligned}
        \mathcal{H}_\mathrm{oct} &= \frac{3}{4}M_\mathrm{s} H_\mathrm{K} \cos(2\varphi_\mathrm{oct}) + \frac{3}{2}M_\mathrm{s} H_\mathrm{J} m^2_z +3 M_\mathrm{s}H_\mathrm{x,eff} \cos \varphi_\mathrm{oct} = \frac{3}{4} M_\mathrm{s} H_\mathrm{K} - 3M_\mathrm{s} H_\mathrm{x,eff}, 
    \end{aligned}
\end{equation}
\begin{equation}
    m_{\mathrm{oct},z} = \sqrt{\frac{H_\mathrm{K} \sin^2 \varphi_\mathrm{oct}-2H_\mathrm{x,eff}(1+\cos \varphi_\mathrm{oct})}{H_\mathrm{J}}}.
\end{equation}
Then Eq.~(\ref{eq:energy dissipation}) can be written as
\begin{equation}
    \begin{aligned}
        \delta \mathcal{F}_\mathrm{oct} &= \Bigg| \int_{\pi/2}^{0} -3 \alpha M_\mathrm{s} H_J \sqrt{\frac{H_\mathrm{K} \sin^2 \varphi_\mathrm{oct}-2H_\mathrm{x,eff}(1+\cos \varphi_\mathrm{oct})}{H_\mathrm{J}}} \mathrm{d} \varphi_\mathrm{oct} \\
        &\times \int_{\pi/2}^{0} \alpha (-3 M_\mathrm{s} H_\mathrm{K} \sin \varphi_\mathrm{oct} \cos \varphi_\mathrm{oct} - 3 M_\mathrm{s} H_\mathrm{x,eff} \sin \varphi_\mathrm{oct} ) \\
        &\times \sqrt{\frac{H_\mathrm{J} }{H_\mathrm{K} \sin^2 \varphi_\mathrm{oct}-2H_\mathrm{x,eff}(1+\cos \varphi_\mathrm{oct})}}(2H_\mathrm{K}\sin \varphi_\mathrm{oct} \cos \varphi_\mathrm{oct} + 2H_\mathrm{x,eff} \sin \varphi_\mathrm{oct}) \mathrm{d} \varphi_\mathrm{oct} \Bigg|.
    \end{aligned}
    \label{eq: deltaE x1}
\end{equation}
{Even in the case of $H_\mathrm{ax}$-assisted dynamics, }$\qty[A\qty(\frac{\delta E}{k_\mathrm{B}T})]^{-1} \approx 1$ with an error of less than $1\%$ when $\Delta E_0 \geq 3k_\mathrm{B}T$, as shown in Fig.~\ref{fig:Ax}(b).



\newpage
\subsection{Numerical simulation of escape time}
\label{sec:numerical}

{\subsubsection{Details and benchmarking of LLG}
\label{sec:benchmark}
The numerical simulation results presented in the main text utilize our in-house developed solver on the CUDA platform. The solver is used to run more than 1000 simulations allows us to collect statistically reliable data for analyzing the effect of thermal noise on the octupole dynamics. 
The LLG equation for sublattice $i= 1,2,3$ is given as
\begin{equation}
    \pdv{\vb{m}_i}{t} = -\gamma (\vb{m}_i \cross \vb{H}_i^{\mathrm{eff}}) +\alpha \qty(\vb{m}_i \times \pdv{\vb{m}_i}{t}),
\end{equation}
where $\vb{H}_i^{\mathrm{eff}}$, the effective field for sublattice $i$, is defined as
\begin{equation}
    \vb{H}_i^{\mathrm{eff}} = -\frac{1}{M_\mathrm{s}}\frac{\partial \mathcal{H}(\vb{m})}{\partial \vb{m}_i} + \vb{H}_{\mathrm{thermal},i}.
\end{equation}
The effective field includes internal fields due to exchange and DM interactions between sublattices, field due to uniaxial anisotropy along respective easy axes, externally applied magnetic field, and the thermal field $\vb{H}_{\mathrm{thermal},i} = \sqrt{\frac{2\alpha k_\mathrm{B} T}{\gamma \mu_0 M_\mathrm{s}^2 \mathcal{V} \Delta t}} \boldsymbol{\xi}^i_t$,  where $\boldsymbol{\xi}^i_t \sim \mathcal{N}(0,1)$. Here, $\gamma$ and $\mu_0$ denote the gyromagnetic ratio and vacuum permeability, respectively, while $\Delta t = 10^{-15}~\mathrm{s}$ is the time step used in the simulation. The LLG equation of each sublattice is numerically solved with finite different Heun method~\cite{ament2016solving}.

To verify the results of our numerical implementation, firstly, we benchmarked our extracted data (in the absence of thermal noise) against previously published results~\cite{shukla2024impact} as depicted in Fig.~\ref{fig:benchmark}(a), Secondly, we benchmarked our extracted numerical data (with thermal noise) against Eq.~(7) of Ref.~[44], as depicted in Fig.~\ref{fig:benchmark}(b) by symbols and line, respectively.
We then proceeded to investigate the field-driven dynamics in single-domain Mn$_3$Sn in the presence of thermal noise.

\begin{figure}[h]
    \centering
    \includegraphics[width=0.8\linewidth]{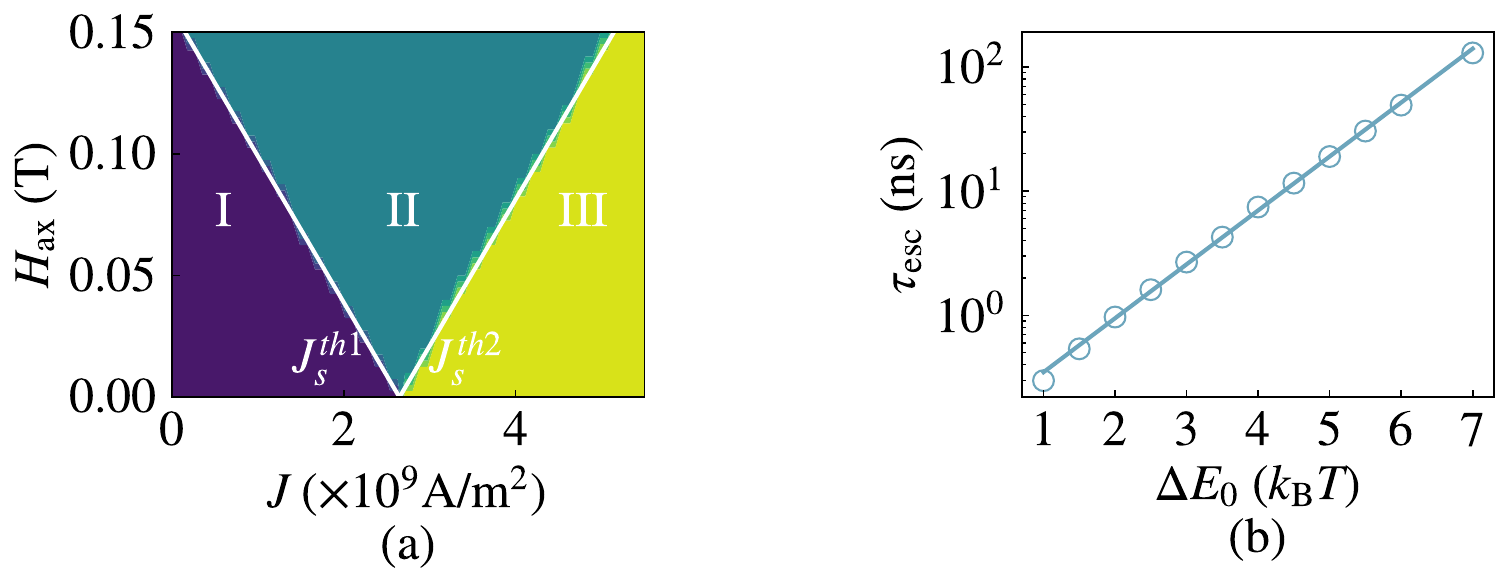}
    \caption{Comparison of our in-house LLG solver against published analytical predictions. (a) Comparison against field-assisted SOT-driven dynamics without thermal noise~\cite{shukla2024impact}. The white lines are predicted threshold current density between non-switching ($\mathrm{I}$), switching ($\mathrm{II}$) and rotation ($\mathrm{III}$). (b) Comparison against field-free thermal fluctuations~\cite{konakanchi2025electrically}. The symbols represent the numerically simulated escape time without external field and the solid line is the theoretical prediction.}
    \label{fig:benchmark}
\end{figure}
}

\subsubsection{Extraction of escape time from LLG}
\label{sec:extraction}
\begin{figure}[h]
    \centering
    \includegraphics[width=0.98\linewidth]{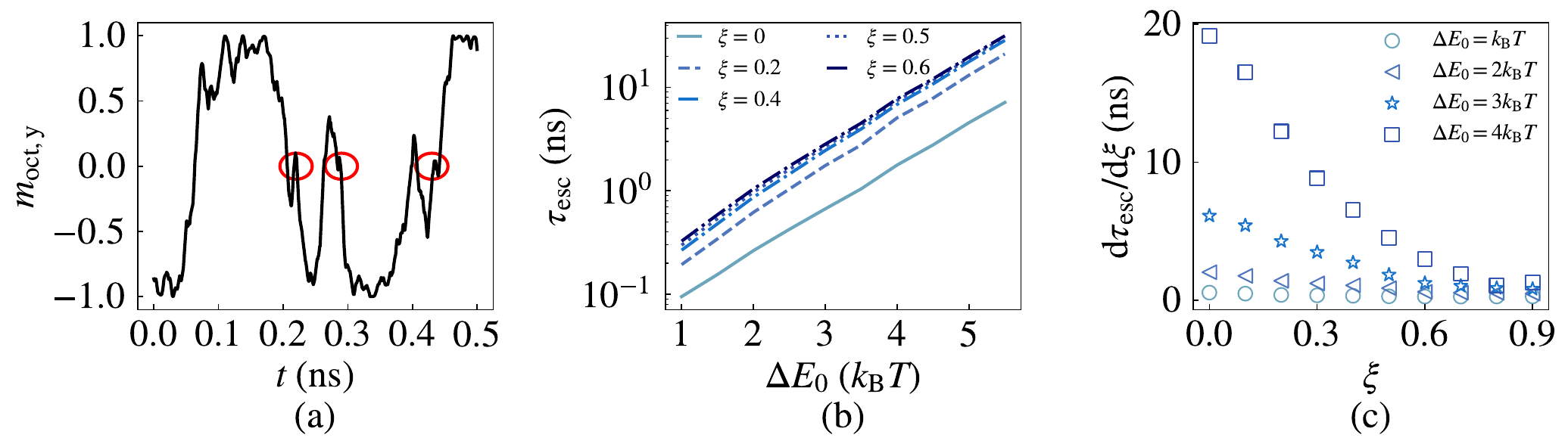}
    \caption{(a) Temporal evolution of $m_\mathrm{oct,y}$ and red circles highlight examples of back-and-forth fluctuations around $m_\mathrm{oct,y} = 0$.  (b) Field-free escape time as a function of $\Delta E_0$ with various $\xi$. (c) The rate of change of field-free $\tau_\mathrm{esc}$ with $\xi$.  }
    \label{fig:escape time}
\end{figure}

The escape time from the up state to the down state is defined as the average time it takes to navigate from the up-state energy well to the down-state energy well. Each simulation is 1~ms long, and the ensemble size is chosen 8, which jointly ensures that at least $10^3$ inter-well switching events occur during each measurement. If the switching is defined corresponding to when $\vb{m}_\mathrm{oct,y}$ crosses the saddle point, as depicted in Fig.~\ref{fig:escape time}(a), it would over-estimate the number of switching events, thus underestimating the escape time. This is because $\vb{m}_\mathrm{oct,y}$ crosses the zero point multiple times within a single state switching event. For physically meaningful extraction, we introduce a parameter $\xi(>0)$, defined below, to obtain the escape time for the thermally activated process:
\begin{itemize}
    \item $\vb{m}_\mathrm{oct}$ switching from up state to down state is assumed to occur when $\vb{m}_\mathrm{oct,y}$ crosses $-\xi$,
    \item $\vb{m}_\mathrm{oct}$ switching from down state to up state is assumed to occur when $\vb{m}_\mathrm{oct,y}$ crosses $\xi$.
\end{itemize}
Figure~\ref{fig:escape time}(b) shows the escape time of up state extracted from the LLG simulations with various $\xi$, while Fig.~\ref{fig:escape time}(c) shows the sensitivity of $\tau_\mathrm{esc}$ to $\xi$. In the field-free thermal fluctuations case, the escape time has a positive dependence on $\xi$, whereas the dependence shrinks as $\xi$ increases. We select $\xi = 0.5$ as the criterion for switching because this ensures $\tau_\mathrm{esc}$ has a reduced dependence on $\xi$ and the result matches well with reported relaxation time~\cite{konakanchi2025electrically} extracted from the magnetization relaxation process. 

\subsubsection{Two Kagome-plane versus one Kagome-plane model of spin configuration}
\label{sec:six spin}
The unit cell of $\mathrm{Mn_3Sn}$ is composed of 6 $\mathrm{Mn}$ spins in two stacked Kagome planes {but the one-Kagome-plane is also widely adopted in theoretical and numerical analysis~\cite{shukla2022spin, he2024magnetic} and it has the same accuracy within the tolerance of error for the following reasons.First of all, the order parameter itself is intra-layer, which means $\vb{m}_\mathrm{oct}$ is built from one triangle of Mn spins in one Kagome plane and this already gives the correct symmetry that underlines the large Berry curvature~\cite{suzuki2017cluster,chen2023octupole}. Secondly, the \textbf{ABAB} stacked Kagome planes share the same octupolar chirality so most in-plane transport and switching results does not qualitatively change between one-plane and two-plane models. Thirdly, the inter-layer interactions is modest compared with intra-layer interactions. To validate for this, we conduct a comparison through numerical simulations.}
{The} complete form of the Hamiltonian consisting 6 spins is~\cite{liu2017anomalous}:
\begin{equation}
    \begin{aligned}
        \mathcal{H}(\vb{m}) &= J_\mathrm{E}((1+\delta_\mathrm{E})\vb{m}_1 \cdot \vb{m}_2+\vb{m}_2\cdot \vb{m}_3+\vb{m}_3\cdot \vb{m}_1)
        +J_\mathrm{E}((1+\delta_\mathrm{E})\vb{m}_4 \cdot \vb{m}_5+\vb{m}_5\cdot \vb{m}_6+\vb{m}_6\cdot \vb{m}_4))\\
        &-J_\mathrm{E}(\vb{m}_1 \cdot \vb{m}_4+\vb{m}_2 \cdot \vb{m}_6+\vb{m}_3 \cdot \vb{m}_6)-\sum_{i=1}^{6}{\left[K_\mathrm{u}(\vb{m}_i\cdot \vb{u}_i)^2+M_\mathrm{s} \vb{H}_\mathrm{a}\cdot \vb{m}_i)\right]}\\
        &+D_\mathrm{M} \vb{z}\cdot (\vb{m}_1\times \vb{m}_2+\vb{m}_2\times \vb{m}_3+\vb{m}_3\times \vb{m}_1)
        +D_\mathrm{M} \vb{z}\cdot (\vb{m}_4\times \vb{m}_5+\vb{m}_5\times \vb{m}_6+\vb{m}_6\times \vb{m}_4),
    \end{aligned}
    \label{eq: free energy6}
\end{equation}
where suffix $i = 4,5,6$ represents another three sublattices in the second kagome plane. The Hamiltonian includes both intra-plane and inter-plane isotropic exchange interactions and the easy axes of these sublattices are $\vb{u}_4 =\vb{u}_1 = -1/2 \vb{x} +\sqrt{3}/2 \vb{y}$, $\vb{u}_5 =\vb{u}_2 = -1/2 \vb{x} -\sqrt{3}/2 \vb{y}$ and $\vb{u}_6 =\vb{u}_3 = \vb{x}$. 

\begin{figure}[h]
    \centering
    \includegraphics[width=0.9\linewidth]{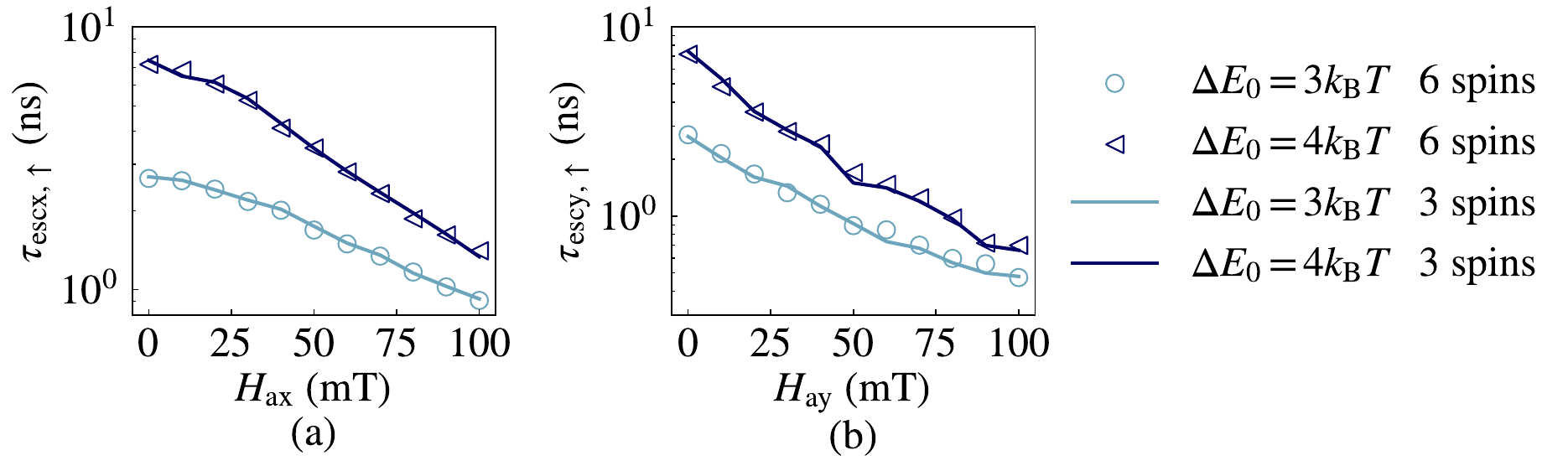}
    \caption{Escape time collected from 3-spin and 6-spin configurations with (a) $H_\mathrm{ax}$ and (b) $H_\mathrm{ay}$.}
    \label{fig:spin compare}
\end{figure}

{We perform coupled LLG equation simulations using Eq.~\ref{eq: free energy6}. 
As shown in Fig.~\ref{fig:spin compare}, the escape times from the two-Kagome-plane model 
overlap with those from the one-Kagome-plane model used in the main text. 
These numerical results confirm that the one-Kagome-plane model is sufficient for 
investigating the thermal dynamics of Mn$_3$Sn.
}

\newpage
{\subsection{Additional analysis of escape time}

{\subsubsection{Differences between numerical and analytical models of escape time}
\label{sec:deviation}

As depicted in Fig.~(3) of the manuscript, theoretical predictions of Eq.~(5)-(8) deviate from the numerical simulation results at high magnetic fields. 
To analyze this deviation, we rewrite Eq.~(5) in the form $\tau_\mathrm{esc} = \tau_0 \exp(\Delta E/k_\mathrm{B}T)$ and analyze $\tau_0$ and $\exp(\Delta E/k_\mathrm{B}T)$ separately. 
For the $H_\mathrm{ax}$-assisted up-to-down escape, the dominant pathway is the escape from $\varphi_\mathrm{oct} = \pi$, and we therefore focus on this process. 
As shown in Fig.~\ref{fig:tau0x}, $\tau_0$ and $\exp(\Delta E_x/k_\mathrm{B}T)$ exhibit opposite dependencies on $H_\mathrm{ax}$. 
With increasing field, the reduction in the exponential factor may not compensate for the growth in $\tau_0$, leading to an increase in the predicted $\tau_\mathrm{esc}$. 
This trend is more pronounced in the case of $H_\mathrm{ay}$-assisted escape (Fig.~\ref{fig:tau0y}), where $\tau_0$ grows more rapidly when the field is applied along the $-y$ axis. 
This explains the abnormal increase in escape time with increasing field strength, which underlies the deviation of the theoretical predictions from numerical simulations.
This deviation warrants further investigation, as the applicability of the perturbation theory under increasing external field strength remains uncertain. Both the linearization of the LLG equation and the eigenvalue analysis of the Hessian matrix rely on a Hamiltonian that depends on $\vb{m}_\mathrm{oct}$, which is itself derived from perturbative assumptions. If a strong external field disrupts the rigid 120$^\circ$ spin configuration characteristic of Mn$_3$Sn, the perturbative expression in Eq.~(2) may significantly deviate from the full Hamiltonian in Eq.~(1). Consequently, this deviation can lead to larger errors in the estimation of both the attempt time $\tau_0$ and the energy barrier $\Delta E$.

\begin{figure}[h!]
    \centering
    \includegraphics[width=0.9\linewidth]{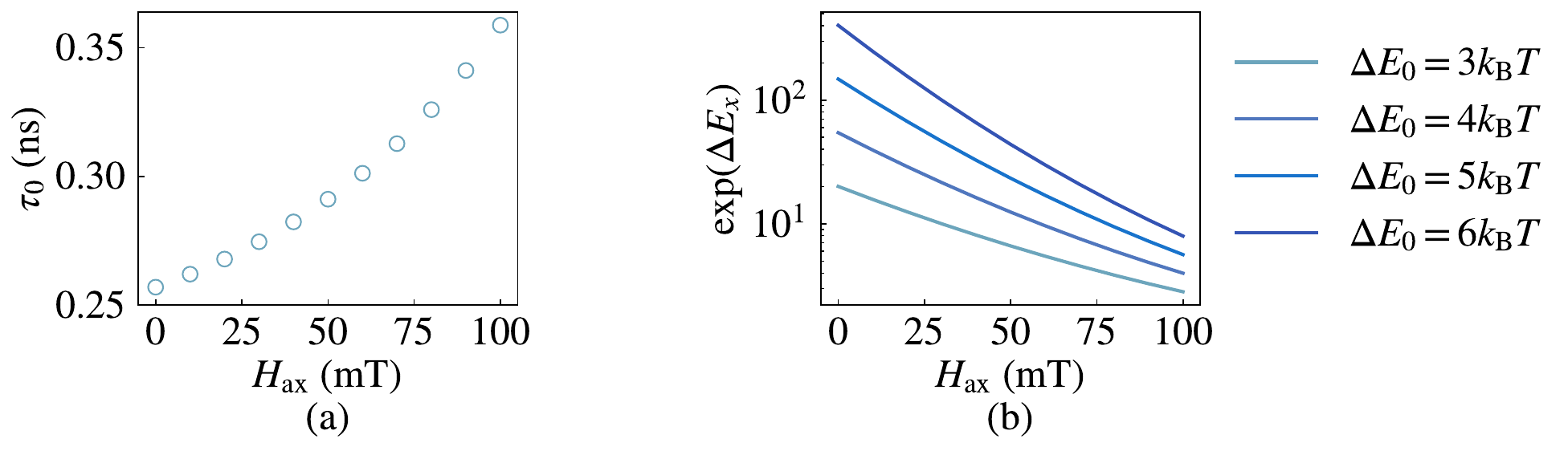}
    \caption{{(a)$\tau_0$ as a function of $H_\mathrm{ax}$ for up-to-down escape at $\varphi_\mathrm{oct} = \pi$. (b)$\exp(\Delta E_{x,2})$ as a function of $H_\mathrm{ax}$. }}
    \label{fig:tau0x}
\end{figure}

\begin{figure}[h!]
    \centering
    \includegraphics[width=0.9\linewidth]{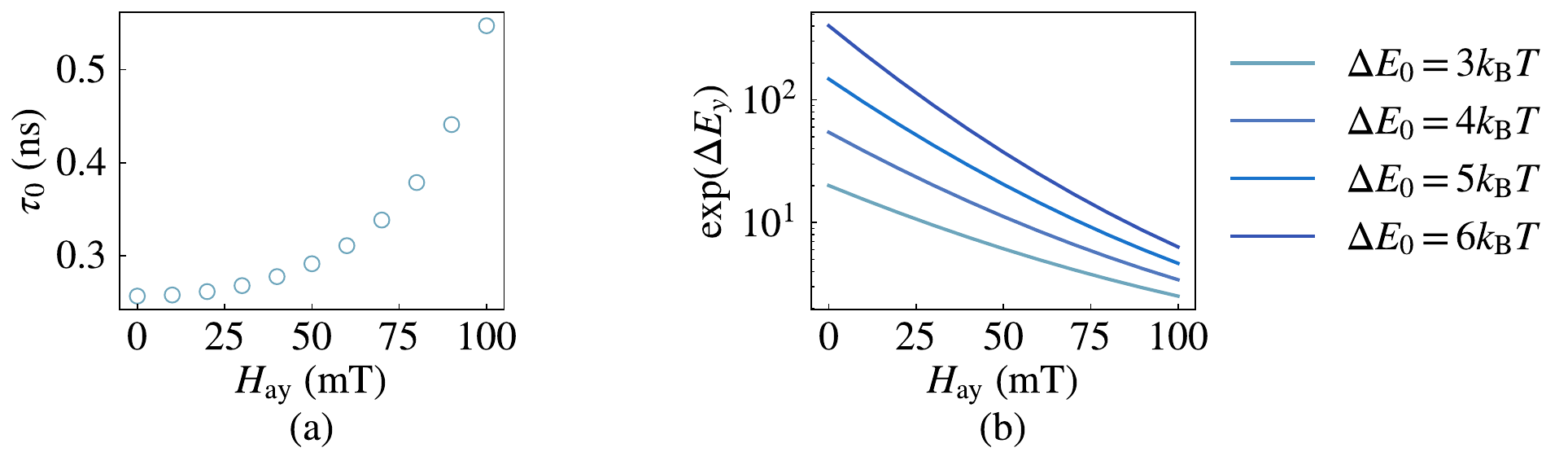}
    \caption{{(a)$\tau_0$ as a function of $H_\mathrm{ay}$ for up-to-down escape. (b)$\exp(\Delta E_{y,1})$ as a function of $H_\mathrm{ay}$.}}
    \label{fig:tau0y}
\end{figure}


}

\subsubsection{Sensitivity of escape time to variations in material parameters}
\label{sec:sensitivity}

The up-to-down escape time is analytically derived from Eqs.~(3)–(8) in the main text, making it difficult to directly assess its dependence on the magnetic parameters listed in Table~I of the main text. To address this, we plot in Fig.~\ref{fig:parameter var fixE} a heat map of the analytically obtained up-to-down escape time by sweeping the external field and other magnetic parameters, while fixing $\Delta E_0 = 4~k_\mathrm{B}T$ in Fig.~\ref{fig:parameter var fixE} and fixing $\mathcal{V} = 80 \times 80 \times 8~\mathrm{nm^3}$ in Fig.~\ref{fig:parameter var fixD}.

\begin{figure}
    \centering
    \includegraphics[width=0.8\linewidth]{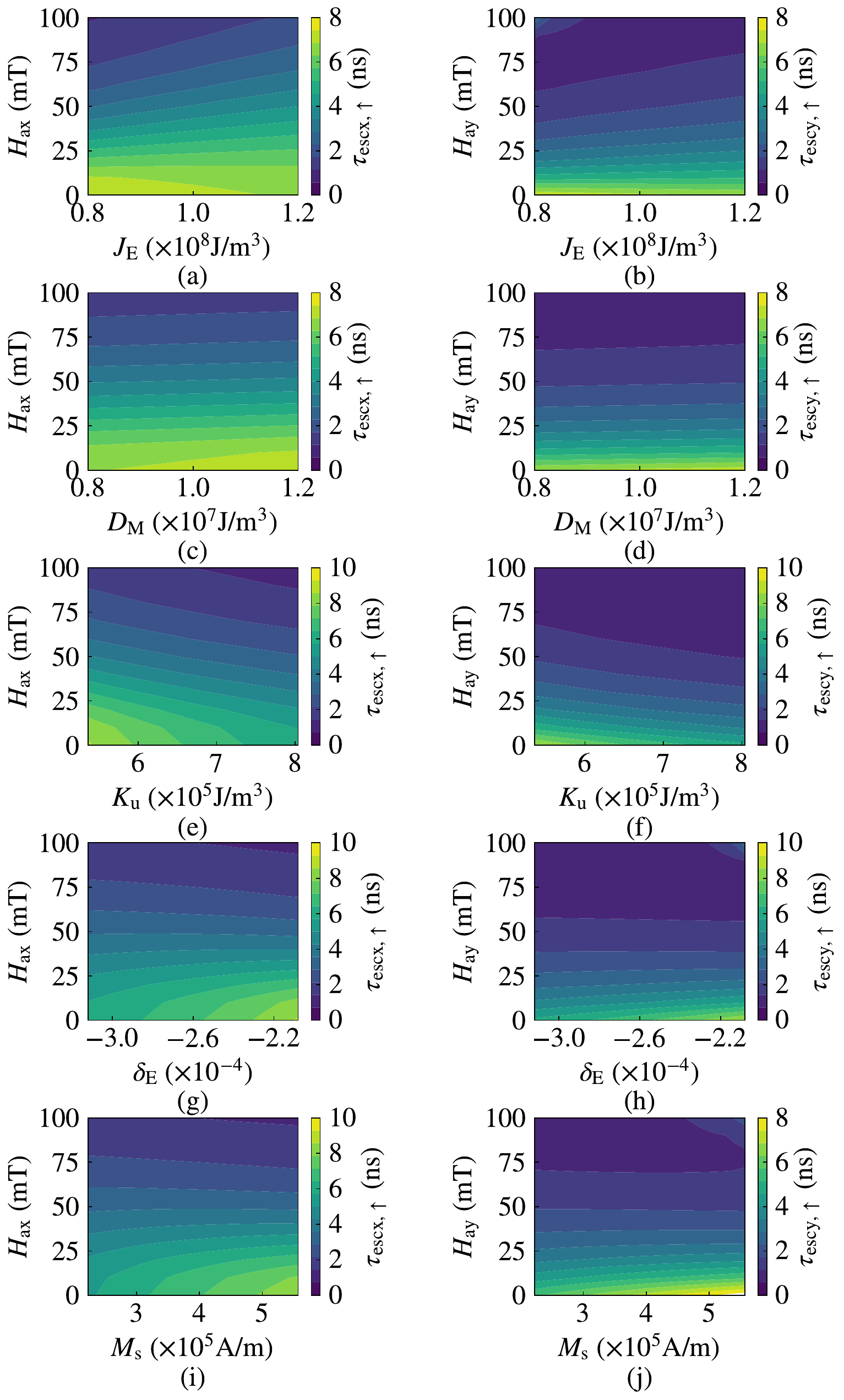}
    \caption{The up-to-down escape time with $\Delta E_0 = 4 k_\mathrm{B}T$ by sweeping external field and magnetic paramters (a)\&(b) $J_\mathrm{E}$, (c)\&(d) $D_\mathrm{M}$, (e)\&(f) $K_\mathrm{u}$, (g)\&(h) $\delta_\mathrm{E}$, (i)\&(j) $M_\mathrm{s}$, in Table~$\mathrm{I}$ in the main text. The left column corresponds to $H_\mathrm{ax}$, while the right column corresponds to $H_\mathrm{ay}$. }
    \label{fig:parameter var fixE}
\end{figure}

\begin{figure}
    \centering
    \includegraphics[width=0.8\linewidth]{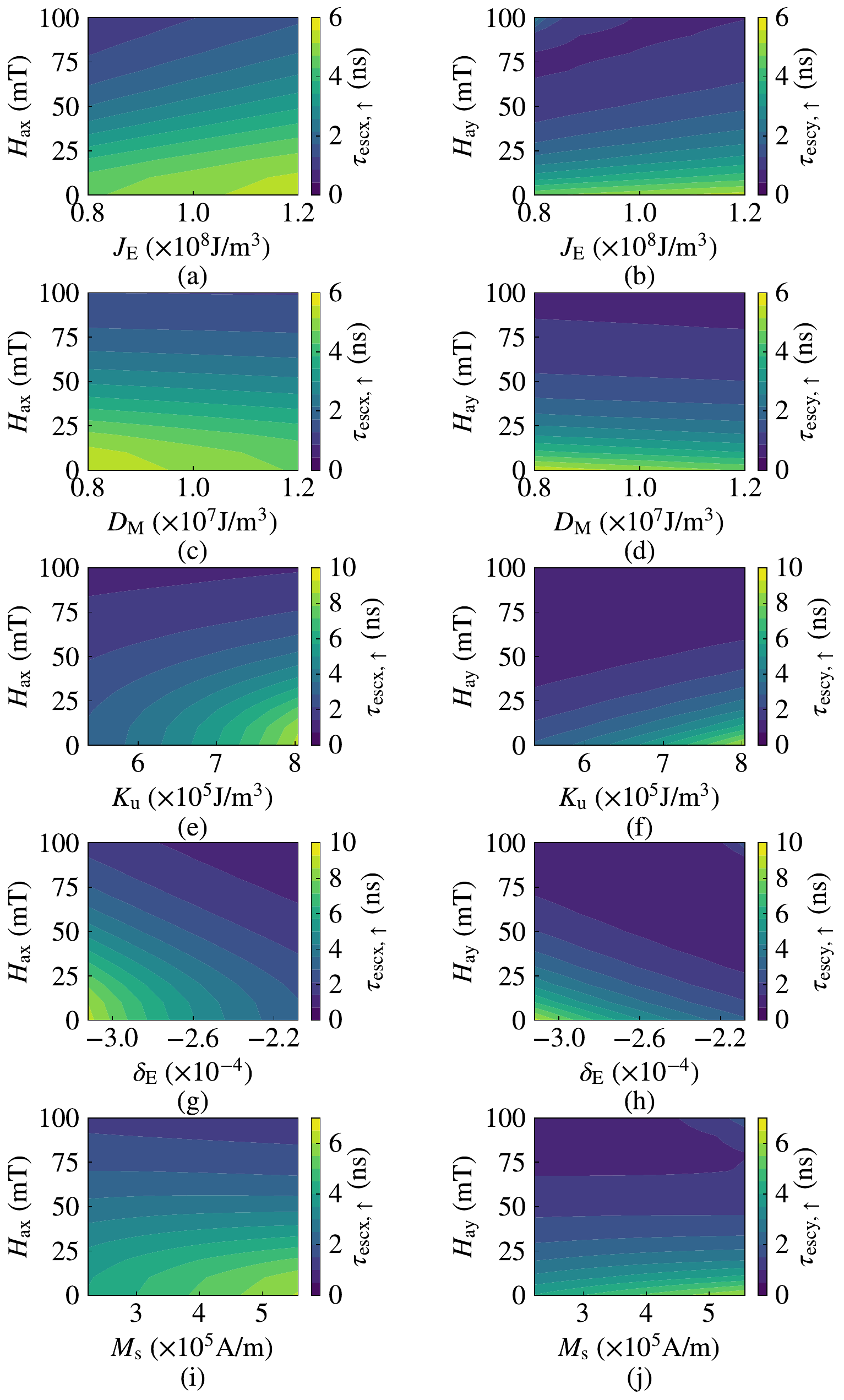}
    \caption{The up-to-down escape time with $\mathcal{V} = 80 \times 80 \times 8~\mathrm{nm^3}$ by sweeping external field and magnetic paramters (a)\&(b) $J_\mathrm{E}$, (c)\&(d) $D_\mathrm{M}$, (e)\&(f) $K_\mathrm{u}$, (g)\&(h) $\delta_\mathrm{E}$, (i)\&(j) $M_\mathrm{s}$, in Table~$\mathrm{I}$ in the main text. The left column corresponds to $H_\mathrm{ax}$, while the right column corresponds to $H_\mathrm{ay}$.}
    \label{fig:parameter var fixD}
\end{figure}

$H_\mathrm{ax}$-assisted and $H_\mathrm{ay}$-assisted thermal escape time exhibit qualitatively similar dependence trends under identical parameter variations, while $\tau_\mathrm{esc}$ is quantitatively more sensitive in the $H_\mathrm{ay}$-assisted case. 

As shown in Figs.~\ref{fig:parameter var fixE}(a)\&(b), increasing $J_\mathrm{E}$ reduces $\tau_\mathrm{esc}$ in the low-field region, indicating that the strong exchange interaction governs the ultrafast thermal fluctuation frequency of Mn$_3$Sn. At higher fields, however, the dependence of $\tau_\mathrm{esc}$ on $J_\mathrm{E}$ reverses, since the attempt frequency in Eq.~(7) is not monotonically dependent on $J_\mathrm{E}$. In contrast, Figs.~\ref{fig:parameter var fixD}(a)\&(b) show that $\tau_\mathrm{esc}$ increases monotonically with $J_\mathrm{E}$, which can be attributed to its contribution to the energy barrier through $H_\mathrm{K}$.In contrast, $D_\mathrm{M}$ exerts only a limited influence on $\tau_\mathrm{esc}$, since it
is about an order of magnitude smaller than $J_\mathrm{E}$; consequently, the effect of $D_\mathrm{M}$ is suppressed by $J_\mathrm{E}$. 

According to Eq.~(7), the attempt frequency is positively related to the magnitude of the uniaxial anisotropy $H_\mathrm{K}$, such that, when $\Delta E_0$ is fixed, the escape time decreases with increasing $H_\mathrm{K}$. Since $H_\mathrm{K}$ is proportional to both $K_\mathrm{u}$ and $\delta_\mathrm{E}$, larger values of these parameters accelerate the dynamics, as shown in Fig.~\ref{fig:parameter var fixE}(e)–(h). Although a larger $K_\mathrm{u}$ corresponds to a stronger uniaxial field that favors alignment of the sublattices along their easy axes, maintaining a constant $\Delta E_0$ requires a reduction in size. This reduction enhances the thermal field, thereby further accelerating the dynamics. In contrast, $M_\mathrm{s}$ exerts the opposite influence: both $H_\mathrm{K}$ and the thermal field decrease with increasing $M_\mathrm{s}$, leading to longer escape times for larger $M_\mathrm{s}$.

Despite the positive dependence of $K_\mathrm{u}$ and $|\delta_\mathrm{E}|$ on the attempt frequency, Figs.~\ref{fig:parameter var fixD}(e)\&(h) show that $\tau_\mathrm{esc}$ increases with these parameters. This is because, under the fixed-volume condition, $\tau_\mathrm{esc}$ is predominantly governed by $\Delta E$. Eqs.~(3)\&(4) clearly indicate that $\Delta E$ is positively dependent on $H_\mathrm{K}$, and therefore also on $K_\mathrm{u}$ and $|\delta_\mathrm{E}|$. In contrast, $M_\mathrm{s}$ is positively dependent on $\Delta E$ despite its negative dependence on $H_\mathrm{K}$, which explains why $\tau_\mathrm{esc}$ in Figs.~\ref{fig:parameter var fixD}(i)\&(j) increases with $M_\mathrm{s}$.

}

\vspace{-10pt}
\subsubsection{Impact of damping coefficient}
\label{sec:damping}
\begin{figure}[h]
    \centering    
    \includegraphics[width=0.9\linewidth]{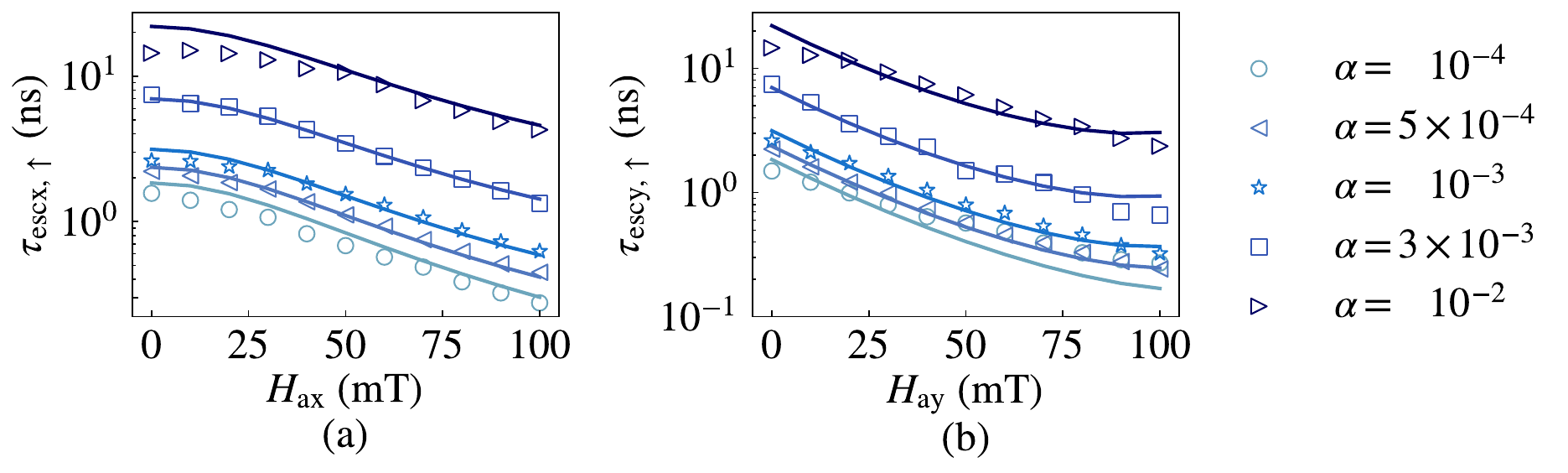}
    \vspace{-10pt}
    \caption{Field-assisted escape time with $\alpha$ ranging from $10^{-4}$ to $10^{-2}$ and $\Delta E_0 = 4k_\mathrm{B}T.$ }
    \label{fig:Alpha}
\end{figure}

\vspace{-10pt}
To investigate the intermediate-to-high damping regime and  validate the applicable range of our theoretical model, we conducted numerical simulations over a broad range of damping coefficients $\alpha$ from $10^{-4}$ to $10^{-2}$. As shown in Fig.~\ref{fig:Alpha}, the theoretical predictions for $H_\mathrm{ax}$-assisted escape remain valid across this entire range, whereas the theory fails for $H_\mathrm{ay}$ when $\alpha < 5\times 10^{-3}$. {As $\alpha$ decreases below $5\times 10^{-3}$, $\tau_\mathrm{escy,\uparrow}$ for $\alpha = 10^{-4}$ becomes even larger than that for $\alpha = 5\times 10^{-4}$. The mechanism behind this behavior remains unclear, but one possible explanation is that, similar to the case of superparamagnets, energy exchange with the thermal bath becomes slow~\cite{kramers1940brownian}, while precessions along the effective field increase~\cite{brown1963thermal}; both effects act to delay the escape process.
}

\subsubsection{Escape time for down-to-up dynamics}
\label{sec:downup}
\begin{figure}[h!]
\vspace{-10pt}
    \centering
    \includegraphics[width=0.95\linewidth]{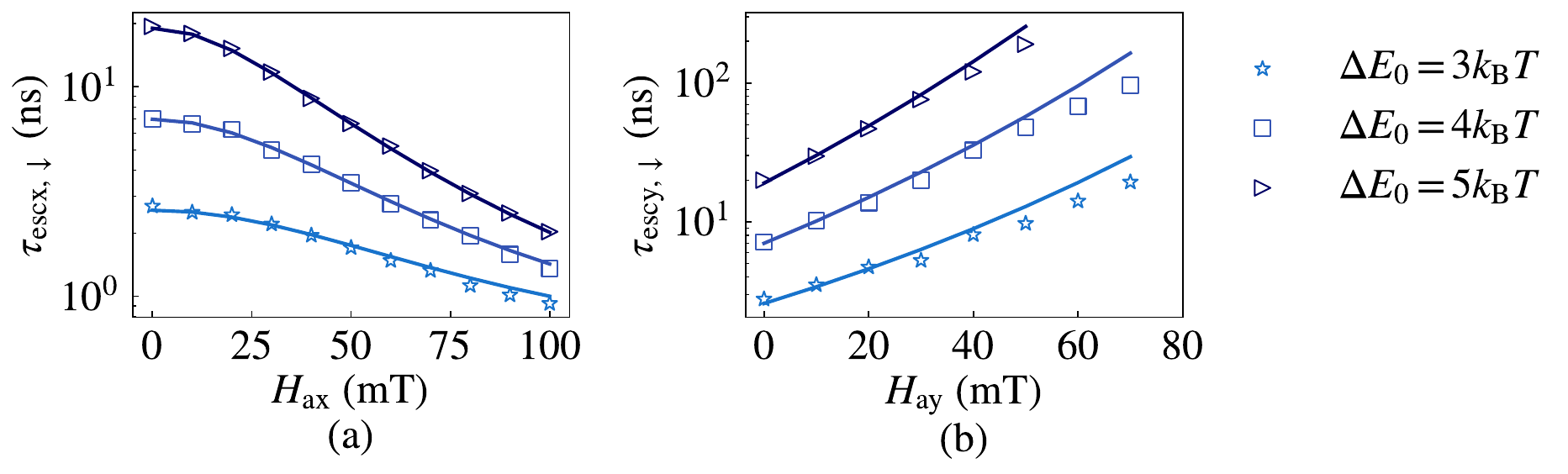}
    \vspace{-10pt}
    \caption{The schematic of escape time from down state to up state while numerical extracted data is in markers and analytical predictions are plotted with lines. $\Delta E_0$ ranges from  {$3k_\mathrm{B}T$} to $5k_\mathrm{B}T$ and field is applied along (a) $x$-axis and (b) $y$-axis.}
    \label{fig:esc down}
\end{figure}

According to the symmetry of the energy landscape, the escape times from down state to up state are $\tau_\mathrm{escx,\downarrow} = \tau_\mathrm{escx,\uparrow}$ and $\tau_\mathrm{escy,\downarrow} = \tau_\mathrm{escy,\uparrow} \exp((\Delta E_\mathrm{y,2}-\Delta E_\mathrm{y,1})k_\mathrm{B}T)$. Fig.~\ref{fig:esc down} shows that our analytic model fits well
with the numerical LLG data. It is worth mentioning that $\tau_\mathrm{escy,\downarrow}$ increases rapidly with $H_\mathrm{ay}$, so we restrict the maximum external field to $80~\mathrm{mT}$ for analysis (the field is restricted to 
$50~\mathrm{mT}$ with $\Delta E_0 = 5k_\mathrm{B}T$).

\nocite{*}
\FloatBarrier
\bibliography{aipsamp}

\begin{thebibliography}{63}%
\makeatletter
\providecommand \@ifxundefined [1]{%
 \@ifx{#1\undefined}
}%
\providecommand \@ifnum [1]{%
 \ifnum #1\expandafter \@firstoftwo
 \else \expandafter \@secondoftwo
 \fi
}%
\providecommand \@ifx [1]{%
 \ifx #1\expandafter \@firstoftwo
 \else \expandafter \@secondoftwo
 \fi
}%
\providecommand \natexlab [1]{#1}%
\providecommand \enquote  [1]{``#1''}%
\providecommand \bibnamefont  [1]{#1}%
\providecommand \bibfnamefont [1]{#1}%
\providecommand \citenamefont [1]{#1}%
\providecommand \href@noop [0]{\@secondoftwo}%
\providecommand \href [0]{\begingroup \@sanitize@url \@href}%
\providecommand \@href[1]{\@@startlink{#1}\@@href}%
\providecommand \@@href[1]{\endgroup#1\@@endlink}%
\providecommand \@sanitize@url [0]{\catcode `\\12\catcode `\$12\catcode `\&12\catcode `\#12\catcode `\^12\catcode `\_12\catcode `\%12\relax}%
\providecommand \@@startlink[1]{}%
\providecommand \@@endlink[0]{}%
\providecommand \url  [0]{\begingroup\@sanitize@url \@url }%
\providecommand \@url [1]{\endgroup\@href {#1}{\urlprefix }}%
\providecommand \urlprefix  [0]{URL }%
\providecommand \Eprint [0]{\href }%
\providecommand \doibase [0]{http://dx.doi.org/}%
\providecommand \selectlanguage [0]{\@gobble}%
\providecommand \bibinfo  [0]{\@secondoftwo}%
\providecommand \bibfield  [0]{\@secondoftwo}%
\providecommand \translation [1]{[#1]}%
\providecommand \BibitemOpen [0]{}%
\providecommand \bibitemStop [0]{}%
\providecommand \bibitemNoStop [0]{.\EOS\space}%
\providecommand \EOS [0]{\spacefactor3000\relax}%
\providecommand \BibitemShut  [1]{\csname bibitem#1\endcsname}%
\let\auto@bib@innerbib\@empty
\bibitem [{\citenamefont {Shukla}, \citenamefont {Qian},\ and\ \citenamefont {Rakheja}(2025)}]{shukla2025spintronic}%
  \BibitemOpen
  \bibfield  {author} {\bibinfo {author} {\bibfnamefont {A.}~\bibnamefont {Shukla}}, \bibinfo {author} {\bibfnamefont {S.}~\bibnamefont {Qian}}, \ and\ \bibinfo {author} {\bibfnamefont {S.}~\bibnamefont {Rakheja}},\ }\bibfield  {title} {\enquote {\bibinfo {title} {Spintronic devices and applications using noncollinear chiral antiferromagnets},}\ }\href@noop {} {\bibfield  {journal} {\bibinfo  {journal} {Nanoscale Horizons}\ } (\bibinfo {year} {2025})}\BibitemShut {NoStop}%
\bibitem [{\citenamefont {Rimmler}, \citenamefont {Pal},\ and\ \citenamefont {Parkin}(2025)}]{rimmler2025non}%
  \BibitemOpen
  \bibfield  {author} {\bibinfo {author} {\bibfnamefont {B.~H.}\ \bibnamefont {Rimmler}}, \bibinfo {author} {\bibfnamefont {B.}~\bibnamefont {Pal}}, \ and\ \bibinfo {author} {\bibfnamefont {S.~S.}\ \bibnamefont {Parkin}},\ }\bibfield  {title} {\enquote {\bibinfo {title} {Non-collinear antiferromagnetic spintronics},}\ }\href@noop {} {\bibfield  {journal} {\bibinfo  {journal} {Nature Reviews Materials}\ }\textbf {\bibinfo {volume} {10}},\ \bibinfo {pages} {109--127} (\bibinfo {year} {2025})}\BibitemShut {NoStop}%
\bibitem [{\citenamefont {Nakatsuji}, \citenamefont {Kiyohara},\ and\ \citenamefont {Higo}(2015)}]{nakatsuji2015large}%
  \BibitemOpen
  \bibfield  {author} {\bibinfo {author} {\bibfnamefont {S.}~\bibnamefont {Nakatsuji}}, \bibinfo {author} {\bibfnamefont {N.}~\bibnamefont {Kiyohara}}, \ and\ \bibinfo {author} {\bibfnamefont {T.}~\bibnamefont {Higo}},\ }\bibfield  {title} {\enquote {\bibinfo {title} {Large anomalous hall effect in a non-collinear antiferromagnet at room temperature},}\ }\href@noop {} {\bibfield  {journal} {\bibinfo  {journal} {Nature}\ }\textbf {\bibinfo {volume} {527}},\ \bibinfo {pages} {212--215} (\bibinfo {year} {2015})}\BibitemShut {NoStop}%
\bibitem [{\citenamefont {Nayak}\ \emph {et~al.}(2016)\citenamefont {Nayak}, \citenamefont {Fischer}, \citenamefont {Sun}, \citenamefont {Yan}, \citenamefont {Karel}, \citenamefont {Komarek}, \citenamefont {Shekhar}, \citenamefont {Kumar}, \citenamefont {Schnelle}, \citenamefont {K{\"u}bler} \emph {et~al.}}]{nayak2016large}%
  \BibitemOpen
  \bibfield  {author} {\bibinfo {author} {\bibfnamefont {A.~K.}\ \bibnamefont {Nayak}}, \bibinfo {author} {\bibfnamefont {J.~E.}\ \bibnamefont {Fischer}}, \bibinfo {author} {\bibfnamefont {Y.}~\bibnamefont {Sun}}, \bibinfo {author} {\bibfnamefont {B.}~\bibnamefont {Yan}}, \bibinfo {author} {\bibfnamefont {J.}~\bibnamefont {Karel}}, \bibinfo {author} {\bibfnamefont {A.~C.}\ \bibnamefont {Komarek}}, \bibinfo {author} {\bibfnamefont {C.}~\bibnamefont {Shekhar}}, \bibinfo {author} {\bibfnamefont {N.}~\bibnamefont {Kumar}}, \bibinfo {author} {\bibfnamefont {W.}~\bibnamefont {Schnelle}}, \bibinfo {author} {\bibfnamefont {J.}~\bibnamefont {K{\"u}bler}},  \emph {et~al.},\ }\bibfield  {title} {\enquote {\bibinfo {title} {Large anomalous hall effect driven by a nonvanishing berry curvature in the noncolinear antiferromagnet {Mn$_3$Ge}},}\ }\href@noop {} {\bibfield  {journal} {\bibinfo  {journal} {Science advances}\ }\textbf {\bibinfo {volume} {2}},\ \bibinfo {pages} {e1501870} (\bibinfo {year} {2016})}\BibitemShut
  {NoStop}%
\bibitem [{\citenamefont {Chen}, \citenamefont {Niu},\ and\ \citenamefont {MacDonald}(2014)}]{chen2014anomalous}%
  \BibitemOpen
  \bibfield  {author} {\bibinfo {author} {\bibfnamefont {H.}~\bibnamefont {Chen}}, \bibinfo {author} {\bibfnamefont {Q.}~\bibnamefont {Niu}}, \ and\ \bibinfo {author} {\bibfnamefont {A.~H.}\ \bibnamefont {MacDonald}},\ }\bibfield  {title} {\enquote {\bibinfo {title} {Anomalous hall effect arising from noncollinear antiferromagnetism},}\ }\href@noop {} {\bibfield  {journal} {\bibinfo  {journal} {Physical review letters}\ }\textbf {\bibinfo {volume} {112}},\ \bibinfo {pages} {017205} (\bibinfo {year} {2014})}\BibitemShut {NoStop}%
\bibitem [{\citenamefont {Zhang}\ \emph {et~al.}(2016)\citenamefont {Zhang}, \citenamefont {Han}, \citenamefont {Yang}, \citenamefont {Sun}, \citenamefont {Zhang}, \citenamefont {Yan},\ and\ \citenamefont {Parkin}}]{zhang2016giant}%
  \BibitemOpen
  \bibfield  {author} {\bibinfo {author} {\bibfnamefont {W.}~\bibnamefont {Zhang}}, \bibinfo {author} {\bibfnamefont {W.}~\bibnamefont {Han}}, \bibinfo {author} {\bibfnamefont {S.-H.}\ \bibnamefont {Yang}}, \bibinfo {author} {\bibfnamefont {Y.}~\bibnamefont {Sun}}, \bibinfo {author} {\bibfnamefont {Y.}~\bibnamefont {Zhang}}, \bibinfo {author} {\bibfnamefont {B.}~\bibnamefont {Yan}}, \ and\ \bibinfo {author} {\bibfnamefont {S.~S.}\ \bibnamefont {Parkin}},\ }\bibfield  {title} {\enquote {\bibinfo {title} {Giant facet-dependent spin-orbit torque and spin hall conductivity in the triangular antiferromagnet irmn3},}\ }\href@noop {} {\bibfield  {journal} {\bibinfo  {journal} {Science Advances}\ }\textbf {\bibinfo {volume} {2}},\ \bibinfo {pages} {e1600759} (\bibinfo {year} {2016})}\BibitemShut {NoStop}%
\bibitem [{\citenamefont {Ikhlas}\ \emph {et~al.}(2017)\citenamefont {Ikhlas}, \citenamefont {Tomita}, \citenamefont {Koretsune}, \citenamefont {Suzuki}, \citenamefont {Nishio-Hamane}, \citenamefont {Arita}, \citenamefont {Otani},\ and\ \citenamefont {Nakatsuji}}]{ikhlas2017large}%
  \BibitemOpen
  \bibfield  {author} {\bibinfo {author} {\bibfnamefont {M.}~\bibnamefont {Ikhlas}}, \bibinfo {author} {\bibfnamefont {T.}~\bibnamefont {Tomita}}, \bibinfo {author} {\bibfnamefont {T.}~\bibnamefont {Koretsune}}, \bibinfo {author} {\bibfnamefont {M.-T.}\ \bibnamefont {Suzuki}}, \bibinfo {author} {\bibfnamefont {D.}~\bibnamefont {Nishio-Hamane}}, \bibinfo {author} {\bibfnamefont {R.}~\bibnamefont {Arita}}, \bibinfo {author} {\bibfnamefont {Y.}~\bibnamefont {Otani}}, \ and\ \bibinfo {author} {\bibfnamefont {S.}~\bibnamefont {Nakatsuji}},\ }\bibfield  {title} {\enquote {\bibinfo {title} {Large anomalous nernst effect at room temperature in a chiral antiferromagnet},}\ }\href@noop {} {\bibfield  {journal} {\bibinfo  {journal} {Nature Physics}\ }\textbf {\bibinfo {volume} {13}},\ \bibinfo {pages} {1085--1090} (\bibinfo {year} {2017})}\BibitemShut {NoStop}%
\bibitem [{\citenamefont {Hong}\ \emph {et~al.}(2020)\citenamefont {Hong}, \citenamefont {Anand}, \citenamefont {Liu}, \citenamefont {Liu}, \citenamefont {Arslan}, \citenamefont {Pearson}, \citenamefont {Bhattacharya},\ and\ \citenamefont {Jiang}}]{hong2020large}%
  \BibitemOpen
  \bibfield  {author} {\bibinfo {author} {\bibfnamefont {D.}~\bibnamefont {Hong}}, \bibinfo {author} {\bibfnamefont {N.}~\bibnamefont {Anand}}, \bibinfo {author} {\bibfnamefont {C.}~\bibnamefont {Liu}}, \bibinfo {author} {\bibfnamefont {H.}~\bibnamefont {Liu}}, \bibinfo {author} {\bibfnamefont {I.}~\bibnamefont {Arslan}}, \bibinfo {author} {\bibfnamefont {J.~E.}\ \bibnamefont {Pearson}}, \bibinfo {author} {\bibfnamefont {A.}~\bibnamefont {Bhattacharya}}, \ and\ \bibinfo {author} {\bibfnamefont {J.}~\bibnamefont {Jiang}},\ }\bibfield  {title} {\enquote {\bibinfo {title} {Large anomalous nernst and inverse spin-hall effects in epitaxial thin films of kagome semimetal {Mn$_3$Ge}},}\ }\href@noop {} {\bibfield  {journal} {\bibinfo  {journal} {Physical Review Materials}\ }\textbf {\bibinfo {volume} {4}},\ \bibinfo {pages} {094201} (\bibinfo {year} {2020})}\BibitemShut {NoStop}%
\bibitem [{\citenamefont {Higo}\ \emph {et~al.}(2018)\citenamefont {Higo}, \citenamefont {Man}, \citenamefont {Gopman}, \citenamefont {Wu}, \citenamefont {Koretsune}, \citenamefont {van’t Erve}, \citenamefont {Kabanov}, \citenamefont {Rees}, \citenamefont {Li}, \citenamefont {Suzuki} \emph {et~al.}}]{higo2018large}%
  \BibitemOpen
  \bibfield  {author} {\bibinfo {author} {\bibfnamefont {T.}~\bibnamefont {Higo}}, \bibinfo {author} {\bibfnamefont {H.}~\bibnamefont {Man}}, \bibinfo {author} {\bibfnamefont {D.~B.}\ \bibnamefont {Gopman}}, \bibinfo {author} {\bibfnamefont {L.}~\bibnamefont {Wu}}, \bibinfo {author} {\bibfnamefont {T.}~\bibnamefont {Koretsune}}, \bibinfo {author} {\bibfnamefont {O.~M.}\ \bibnamefont {van’t Erve}}, \bibinfo {author} {\bibfnamefont {Y.~P.}\ \bibnamefont {Kabanov}}, \bibinfo {author} {\bibfnamefont {D.}~\bibnamefont {Rees}}, \bibinfo {author} {\bibfnamefont {Y.}~\bibnamefont {Li}}, \bibinfo {author} {\bibfnamefont {M.-T.}\ \bibnamefont {Suzuki}},  \emph {et~al.},\ }\bibfield  {title} {\enquote {\bibinfo {title} {Large magneto-optical kerr effect and imaging of magnetic octupole domains in an antiferromagnetic metal},}\ }\href@noop {} {\bibfield  {journal} {\bibinfo  {journal} {Nature photonics}\ }\textbf {\bibinfo {volume} {12}},\ \bibinfo {pages} {73--78} (\bibinfo {year} {2018})}\BibitemShut {NoStop}%
\bibitem [{\citenamefont {Dong}\ \emph {et~al.}(2022)\citenamefont {Dong}, \citenamefont {Li}, \citenamefont {Gurung}, \citenamefont {Zhu}, \citenamefont {Zhang}, \citenamefont {Zheng}, \citenamefont {Tsymbal},\ and\ \citenamefont {Zhang}}]{dong2022tunneling}%
  \BibitemOpen
  \bibfield  {author} {\bibinfo {author} {\bibfnamefont {J.}~\bibnamefont {Dong}}, \bibinfo {author} {\bibfnamefont {X.}~\bibnamefont {Li}}, \bibinfo {author} {\bibfnamefont {G.}~\bibnamefont {Gurung}}, \bibinfo {author} {\bibfnamefont {M.}~\bibnamefont {Zhu}}, \bibinfo {author} {\bibfnamefont {P.}~\bibnamefont {Zhang}}, \bibinfo {author} {\bibfnamefont {F.}~\bibnamefont {Zheng}}, \bibinfo {author} {\bibfnamefont {E.~Y.}\ \bibnamefont {Tsymbal}}, \ and\ \bibinfo {author} {\bibfnamefont {J.}~\bibnamefont {Zhang}},\ }\bibfield  {title} {\enquote {\bibinfo {title} {Tunneling magnetoresistance in noncollinear antiferromagnetic tunnel junctions},}\ }\href@noop {} {\bibfield  {journal} {\bibinfo  {journal} {Physical Review Letters}\ }\textbf {\bibinfo {volume} {128}},\ \bibinfo {pages} {197201} (\bibinfo {year} {2022})}\BibitemShut {NoStop}%
\bibitem [{\citenamefont {Chen}\ \emph {et~al.}(2023)\citenamefont {Chen}, \citenamefont {Higo}, \citenamefont {Tanaka}, \citenamefont {Nomoto}, \citenamefont {Tsai}, \citenamefont {Idzuchi}, \citenamefont {Shiga}, \citenamefont {Sakamoto}, \citenamefont {Ando}, \citenamefont {Kosaki} \emph {et~al.}}]{chen2023octupole}%
  \BibitemOpen
  \bibfield  {author} {\bibinfo {author} {\bibfnamefont {X.}~\bibnamefont {Chen}}, \bibinfo {author} {\bibfnamefont {T.}~\bibnamefont {Higo}}, \bibinfo {author} {\bibfnamefont {K.}~\bibnamefont {Tanaka}}, \bibinfo {author} {\bibfnamefont {T.}~\bibnamefont {Nomoto}}, \bibinfo {author} {\bibfnamefont {H.}~\bibnamefont {Tsai}}, \bibinfo {author} {\bibfnamefont {H.}~\bibnamefont {Idzuchi}}, \bibinfo {author} {\bibfnamefont {M.}~\bibnamefont {Shiga}}, \bibinfo {author} {\bibfnamefont {S.}~\bibnamefont {Sakamoto}}, \bibinfo {author} {\bibfnamefont {R.}~\bibnamefont {Ando}}, \bibinfo {author} {\bibfnamefont {H.}~\bibnamefont {Kosaki}},  \emph {et~al.},\ }\bibfield  {title} {\enquote {\bibinfo {title} {Octupole-driven magnetoresistance in an antiferromagnetic tunnel junction},}\ }\href@noop {} {\bibfield  {journal} {\bibinfo  {journal} {Nature}\ }\textbf {\bibinfo {volume} {613}},\ \bibinfo {pages} {490--495} (\bibinfo {year} {2023})}\BibitemShut {NoStop}%
\bibitem [{\citenamefont {Qin}\ \emph {et~al.}(2023)\citenamefont {Qin}, \citenamefont {Yan}, \citenamefont {Wang}, \citenamefont {Chen}, \citenamefont {Meng}, \citenamefont {Dong}, \citenamefont {Zhu}, \citenamefont {Cai}, \citenamefont {Feng}, \citenamefont {Zhou} \emph {et~al.}}]{qin2023room}%
  \BibitemOpen
  \bibfield  {author} {\bibinfo {author} {\bibfnamefont {P.}~\bibnamefont {Qin}}, \bibinfo {author} {\bibfnamefont {H.}~\bibnamefont {Yan}}, \bibinfo {author} {\bibfnamefont {X.}~\bibnamefont {Wang}}, \bibinfo {author} {\bibfnamefont {H.}~\bibnamefont {Chen}}, \bibinfo {author} {\bibfnamefont {Z.}~\bibnamefont {Meng}}, \bibinfo {author} {\bibfnamefont {J.}~\bibnamefont {Dong}}, \bibinfo {author} {\bibfnamefont {M.}~\bibnamefont {Zhu}}, \bibinfo {author} {\bibfnamefont {J.}~\bibnamefont {Cai}}, \bibinfo {author} {\bibfnamefont {Z.}~\bibnamefont {Feng}}, \bibinfo {author} {\bibfnamefont {X.}~\bibnamefont {Zhou}},  \emph {et~al.},\ }\bibfield  {title} {\enquote {\bibinfo {title} {Room-temperature magnetoresistance in an all-antiferromagnetic tunnel junction},}\ }\href@noop {} {\bibfield  {journal} {\bibinfo  {journal} {Nature}\ }\textbf {\bibinfo {volume} {613}},\ \bibinfo {pages} {485--489} (\bibinfo {year} {2023})}\BibitemShut {NoStop}%
\bibitem [{\citenamefont {Wang}\ \emph {et~al.}(2024)\citenamefont {Wang}, \citenamefont {Bian}, \citenamefont {Zhang},\ and\ \citenamefont {Yu}}]{wang2024mn3sn}%
  \BibitemOpen
  \bibfield  {author} {\bibinfo {author} {\bibfnamefont {Z.}~\bibnamefont {Wang}}, \bibinfo {author} {\bibfnamefont {B.}~\bibnamefont {Bian}}, \bibinfo {author} {\bibfnamefont {L.}~\bibnamefont {Zhang}}, \ and\ \bibinfo {author} {\bibfnamefont {Z.}~\bibnamefont {Yu}},\ }\bibfield  {title} {\enquote {\bibinfo {title} {{Mn$_3$Sn}-based noncollinear antiferromagnetic tunnel junctions with bilayer boron nitride tunnel barriers},}\ }\href@noop {} {\bibfield  {journal} {\bibinfo  {journal} {Applied Physics Letters}\ }\textbf {\bibinfo {volume} {125}} (\bibinfo {year} {2024})}\BibitemShut {NoStop}%
\bibitem [{\citenamefont {Liu}\ \emph {et~al.}(2023{\natexlab{a}})\citenamefont {Liu}, \citenamefont {Zhang}, \citenamefont {Fu}, \citenamefont {Zhao}, \citenamefont {Xie}, \citenamefont {Cao}, \citenamefont {Bai}, \citenamefont {Kang}, \citenamefont {Chen}, \citenamefont {Yan} \emph {et~al.}}]{liu2023anomalous}%
  \BibitemOpen
  \bibfield  {author} {\bibinfo {author} {\bibfnamefont {J.}~\bibnamefont {Liu}}, \bibinfo {author} {\bibfnamefont {Z.}~\bibnamefont {Zhang}}, \bibinfo {author} {\bibfnamefont {M.}~\bibnamefont {Fu}}, \bibinfo {author} {\bibfnamefont {X.}~\bibnamefont {Zhao}}, \bibinfo {author} {\bibfnamefont {R.}~\bibnamefont {Xie}}, \bibinfo {author} {\bibfnamefont {Q.}~\bibnamefont {Cao}}, \bibinfo {author} {\bibfnamefont {L.}~\bibnamefont {Bai}}, \bibinfo {author} {\bibfnamefont {S.}~\bibnamefont {Kang}}, \bibinfo {author} {\bibfnamefont {Y.}~\bibnamefont {Chen}}, \bibinfo {author} {\bibfnamefont {S.}~\bibnamefont {Yan}},  \emph {et~al.},\ }\bibfield  {title} {\enquote {\bibinfo {title} {The anomalous {H}all effect controlled by residual epitaxial strain in antiferromagnetic weyl semimetal {Mn$_3$Sn} thin films grown by molecular beam epitaxy},}\ }\href@noop {} {\bibfield  {journal} {\bibinfo  {journal} {Results in Physics}\ ,\ \bibinfo {pages} {106803}} (\bibinfo {year} {2023}{\natexlab{a}})}\BibitemShut {NoStop}%
\bibitem [{\citenamefont {Yano}\ \emph {et~al.}(2024)\citenamefont {Yano}, \citenamefont {Kihara}, \citenamefont {Yoneda}, \citenamefont {Vu}, \citenamefont {Suto}, \citenamefont {Katayama}, \citenamefont {Yamaguchi}, \citenamefont {Kuwahara}, \citenamefont {Suzuki}, \citenamefont {Saitoh} \emph {et~al.}}]{yano2024giant}%
  \BibitemOpen
  \bibfield  {author} {\bibinfo {author} {\bibfnamefont {R.}~\bibnamefont {Yano}}, \bibinfo {author} {\bibfnamefont {S.}~\bibnamefont {Kihara}}, \bibinfo {author} {\bibfnamefont {M.}~\bibnamefont {Yoneda}}, \bibinfo {author} {\bibfnamefont {H.~T.~N.}\ \bibnamefont {Vu}}, \bibinfo {author} {\bibfnamefont {H.}~\bibnamefont {Suto}}, \bibinfo {author} {\bibfnamefont {N.}~\bibnamefont {Katayama}}, \bibinfo {author} {\bibfnamefont {T.}~\bibnamefont {Yamaguchi}}, \bibinfo {author} {\bibfnamefont {M.}~\bibnamefont {Kuwahara}}, \bibinfo {author} {\bibfnamefont {M.-T.}\ \bibnamefont {Suzuki}}, \bibinfo {author} {\bibfnamefont {K.}~\bibnamefont {Saitoh}},  \emph {et~al.},\ }\bibfield  {title} {\enquote {\bibinfo {title} {Giant impurity effect on anomalous hall effect of {Mn$_3$Sn}},}\ }\href@noop {} {\bibfield  {journal} {\bibinfo  {journal} {The Journal of Chemical Physics}\ }\textbf {\bibinfo {volume} {160}} (\bibinfo {year} {2024})}\BibitemShut {NoStop}%
\bibitem [{\citenamefont {Sung}\ \emph {et~al.}(2018)\citenamefont {Sung}, \citenamefont {Ronning}, \citenamefont {Thompson},\ and\ \citenamefont {Bauer}}]{sung2018magnetic}%
  \BibitemOpen
  \bibfield  {author} {\bibinfo {author} {\bibfnamefont {N.~H.}\ \bibnamefont {Sung}}, \bibinfo {author} {\bibfnamefont {F.}~\bibnamefont {Ronning}}, \bibinfo {author} {\bibfnamefont {J.~D.}\ \bibnamefont {Thompson}}, \ and\ \bibinfo {author} {\bibfnamefont {E.~D.}\ \bibnamefont {Bauer}},\ }\bibfield  {title} {\enquote {\bibinfo {title} {Magnetic phase dependence of the anomalous hall effect in {Mn$_3$Sn} single crystals},}\ }\href@noop {} {\bibfield  {journal} {\bibinfo  {journal} {Applied Physics Letters}\ }\textbf {\bibinfo {volume} {112}} (\bibinfo {year} {2018})}\BibitemShut {NoStop}%
\bibitem [{\citenamefont {Higo}\ \emph {et~al.}(2022)\citenamefont {Higo}, \citenamefont {Kondou}, \citenamefont {Nomoto}, \citenamefont {Shiga}, \citenamefont {Sakamoto}, \citenamefont {Chen}, \citenamefont {Nishio-Hamane}, \citenamefont {Arita}, \citenamefont {Otani}, \citenamefont {Miwa} \emph {et~al.}}]{higo2022perpendicular}%
  \BibitemOpen
  \bibfield  {author} {\bibinfo {author} {\bibfnamefont {T.}~\bibnamefont {Higo}}, \bibinfo {author} {\bibfnamefont {K.}~\bibnamefont {Kondou}}, \bibinfo {author} {\bibfnamefont {T.}~\bibnamefont {Nomoto}}, \bibinfo {author} {\bibfnamefont {M.}~\bibnamefont {Shiga}}, \bibinfo {author} {\bibfnamefont {S.}~\bibnamefont {Sakamoto}}, \bibinfo {author} {\bibfnamefont {X.}~\bibnamefont {Chen}}, \bibinfo {author} {\bibfnamefont {D.}~\bibnamefont {Nishio-Hamane}}, \bibinfo {author} {\bibfnamefont {R.}~\bibnamefont {Arita}}, \bibinfo {author} {\bibfnamefont {Y.}~\bibnamefont {Otani}}, \bibinfo {author} {\bibfnamefont {S.}~\bibnamefont {Miwa}},  \emph {et~al.},\ }\bibfield  {title} {\enquote {\bibinfo {title} {Perpendicular full switching of chiral antiferromagnetic order by current},}\ }\href@noop {} {\bibfield  {journal} {\bibinfo  {journal} {Nature}\ }\textbf {\bibinfo {volume} {607}},\ \bibinfo {pages} {474--479} (\bibinfo {year} {2022})}\BibitemShut {NoStop}%
\bibitem [{\citenamefont {Yoon}\ \emph {et~al.}(2023)\citenamefont {Yoon}, \citenamefont {Zhang}, \citenamefont {Chou}, \citenamefont {Takeuchi}, \citenamefont {Uchimura}, \citenamefont {Hou}, \citenamefont {Han}, \citenamefont {Kanai}, \citenamefont {Ohno}, \citenamefont {Fukami} \emph {et~al.}}]{yoon2023handedness}%
  \BibitemOpen
  \bibfield  {author} {\bibinfo {author} {\bibfnamefont {J.-Y.}\ \bibnamefont {Yoon}}, \bibinfo {author} {\bibfnamefont {P.}~\bibnamefont {Zhang}}, \bibinfo {author} {\bibfnamefont {C.-T.}\ \bibnamefont {Chou}}, \bibinfo {author} {\bibfnamefont {Y.}~\bibnamefont {Takeuchi}}, \bibinfo {author} {\bibfnamefont {T.}~\bibnamefont {Uchimura}}, \bibinfo {author} {\bibfnamefont {J.~T.}\ \bibnamefont {Hou}}, \bibinfo {author} {\bibfnamefont {J.}~\bibnamefont {Han}}, \bibinfo {author} {\bibfnamefont {S.}~\bibnamefont {Kanai}}, \bibinfo {author} {\bibfnamefont {H.}~\bibnamefont {Ohno}}, \bibinfo {author} {\bibfnamefont {S.}~\bibnamefont {Fukami}},  \emph {et~al.},\ }\bibfield  {title} {\enquote {\bibinfo {title} {Handedness anomaly in a non-collinear antiferromagnet under spin--orbit torque},}\ }\href@noop {} {\bibfield  {journal} {\bibinfo  {journal} {Nature Materials}\ }\textbf {\bibinfo {volume} {22}},\ \bibinfo {pages} {1106--1113} (\bibinfo {year} {2023})}\BibitemShut {NoStop}%
\bibitem [{\citenamefont {Tsai}\ \emph {et~al.}(2020)\citenamefont {Tsai}, \citenamefont {Higo}, \citenamefont {Kondou}, \citenamefont {Nomoto}, \citenamefont {Sakai}, \citenamefont {Kobayashi}, \citenamefont {Nakano}, \citenamefont {Yakushiji}, \citenamefont {Arita}, \citenamefont {Miwa} \emph {et~al.}}]{tsai2020electrical}%
  \BibitemOpen
  \bibfield  {author} {\bibinfo {author} {\bibfnamefont {H.}~\bibnamefont {Tsai}}, \bibinfo {author} {\bibfnamefont {T.}~\bibnamefont {Higo}}, \bibinfo {author} {\bibfnamefont {K.}~\bibnamefont {Kondou}}, \bibinfo {author} {\bibfnamefont {T.}~\bibnamefont {Nomoto}}, \bibinfo {author} {\bibfnamefont {A.}~\bibnamefont {Sakai}}, \bibinfo {author} {\bibfnamefont {A.}~\bibnamefont {Kobayashi}}, \bibinfo {author} {\bibfnamefont {T.}~\bibnamefont {Nakano}}, \bibinfo {author} {\bibfnamefont {K.}~\bibnamefont {Yakushiji}}, \bibinfo {author} {\bibfnamefont {R.}~\bibnamefont {Arita}}, \bibinfo {author} {\bibfnamefont {S.}~\bibnamefont {Miwa}},  \emph {et~al.},\ }\bibfield  {title} {\enquote {\bibinfo {title} {Electrical manipulation of a topological antiferromagnetic state},}\ }\href@noop {} {\bibfield  {journal} {\bibinfo  {journal} {Nature}\ }\textbf {\bibinfo {volume} {580}},\ \bibinfo {pages} {608--613} (\bibinfo {year} {2020})}\BibitemShut {NoStop}%
\bibitem [{\citenamefont {Takeuchi}\ \emph {et~al.}(2021)\citenamefont {Takeuchi}, \citenamefont {Yamane}, \citenamefont {Yoon}, \citenamefont {Itoh}, \citenamefont {Jinnai}, \citenamefont {Kanai}, \citenamefont {Ieda}, \citenamefont {Fukami},\ and\ \citenamefont {Ohno}}]{takeuchi2021chiral}%
  \BibitemOpen
  \bibfield  {author} {\bibinfo {author} {\bibfnamefont {Y.}~\bibnamefont {Takeuchi}}, \bibinfo {author} {\bibfnamefont {Y.}~\bibnamefont {Yamane}}, \bibinfo {author} {\bibfnamefont {J.-Y.}\ \bibnamefont {Yoon}}, \bibinfo {author} {\bibfnamefont {R.}~\bibnamefont {Itoh}}, \bibinfo {author} {\bibfnamefont {B.}~\bibnamefont {Jinnai}}, \bibinfo {author} {\bibfnamefont {S.}~\bibnamefont {Kanai}}, \bibinfo {author} {\bibfnamefont {J.}~\bibnamefont {Ieda}}, \bibinfo {author} {\bibfnamefont {S.}~\bibnamefont {Fukami}}, \ and\ \bibinfo {author} {\bibfnamefont {H.}~\bibnamefont {Ohno}},\ }\bibfield  {title} {\enquote {\bibinfo {title} {Chiral-spin rotation of non-collinear antiferromagnet by spin--orbit torque},}\ }\href@noop {} {\bibfield  {journal} {\bibinfo  {journal} {Nature Materials}\ }\textbf {\bibinfo {volume} {20}},\ \bibinfo {pages} {1364--1370} (\bibinfo {year} {2021})}\BibitemShut {NoStop}%
\bibitem [{\citenamefont {Yan}\ \emph {et~al.}(2022)\citenamefont {Yan}, \citenamefont {Li}, \citenamefont {Lu}, \citenamefont {Huang}, \citenamefont {Xiao}, \citenamefont {Wernert}, \citenamefont {Brock}, \citenamefont {Fullerton}, \citenamefont {Chen}, \citenamefont {Wang} \emph {et~al.}}]{yan2022quantum}%
  \BibitemOpen
  \bibfield  {author} {\bibinfo {author} {\bibfnamefont {G.~Q.}\ \bibnamefont {Yan}}, \bibinfo {author} {\bibfnamefont {S.}~\bibnamefont {Li}}, \bibinfo {author} {\bibfnamefont {H.}~\bibnamefont {Lu}}, \bibinfo {author} {\bibfnamefont {M.}~\bibnamefont {Huang}}, \bibinfo {author} {\bibfnamefont {Y.}~\bibnamefont {Xiao}}, \bibinfo {author} {\bibfnamefont {L.}~\bibnamefont {Wernert}}, \bibinfo {author} {\bibfnamefont {J.~A.}\ \bibnamefont {Brock}}, \bibinfo {author} {\bibfnamefont {E.~E.}\ \bibnamefont {Fullerton}}, \bibinfo {author} {\bibfnamefont {H.}~\bibnamefont {Chen}}, \bibinfo {author} {\bibfnamefont {H.}~\bibnamefont {Wang}},  \emph {et~al.},\ }\bibfield  {title} {\enquote {\bibinfo {title} {Quantum sensing and imaging of spin--orbit-torque-driven spin dynamics in the non-collinear antiferromagnet {Mn$_3$Sn}},}\ }\href@noop {} {\bibfield  {journal} {\bibinfo  {journal} {Advanced Materials}\ }\textbf {\bibinfo {volume} {34}},\ \bibinfo {pages} {2200327} (\bibinfo {year} {2022})}\BibitemShut {NoStop}%
\bibitem [{\citenamefont {Pal}\ \emph {et~al.}(2022)\citenamefont {Pal}, \citenamefont {Hazra}, \citenamefont {G{\"o}bel}, \citenamefont {Jeon}, \citenamefont {Pandeya}, \citenamefont {Chakraborty}, \citenamefont {Busch}, \citenamefont {Srivastava}, \citenamefont {Deniz}, \citenamefont {Taylor} \emph {et~al.}}]{pal2022setting}%
  \BibitemOpen
  \bibfield  {author} {\bibinfo {author} {\bibfnamefont {B.}~\bibnamefont {Pal}}, \bibinfo {author} {\bibfnamefont {B.~K.}\ \bibnamefont {Hazra}}, \bibinfo {author} {\bibfnamefont {B.}~\bibnamefont {G{\"o}bel}}, \bibinfo {author} {\bibfnamefont {J.-C.}\ \bibnamefont {Jeon}}, \bibinfo {author} {\bibfnamefont {A.~K.}\ \bibnamefont {Pandeya}}, \bibinfo {author} {\bibfnamefont {A.}~\bibnamefont {Chakraborty}}, \bibinfo {author} {\bibfnamefont {O.}~\bibnamefont {Busch}}, \bibinfo {author} {\bibfnamefont {A.~K.}\ \bibnamefont {Srivastava}}, \bibinfo {author} {\bibfnamefont {H.}~\bibnamefont {Deniz}}, \bibinfo {author} {\bibfnamefont {J.~M.}\ \bibnamefont {Taylor}},  \emph {et~al.},\ }\bibfield  {title} {\enquote {\bibinfo {title} {Setting of the magnetic structure of chiral kagome antiferromagnets by a seeded spin-orbit torque},}\ }\href@noop {} {\bibfield  {journal} {\bibinfo  {journal} {Science Advances}\ }\textbf {\bibinfo {volume} {8}},\ \bibinfo {pages} {eabo5930} (\bibinfo {year} {2022})}\BibitemShut
  {NoStop}%
\bibitem [{\citenamefont {Krishnaswamy}\ \emph {et~al.}(2022)\citenamefont {Krishnaswamy}, \citenamefont {Sala}, \citenamefont {Jacot}, \citenamefont {Lambert}, \citenamefont {Schlitz}, \citenamefont {Rossell}, \citenamefont {N{\"o}el},\ and\ \citenamefont {Gambardella}}]{krishnaswamy2022time}%
  \BibitemOpen
  \bibfield  {author} {\bibinfo {author} {\bibfnamefont {G.~K.}\ \bibnamefont {Krishnaswamy}}, \bibinfo {author} {\bibfnamefont {G.}~\bibnamefont {Sala}}, \bibinfo {author} {\bibfnamefont {B.}~\bibnamefont {Jacot}}, \bibinfo {author} {\bibfnamefont {C.-H.}\ \bibnamefont {Lambert}}, \bibinfo {author} {\bibfnamefont {R.}~\bibnamefont {Schlitz}}, \bibinfo {author} {\bibfnamefont {M.~D.}\ \bibnamefont {Rossell}}, \bibinfo {author} {\bibfnamefont {P.}~\bibnamefont {N{\"o}el}}, \ and\ \bibinfo {author} {\bibfnamefont {P.}~\bibnamefont {Gambardella}},\ }\bibfield  {title} {\enquote {\bibinfo {title} {Time-dependent multistate switching of topological antiferromagnetic order in {Mn$_3$Sn}},}\ }\href@noop {} {\bibfield  {journal} {\bibinfo  {journal} {Physical Review Applied}\ }\textbf {\bibinfo {volume} {18}},\ \bibinfo {pages} {024064} (\bibinfo {year} {2022})}\BibitemShut {NoStop}%
\bibitem [{\citenamefont {Xu}\ \emph {et~al.}(2023)\citenamefont {Xu}, \citenamefont {Bai}, \citenamefont {Dong}, \citenamefont {Zhao}, \citenamefont {Zhou}, \citenamefont {Zhang}, \citenamefont {Zhang},\ and\ \citenamefont {Jiang}}]{xu2023robust}%
  \BibitemOpen
  \bibfield  {author} {\bibinfo {author} {\bibfnamefont {T.}~\bibnamefont {Xu}}, \bibinfo {author} {\bibfnamefont {H.}~\bibnamefont {Bai}}, \bibinfo {author} {\bibfnamefont {Y.}~\bibnamefont {Dong}}, \bibinfo {author} {\bibfnamefont {L.}~\bibnamefont {Zhao}}, \bibinfo {author} {\bibfnamefont {H.-A.}\ \bibnamefont {Zhou}}, \bibinfo {author} {\bibfnamefont {J.}~\bibnamefont {Zhang}}, \bibinfo {author} {\bibfnamefont {X.-X.}\ \bibnamefont {Zhang}}, \ and\ \bibinfo {author} {\bibfnamefont {W.}~\bibnamefont {Jiang}},\ }\bibfield  {title} {\enquote {\bibinfo {title} {Robust spin torque switching of noncollinear antiferromagnet {Mn$_3$Sn}},}\ }\href@noop {} {\bibfield  {journal} {\bibinfo  {journal} {APL Materials}\ }\textbf {\bibinfo {volume} {11}} (\bibinfo {year} {2023})}\BibitemShut {NoStop}%
\bibitem [{\citenamefont {Yoo}\ \emph {et~al.}(2024)\citenamefont {Yoo}, \citenamefont {Lorenz}, \citenamefont {Hoffmann},\ and\ \citenamefont {Cahill}}]{yoo2024thermal}%
  \BibitemOpen
  \bibfield  {author} {\bibinfo {author} {\bibfnamefont {M.-W.}\ \bibnamefont {Yoo}}, \bibinfo {author} {\bibfnamefont {V.~O.}\ \bibnamefont {Lorenz}}, \bibinfo {author} {\bibfnamefont {A.}~\bibnamefont {Hoffmann}}, \ and\ \bibinfo {author} {\bibfnamefont {D.~G.}\ \bibnamefont {Cahill}},\ }\bibfield  {title} {\enquote {\bibinfo {title} {{Thermal contribution to current-driven antiferromagnetic-order switching}},}\ }\href@noop {} {\bibfield  {journal} {\bibinfo  {journal} {APL Materials}\ }\textbf {\bibinfo {volume} {12}},\ \bibinfo {pages} {081107} (\bibinfo {year} {2024})}\BibitemShut {NoStop}%
\bibitem [{\citenamefont {Zheng}\ \emph {et~al.}(2025)\citenamefont {Zheng}, \citenamefont {Jia}, \citenamefont {Zhang}, \citenamefont {Shen}, \citenamefont {Zhou}, \citenamefont {Cui}, \citenamefont {Ren}, \citenamefont {Chen}, \citenamefont {Jamaludin}, \citenamefont {Zhao} \emph {et~al.}}]{zheng2025all}%
  \BibitemOpen
  \bibfield  {author} {\bibinfo {author} {\bibfnamefont {Z.}~\bibnamefont {Zheng}}, \bibinfo {author} {\bibfnamefont {L.}~\bibnamefont {Jia}}, \bibinfo {author} {\bibfnamefont {Z.}~\bibnamefont {Zhang}}, \bibinfo {author} {\bibfnamefont {Q.}~\bibnamefont {Shen}}, \bibinfo {author} {\bibfnamefont {G.}~\bibnamefont {Zhou}}, \bibinfo {author} {\bibfnamefont {Z.}~\bibnamefont {Cui}}, \bibinfo {author} {\bibfnamefont {L.}~\bibnamefont {Ren}}, \bibinfo {author} {\bibfnamefont {Z.}~\bibnamefont {Chen}}, \bibinfo {author} {\bibfnamefont {N.~F.}\ \bibnamefont {Jamaludin}}, \bibinfo {author} {\bibfnamefont {T.}~\bibnamefont {Zhao}},  \emph {et~al.},\ }\bibfield  {title} {\enquote {\bibinfo {title} {All-electrical perpendicular switching of chiral antiferromagnetic order},}\ }\href@noop {} {\bibfield  {journal} {\bibinfo  {journal} {Nature Materials}\ ,\ \bibinfo {pages} {1--7}} (\bibinfo {year} {2025})}\BibitemShut {NoStop}%
\bibitem [{\citenamefont {Lee}\ \emph {et~al.}(2025)\citenamefont {Lee}, \citenamefont {Hwang}, \citenamefont {Ko}, \citenamefont {Park}, \citenamefont {Lee},\ and\ \citenamefont {Choi}}]{lee2025spin}%
  \BibitemOpen
  \bibfield  {author} {\bibinfo {author} {\bibfnamefont {W.-B.}\ \bibnamefont {Lee}}, \bibinfo {author} {\bibfnamefont {S.}~\bibnamefont {Hwang}}, \bibinfo {author} {\bibfnamefont {H.-W.}\ \bibnamefont {Ko}}, \bibinfo {author} {\bibfnamefont {B.-G.}\ \bibnamefont {Park}}, \bibinfo {author} {\bibfnamefont {K.-J.}\ \bibnamefont {Lee}}, \ and\ \bibinfo {author} {\bibfnamefont {G.-M.}\ \bibnamefont {Choi}},\ }\bibfield  {title} {\enquote {\bibinfo {title} {Spin-torque-driven gigahertz magnetization dynamics in the non-collinear antiferromagnet {Mn$_3$Sn}},}\ }\href@noop {} {\bibfield  {journal} {\bibinfo  {journal} {Nature Nanotechnology}\ ,\ \bibinfo {pages} {1--7}} (\bibinfo {year} {2025})}\BibitemShut {NoStop}%
\bibitem [{\citenamefont {Deng}\ \emph {et~al.}(2023)\citenamefont {Deng}, \citenamefont {Liu}, \citenamefont {Chen}, \citenamefont {Du}, \citenamefont {Jiang}, \citenamefont {Shen}, \citenamefont {Zhang}, \citenamefont {Zheng}, \citenamefont {Lu},\ and\ \citenamefont {Wang}}]{deng2023all}%
  \BibitemOpen
  \bibfield  {author} {\bibinfo {author} {\bibfnamefont {Y.}~\bibnamefont {Deng}}, \bibinfo {author} {\bibfnamefont {X.}~\bibnamefont {Liu}}, \bibinfo {author} {\bibfnamefont {Y.}~\bibnamefont {Chen}}, \bibinfo {author} {\bibfnamefont {Z.}~\bibnamefont {Du}}, \bibinfo {author} {\bibfnamefont {N.}~\bibnamefont {Jiang}}, \bibinfo {author} {\bibfnamefont {C.}~\bibnamefont {Shen}}, \bibinfo {author} {\bibfnamefont {E.}~\bibnamefont {Zhang}}, \bibinfo {author} {\bibfnamefont {H.}~\bibnamefont {Zheng}}, \bibinfo {author} {\bibfnamefont {H.-Z.}\ \bibnamefont {Lu}}, \ and\ \bibinfo {author} {\bibfnamefont {K.}~\bibnamefont {Wang}},\ }\bibfield  {title} {\enquote {\bibinfo {title} {All-electrical switching of a topological non-collinear antiferromagnet at room temperature},}\ }\href@noop {} {\bibfield  {journal} {\bibinfo  {journal} {National Science Review}\ }\textbf {\bibinfo {volume} {10}},\ \bibinfo {pages} {nwac154} (\bibinfo {year} {2023})}\BibitemShut {NoStop}%
\bibitem [{\citenamefont {Liu}\ \emph {et~al.}(2023{\natexlab{b}})\citenamefont {Liu}, \citenamefont {Feng}, \citenamefont {Zhang}, \citenamefont {Deng}, \citenamefont {Dong}, \citenamefont {Zhang}, \citenamefont {Li}, \citenamefont {Lu}, \citenamefont {Chang},\ and\ \citenamefont {Wang}}]{liu2023topological}%
  \BibitemOpen
  \bibfield  {author} {\bibinfo {author} {\bibfnamefont {X.}~\bibnamefont {Liu}}, \bibinfo {author} {\bibfnamefont {Q.}~\bibnamefont {Feng}}, \bibinfo {author} {\bibfnamefont {D.}~\bibnamefont {Zhang}}, \bibinfo {author} {\bibfnamefont {Y.}~\bibnamefont {Deng}}, \bibinfo {author} {\bibfnamefont {S.}~\bibnamefont {Dong}}, \bibinfo {author} {\bibfnamefont {E.}~\bibnamefont {Zhang}}, \bibinfo {author} {\bibfnamefont {W.}~\bibnamefont {Li}}, \bibinfo {author} {\bibfnamefont {Q.}~\bibnamefont {Lu}}, \bibinfo {author} {\bibfnamefont {K.}~\bibnamefont {Chang}}, \ and\ \bibinfo {author} {\bibfnamefont {K.}~\bibnamefont {Wang}},\ }\bibfield  {title} {\enquote {\bibinfo {title} {Topological spin textures in a non-collinear antiferromagnet system},}\ }\href@noop {} {\bibfield  {journal} {\bibinfo  {journal} {Advanced Materials}\ }\textbf {\bibinfo {volume} {35}},\ \bibinfo {pages} {2211634} (\bibinfo {year} {2023}{\natexlab{b}})}\BibitemShut {NoStop}%
\bibitem [{\citenamefont {Ikeda}\ \emph {et~al.}(2018)\citenamefont {Ikeda}, \citenamefont {Tsunoda}, \citenamefont {Oogane}, \citenamefont {Oh}, \citenamefont {Morita},\ and\ \citenamefont {Ando}}]{ikeda2018anomalous}%
  \BibitemOpen
  \bibfield  {author} {\bibinfo {author} {\bibfnamefont {T.}~\bibnamefont {Ikeda}}, \bibinfo {author} {\bibfnamefont {M.}~\bibnamefont {Tsunoda}}, \bibinfo {author} {\bibfnamefont {M.}~\bibnamefont {Oogane}}, \bibinfo {author} {\bibfnamefont {S.}~\bibnamefont {Oh}}, \bibinfo {author} {\bibfnamefont {T.}~\bibnamefont {Morita}}, \ and\ \bibinfo {author} {\bibfnamefont {Y.}~\bibnamefont {Ando}},\ }\bibfield  {title} {\enquote {\bibinfo {title} {Anomalous hall effect in polycrystalline {Mn$_3$Sn} thin films},}\ }\href@noop {} {\bibfield  {journal} {\bibinfo  {journal} {Applied Physics Letters}\ }\textbf {\bibinfo {volume} {113}} (\bibinfo {year} {2018})}\BibitemShut {NoStop}%
\bibitem [{\citenamefont {Xie}\ \emph {et~al.}(2022)\citenamefont {Xie}, \citenamefont {Chen}, \citenamefont {Zhang}, \citenamefont {Mu}, \citenamefont {Zhang}, \citenamefont {Yan},\ and\ \citenamefont {Wu}}]{xie2022magnetization}%
  \BibitemOpen
  \bibfield  {author} {\bibinfo {author} {\bibfnamefont {H.}~\bibnamefont {Xie}}, \bibinfo {author} {\bibfnamefont {X.}~\bibnamefont {Chen}}, \bibinfo {author} {\bibfnamefont {Q.}~\bibnamefont {Zhang}}, \bibinfo {author} {\bibfnamefont {Z.}~\bibnamefont {Mu}}, \bibinfo {author} {\bibfnamefont {X.}~\bibnamefont {Zhang}}, \bibinfo {author} {\bibfnamefont {B.}~\bibnamefont {Yan}}, \ and\ \bibinfo {author} {\bibfnamefont {Y.}~\bibnamefont {Wu}},\ }\bibfield  {title} {\enquote {\bibinfo {title} {Magnetization switching in polycrystalline {Mn$_3$Sn} thin film induced by self-generated spin-polarized current},}\ }\href@noop {} {\bibfield  {journal} {\bibinfo  {journal} {Nature communications}\ }\textbf {\bibinfo {volume} {13}},\ \bibinfo {pages} {5744} (\bibinfo {year} {2022})}\BibitemShut {NoStop}%
\bibitem [{\citenamefont {Zhao}, \citenamefont {He},\ and\ \citenamefont {Cai}(2021)}]{zhao2021terahertz}%
  \BibitemOpen
  \bibfield  {author} {\bibinfo {author} {\bibfnamefont {D.-Y.}\ \bibnamefont {Zhao}}, \bibinfo {author} {\bibfnamefont {P.-B.}\ \bibnamefont {He}}, \ and\ \bibinfo {author} {\bibfnamefont {M.-Q.}\ \bibnamefont {Cai}},\ }\bibfield  {title} {\enquote {\bibinfo {title} {Terahertz oscillation in a noncollinear antiferromagnet under spin-orbit torques},}\ }\href@noop {} {\bibfield  {journal} {\bibinfo  {journal} {Physical Review B}\ }\textbf {\bibinfo {volume} {104}},\ \bibinfo {pages} {214423} (\bibinfo {year} {2021})}\BibitemShut {NoStop}%
\bibitem [{\citenamefont {Shukla}\ and\ \citenamefont {Rakheja}(2022)}]{shukla2022spin}%
  \BibitemOpen
  \bibfield  {author} {\bibinfo {author} {\bibfnamefont {A.}~\bibnamefont {Shukla}}\ and\ \bibinfo {author} {\bibfnamefont {S.}~\bibnamefont {Rakheja}},\ }\bibfield  {title} {\enquote {\bibinfo {title} {Spin-torque-driven terahertz auto-oscillations in noncollinear coplanar antiferromagnets},}\ }\href@noop {} {\bibfield  {journal} {\bibinfo  {journal} {Physical Review Applied}\ }\textbf {\bibinfo {volume} {17}},\ \bibinfo {pages} {034037} (\bibinfo {year} {2022})}\BibitemShut {NoStop}%
\bibitem [{\citenamefont {Lund}\ \emph {et~al.}(2023)\citenamefont {Lund}, \citenamefont {Rodrigues}, \citenamefont {Everschor-Sitte},\ and\ \citenamefont {Hals}}]{lund2023voltage}%
  \BibitemOpen
  \bibfield  {author} {\bibinfo {author} {\bibfnamefont {M.~A.}\ \bibnamefont {Lund}}, \bibinfo {author} {\bibfnamefont {D.~R.}\ \bibnamefont {Rodrigues}}, \bibinfo {author} {\bibfnamefont {K.}~\bibnamefont {Everschor-Sitte}}, \ and\ \bibinfo {author} {\bibfnamefont {K.~M.}\ \bibnamefont {Hals}},\ }\bibfield  {title} {\enquote {\bibinfo {title} {Voltage-controlled high-bandwidth terahertz oscillators based on antiferromagnets},}\ }\href@noop {} {\bibfield  {journal} {\bibinfo  {journal} {Physical Review Letters}\ }\textbf {\bibinfo {volume} {131}},\ \bibinfo {pages} {156704} (\bibinfo {year} {2023})}\BibitemShut {NoStop}%
\bibitem [{\citenamefont {Shukla}, \citenamefont {Qian},\ and\ \citenamefont {Rakheja}(2023)}]{shukla2023order}%
  \BibitemOpen
  \bibfield  {author} {\bibinfo {author} {\bibfnamefont {A.}~\bibnamefont {Shukla}}, \bibinfo {author} {\bibfnamefont {S.}~\bibnamefont {Qian}}, \ and\ \bibinfo {author} {\bibfnamefont {S.}~\bibnamefont {Rakheja}},\ }\bibfield  {title} {\enquote {\bibinfo {title} {Order parameter dynamics in {Mn$_3$Sn} driven by dc and pulsed spin--orbit torques},}\ }\href@noop {} {\bibfield  {journal} {\bibinfo  {journal} {APL Materials}\ }\textbf {\bibinfo {volume} {11}} (\bibinfo {year} {2023})}\BibitemShut {NoStop}%
\bibitem [{\citenamefont {Xu}\ \emph {et~al.}(2024)\citenamefont {Xu}, \citenamefont {Zhang}, \citenamefont {Qiao}, \citenamefont {Liang}, \citenamefont {Shi},\ and\ \citenamefont {Zhu}}]{xu2024deterministic}%
  \BibitemOpen
  \bibfield  {author} {\bibinfo {author} {\bibfnamefont {Z.}~\bibnamefont {Xu}}, \bibinfo {author} {\bibfnamefont {X.}~\bibnamefont {Zhang}}, \bibinfo {author} {\bibfnamefont {Y.}~\bibnamefont {Qiao}}, \bibinfo {author} {\bibfnamefont {G.}~\bibnamefont {Liang}}, \bibinfo {author} {\bibfnamefont {S.}~\bibnamefont {Shi}}, \ and\ \bibinfo {author} {\bibfnamefont {Z.}~\bibnamefont {Zhu}},\ }\bibfield  {title} {\enquote {\bibinfo {title} {Deterministic spin-orbit torque switching including the interplay between spin polarization and kagome plane in {Mn$_3$Sn}},}\ }\href@noop {} {\bibfield  {journal} {\bibinfo  {journal} {Physical Review B}\ }\textbf {\bibinfo {volume} {109}},\ \bibinfo {pages} {134433} (\bibinfo {year} {2024})}\BibitemShut {NoStop}%
\bibitem [{\citenamefont {Shukla}, \citenamefont {Qian},\ and\ \citenamefont {Rakheja}(2024)}]{shukla2024impact}%
  \BibitemOpen
  \bibfield  {author} {\bibinfo {author} {\bibfnamefont {A.}~\bibnamefont {Shukla}}, \bibinfo {author} {\bibfnamefont {S.}~\bibnamefont {Qian}}, \ and\ \bibinfo {author} {\bibfnamefont {S.}~\bibnamefont {Rakheja}},\ }\bibfield  {title} {\enquote {\bibinfo {title} {Impact of strain on the sot-driven dynamics of thin film {Mn$_3$Sn}},}\ }\href@noop {} {\bibfield  {journal} {\bibinfo  {journal} {Journal of Applied Physics}\ }\textbf {\bibinfo {volume} {135}} (\bibinfo {year} {2024})}\BibitemShut {NoStop}%
\bibitem [{\citenamefont {Zhao}\ \emph {et~al.}(2025)\citenamefont {Zhao}, \citenamefont {Xu}, \citenamefont {Zhang}, \citenamefont {Kong}, \citenamefont {Shi},\ and\ \citenamefont {Zhu}}]{zhao2025investigation}%
  \BibitemOpen
  \bibfield  {author} {\bibinfo {author} {\bibfnamefont {B.}~\bibnamefont {Zhao}}, \bibinfo {author} {\bibfnamefont {Z.}~\bibnamefont {Xu}}, \bibinfo {author} {\bibfnamefont {X.}~\bibnamefont {Zhang}}, \bibinfo {author} {\bibfnamefont {Z.}~\bibnamefont {Kong}}, \bibinfo {author} {\bibfnamefont {S.}~\bibnamefont {Shi}}, \ and\ \bibinfo {author} {\bibfnamefont {Z.}~\bibnamefont {Zhu}},\ }\bibfield  {title} {\enquote {\bibinfo {title} {Investigation of sub-configurations reveals stable spin-orbit torque switching polarity in polycrystalline {Mn$_3$Sn}},}\ }\href@noop {} {\bibfield  {journal} {\bibinfo  {journal} {arXiv preprint arXiv:2501.15815}\ } (\bibinfo {year} {2025})}\BibitemShut {NoStop}%
\bibitem [{\citenamefont {Coffey}\ and\ \citenamefont {Kalmykov}(2012)}]{coffey2012thermal}%
  \BibitemOpen
  \bibfield  {author} {\bibinfo {author} {\bibfnamefont {W.~T.}\ \bibnamefont {Coffey}}\ and\ \bibinfo {author} {\bibfnamefont {Y.~P.}\ \bibnamefont {Kalmykov}},\ }\bibfield  {title} {\enquote {\bibinfo {title} {Thermal fluctuations of magnetic nanoparticles: Fifty years after {B}rown},}\ }\href@noop {} {\bibfield  {journal} {\bibinfo  {journal} {Journal of Applied Physics}\ }\textbf {\bibinfo {volume} {112}},\ \bibinfo {pages} {121301} (\bibinfo {year} {2012})}\BibitemShut {NoStop}%
\bibitem [{\citenamefont {Kaiser}\ \emph {et~al.}(2019)\citenamefont {Kaiser}, \citenamefont {Rustagi}, \citenamefont {Camsari}, \citenamefont {Sun}, \citenamefont {Datta},\ and\ \citenamefont {Upadhyaya}}]{kaiser2019subnanosecond}%
  \BibitemOpen
  \bibfield  {author} {\bibinfo {author} {\bibfnamefont {J.}~\bibnamefont {Kaiser}}, \bibinfo {author} {\bibfnamefont {A.}~\bibnamefont {Rustagi}}, \bibinfo {author} {\bibfnamefont {K.~Y.}\ \bibnamefont {Camsari}}, \bibinfo {author} {\bibfnamefont {J.~Z.}\ \bibnamefont {Sun}}, \bibinfo {author} {\bibfnamefont {S.}~\bibnamefont {Datta}}, \ and\ \bibinfo {author} {\bibfnamefont {P.}~\bibnamefont {Upadhyaya}},\ }\bibfield  {title} {\enquote {\bibinfo {title} {Subnanosecond fluctuations in low-barrier nanomagnets},}\ }\href@noop {} {\bibfield  {journal} {\bibinfo  {journal} {Physical Review Applied}\ }\textbf {\bibinfo {volume} {12}},\ \bibinfo {pages} {054056} (\bibinfo {year} {2019})}\BibitemShut {NoStop}%
\bibitem [{\citenamefont {Shukla}\ \emph {et~al.}(2023)\citenamefont {Shukla}, \citenamefont {Heller}, \citenamefont {Morshed}, \citenamefont {Rehm}, \citenamefont {Ghosh}, \citenamefont {Kent},\ and\ \citenamefont {Rakheja}}]{shukla2023true}%
  \BibitemOpen
  \bibfield  {author} {\bibinfo {author} {\bibfnamefont {A.}~\bibnamefont {Shukla}}, \bibinfo {author} {\bibfnamefont {L.}~\bibnamefont {Heller}}, \bibinfo {author} {\bibfnamefont {M.~G.}\ \bibnamefont {Morshed}}, \bibinfo {author} {\bibfnamefont {L.}~\bibnamefont {Rehm}}, \bibinfo {author} {\bibfnamefont {A.~W.}\ \bibnamefont {Ghosh}}, \bibinfo {author} {\bibfnamefont {A.~D.}\ \bibnamefont {Kent}}, \ and\ \bibinfo {author} {\bibfnamefont {S.}~\bibnamefont {Rakheja}},\ }\bibfield  {title} {\enquote {\bibinfo {title} {A true random number generator for probabilistic computing using stochastic magnetic actuated random transducer devices},}\ }in\ \href@noop {} {\emph {\bibinfo {booktitle} {2023 24th International Symposium on Quality Electronic Design (ISQED)}}}\ (\bibinfo {organization} {IEEE},\ \bibinfo {year} {2023})\ pp.\ \bibinfo {pages} {1--10}\BibitemShut {NoStop}%
\bibitem [{\citenamefont {Chun}\ \emph {et~al.}(2015)\citenamefont {Chun}, \citenamefont {Lee}, \citenamefont {Hara}, \citenamefont {Park},\ and\ \citenamefont {Kim}}]{chun2015high}%
  \BibitemOpen
  \bibfield  {author} {\bibinfo {author} {\bibfnamefont {S.}~\bibnamefont {Chun}}, \bibinfo {author} {\bibfnamefont {S.-B.}\ \bibnamefont {Lee}}, \bibinfo {author} {\bibfnamefont {M.}~\bibnamefont {Hara}}, \bibinfo {author} {\bibfnamefont {W.}~\bibnamefont {Park}}, \ and\ \bibinfo {author} {\bibfnamefont {S.-J.}\ \bibnamefont {Kim}},\ }\bibfield  {title} {\enquote {\bibinfo {title} {High-density physical random number generator using spin signals in multidomain ferromagnetic layer},}\ }\href@noop {} {\bibfield  {journal} {\bibinfo  {journal} {Advances in Condensed Matter Physics}\ }\textbf {\bibinfo {volume} {2015}},\ \bibinfo {pages} {251819} (\bibinfo {year} {2015})}\BibitemShut {NoStop}%
\bibitem [{\citenamefont {Shao}\ \emph {et~al.}(2023)\citenamefont {Shao}, \citenamefont {Duffee}, \citenamefont {Raimondo}, \citenamefont {Davila}, \citenamefont {Lopez-Dominguez}, \citenamefont {Katine}, \citenamefont {Finocchio},\ and\ \citenamefont {Amiri}}]{shao2023probabilistic}%
  \BibitemOpen
  \bibfield  {author} {\bibinfo {author} {\bibfnamefont {Y.}~\bibnamefont {Shao}}, \bibinfo {author} {\bibfnamefont {C.}~\bibnamefont {Duffee}}, \bibinfo {author} {\bibfnamefont {E.}~\bibnamefont {Raimondo}}, \bibinfo {author} {\bibfnamefont {N.}~\bibnamefont {Davila}}, \bibinfo {author} {\bibfnamefont {V.}~\bibnamefont {Lopez-Dominguez}}, \bibinfo {author} {\bibfnamefont {J.~A.}\ \bibnamefont {Katine}}, \bibinfo {author} {\bibfnamefont {G.}~\bibnamefont {Finocchio}}, \ and\ \bibinfo {author} {\bibfnamefont {P.~K.}\ \bibnamefont {Amiri}},\ }\bibfield  {title} {\enquote {\bibinfo {title} {Probabilistic computing with voltage-controlled dynamics in magnetic tunnel junctions},}\ }\href@noop {} {\bibfield  {journal} {\bibinfo  {journal} {Nanotechnology}\ }\textbf {\bibinfo {volume} {34}},\ \bibinfo {pages} {495203} (\bibinfo {year} {2023})}\BibitemShut {NoStop}%
\bibitem [{\citenamefont {Sato}\ \emph {et~al.}(2023)\citenamefont {Sato}, \citenamefont {Takeuchi}, \citenamefont {Yamane}, \citenamefont {Yoon}, \citenamefont {Kanai}, \citenamefont {Ieda}, \citenamefont {Ohno},\ and\ \citenamefont {Fukami}}]{sato2023thermal}%
  \BibitemOpen
  \bibfield  {author} {\bibinfo {author} {\bibfnamefont {Y.}~\bibnamefont {Sato}}, \bibinfo {author} {\bibfnamefont {Y.}~\bibnamefont {Takeuchi}}, \bibinfo {author} {\bibfnamefont {Y.}~\bibnamefont {Yamane}}, \bibinfo {author} {\bibfnamefont {J.-Y.}\ \bibnamefont {Yoon}}, \bibinfo {author} {\bibfnamefont {S.}~\bibnamefont {Kanai}}, \bibinfo {author} {\bibfnamefont {J.}~\bibnamefont {Ieda}}, \bibinfo {author} {\bibfnamefont {H.}~\bibnamefont {Ohno}}, \ and\ \bibinfo {author} {\bibfnamefont {S.}~\bibnamefont {Fukami}},\ }\bibfield  {title} {\enquote {\bibinfo {title} {Thermal stability of non-collinear antiferromagnetic {Mn$_3$Sn} nanodot},}\ }\href@noop {} {\bibfield  {journal} {\bibinfo  {journal} {Applied Physics Letters}\ }\textbf {\bibinfo {volume} {122}} (\bibinfo {year} {2023})}\BibitemShut {NoStop}%
\bibitem [{\citenamefont {Kobayashi}\ \emph {et~al.}(2023)\citenamefont {Kobayashi}, \citenamefont {Shiota}, \citenamefont {Narita}, \citenamefont {Ono},\ and\ \citenamefont {Moriyama}}]{kobayashi2023pulse}%
  \BibitemOpen
  \bibfield  {author} {\bibinfo {author} {\bibfnamefont {Y.}~\bibnamefont {Kobayashi}}, \bibinfo {author} {\bibfnamefont {Y.}~\bibnamefont {Shiota}}, \bibinfo {author} {\bibfnamefont {H.}~\bibnamefont {Narita}}, \bibinfo {author} {\bibfnamefont {T.}~\bibnamefont {Ono}}, \ and\ \bibinfo {author} {\bibfnamefont {T.}~\bibnamefont {Moriyama}},\ }\bibfield  {title} {\enquote {\bibinfo {title} {Pulse-width dependence of spin--orbit torque switching in {Mn$_3$Sn}/{P}t thin films},}\ }\href@noop {} {\bibfield  {journal} {\bibinfo  {journal} {Applied Physics Letters}\ }\textbf {\bibinfo {volume} {122}} (\bibinfo {year} {2023})}\BibitemShut {NoStop}%
\bibitem [{\citenamefont {Konakanchi}\ \emph {et~al.}(2025)\citenamefont {Konakanchi}, \citenamefont {Banerjee}, \citenamefont {Rahman}, \citenamefont {Yamane}, \citenamefont {Kanai}, \citenamefont {Fukami},\ and\ \citenamefont {Upadhyaya}}]{konakanchi2025electrically}%
  \BibitemOpen
  \bibfield  {author} {\bibinfo {author} {\bibfnamefont {S.~T.}\ \bibnamefont {Konakanchi}}, \bibinfo {author} {\bibfnamefont {S.}~\bibnamefont {Banerjee}}, \bibinfo {author} {\bibfnamefont {M.~M.}\ \bibnamefont {Rahman}}, \bibinfo {author} {\bibfnamefont {Y.}~\bibnamefont {Yamane}}, \bibinfo {author} {\bibfnamefont {S.}~\bibnamefont {Kanai}}, \bibinfo {author} {\bibfnamefont {S.}~\bibnamefont {Fukami}}, \ and\ \bibinfo {author} {\bibfnamefont {P.}~\bibnamefont {Upadhyaya}},\ }\bibfield  {title} {\enquote {\bibinfo {title} {Electrically tunable picosecond-scale octupole fluctuations in chiral antiferromagnets},}\ }\href@noop {} {\bibfield  {journal} {\bibinfo  {journal} {arXiv preprint arXiv:2501.18978}\ } (\bibinfo {year} {2025})}\BibitemShut {NoStop}%
\bibitem [{\citenamefont {Tomiyoshi}\ and\ \citenamefont {Yamaguchi}(1982)}]{tomiyoshi1982magnetic}%
  \BibitemOpen
  \bibfield  {author} {\bibinfo {author} {\bibfnamefont {S.}~\bibnamefont {Tomiyoshi}}\ and\ \bibinfo {author} {\bibfnamefont {Y.}~\bibnamefont {Yamaguchi}},\ }\bibfield  {title} {\enquote {\bibinfo {title} {Magnetic structure and weak ferromagnetism of {Mn$_3$Sn} studied by polarized neutron diffraction},}\ }\href@noop {} {\bibfield  {journal} {\bibinfo  {journal} {Journal of the Physical Society of Japan}\ }\textbf {\bibinfo {volume} {51}},\ \bibinfo {pages} {2478--2486} (\bibinfo {year} {1982})}\BibitemShut {NoStop}%
\bibitem [{\citenamefont {Park}\ \emph {et~al.}(2018)\citenamefont {Park}, \citenamefont {Oh}, \citenamefont {Uhl{\'\i}{\v{r}}ov{\'a}}, \citenamefont {Jackson}, \citenamefont {De{\'a}k}, \citenamefont {Szunyogh}, \citenamefont {Lee}, \citenamefont {Cho}, \citenamefont {Kim}, \citenamefont {Walker} \emph {et~al.}}]{park2018magnetic}%
  \BibitemOpen
  \bibfield  {author} {\bibinfo {author} {\bibfnamefont {P.}~\bibnamefont {Park}}, \bibinfo {author} {\bibfnamefont {J.}~\bibnamefont {Oh}}, \bibinfo {author} {\bibfnamefont {K.}~\bibnamefont {Uhl{\'\i}{\v{r}}ov{\'a}}}, \bibinfo {author} {\bibfnamefont {J.}~\bibnamefont {Jackson}}, \bibinfo {author} {\bibfnamefont {A.}~\bibnamefont {De{\'a}k}}, \bibinfo {author} {\bibfnamefont {L.}~\bibnamefont {Szunyogh}}, \bibinfo {author} {\bibfnamefont {K.~H.}\ \bibnamefont {Lee}}, \bibinfo {author} {\bibfnamefont {H.}~\bibnamefont {Cho}}, \bibinfo {author} {\bibfnamefont {H.-L.}\ \bibnamefont {Kim}}, \bibinfo {author} {\bibfnamefont {H.~C.}\ \bibnamefont {Walker}},  \emph {et~al.},\ }\bibfield  {title} {\enquote {\bibinfo {title} {Magnetic excitations in non-collinear antiferromagnetic weyl semimetal {Mn$_3$Sn}},}\ }\href@noop {} {\bibfield  {journal} {\bibinfo  {journal} {npj Quantum Materials}\ }\textbf {\bibinfo {volume} {3}},\ \bibinfo {pages} {63} (\bibinfo {year} {2018})}\BibitemShut {NoStop}%
\bibitem [{\citenamefont {He}\ and\ \citenamefont {Liu}(2024)}]{he2024magnetic}%
  \BibitemOpen
  \bibfield  {author} {\bibinfo {author} {\bibfnamefont {Z.}~\bibnamefont {He}}\ and\ \bibinfo {author} {\bibfnamefont {L.}~\bibnamefont {Liu}},\ }\bibfield  {title} {\enquote {\bibinfo {title} {Magnetic dynamics of strained non-collinear antiferromagnet},}\ }\href@noop {} {\bibfield  {journal} {\bibinfo  {journal} {Journal of Applied Physics}\ }\textbf {\bibinfo {volume} {135}} (\bibinfo {year} {2024})}\BibitemShut {NoStop}%
\bibitem [{\citenamefont {Morse}\ and\ \citenamefont {Feshbach}(1946)}]{morse1946methods}%
  \BibitemOpen
  \bibfield  {author} {\bibinfo {author} {\bibfnamefont {P.~M.}\ \bibnamefont {Morse}}\ and\ \bibinfo {author} {\bibfnamefont {H.}~\bibnamefont {Feshbach}},\ }\href@noop {} {\emph {\bibinfo {title} {Methods of theoretical physics}}}\ (\bibinfo  {publisher} {Technology Press},\ \bibinfo {year} {1946})\BibitemShut {NoStop}%
\bibitem [{\citenamefont {R{\'o}zsa}\ \emph {et~al.}(2019)\citenamefont {R{\'o}zsa}, \citenamefont {Selzer}, \citenamefont {Birk}, \citenamefont {Atxitia},\ and\ \citenamefont {Nowak}}]{rozsa2019reduced}%
  \BibitemOpen
  \bibfield  {author} {\bibinfo {author} {\bibfnamefont {L.}~\bibnamefont {R{\'o}zsa}}, \bibinfo {author} {\bibfnamefont {S.}~\bibnamefont {Selzer}}, \bibinfo {author} {\bibfnamefont {T.}~\bibnamefont {Birk}}, \bibinfo {author} {\bibfnamefont {U.}~\bibnamefont {Atxitia}}, \ and\ \bibinfo {author} {\bibfnamefont {U.}~\bibnamefont {Nowak}},\ }\bibfield  {title} {\enquote {\bibinfo {title} {Reduced thermal stability of antiferromagnetic nanostructures},}\ }\href@noop {} {\bibfield  {journal} {\bibinfo  {journal} {Physical Review B}\ }\textbf {\bibinfo {volume} {100}},\ \bibinfo {pages} {064422} (\bibinfo {year} {2019})}\BibitemShut {NoStop}%
\bibitem [{\citenamefont {Mayergoyz}, \citenamefont {Bertotti},\ and\ \citenamefont {Serpico}(2009)}]{mayergoyz2009nonlinear}%
  \BibitemOpen
  \bibfield  {author} {\bibinfo {author} {\bibfnamefont {I.~D.}\ \bibnamefont {Mayergoyz}}, \bibinfo {author} {\bibfnamefont {G.}~\bibnamefont {Bertotti}}, \ and\ \bibinfo {author} {\bibfnamefont {C.}~\bibnamefont {Serpico}},\ }\href@noop {} {\emph {\bibinfo {title} {Nonlinear magnetization dynamics in nanosystems}}}\ (\bibinfo  {publisher} {Elsevier},\ \bibinfo {year} {2009})\BibitemShut {NoStop}%
\bibitem [{\citenamefont {Go}\ \emph {et~al.}(2022)\citenamefont {Go}, \citenamefont {Sallermann}, \citenamefont {Lux}, \citenamefont {Bl{\"u}gel}, \citenamefont {Gomonay},\ and\ \citenamefont {Mokrousov}}]{go2022noncollinear}%
  \BibitemOpen
  \bibfield  {author} {\bibinfo {author} {\bibfnamefont {D.}~\bibnamefont {Go}}, \bibinfo {author} {\bibfnamefont {M.}~\bibnamefont {Sallermann}}, \bibinfo {author} {\bibfnamefont {F.~R.}\ \bibnamefont {Lux}}, \bibinfo {author} {\bibfnamefont {S.}~\bibnamefont {Bl{\"u}gel}}, \bibinfo {author} {\bibfnamefont {O.}~\bibnamefont {Gomonay}}, \ and\ \bibinfo {author} {\bibfnamefont {Y.}~\bibnamefont {Mokrousov}},\ }\bibfield  {title} {\enquote {\bibinfo {title} {Noncollinear spin current for switching of chiral magnetic textures},}\ }\href@noop {} {\bibfield  {journal} {\bibinfo  {journal} {Physical review letters}\ }\textbf {\bibinfo {volume} {129}},\ \bibinfo {pages} {097204} (\bibinfo {year} {2022})}\BibitemShut {NoStop}%
\bibitem [{\citenamefont {Ament}\ \emph {et~al.}(2016)\citenamefont {Ament}, \citenamefont {Rangarajan}, \citenamefont {Parthasarathy},\ and\ \citenamefont {Rakheja}}]{ament2016solving}%
  \BibitemOpen
  \bibfield  {author} {\bibinfo {author} {\bibfnamefont {S.}~\bibnamefont {Ament}}, \bibinfo {author} {\bibfnamefont {N.}~\bibnamefont {Rangarajan}}, \bibinfo {author} {\bibfnamefont {A.}~\bibnamefont {Parthasarathy}}, \ and\ \bibinfo {author} {\bibfnamefont {S.}~\bibnamefont {Rakheja}},\ }\bibfield  {title} {\enquote {\bibinfo {title} {Solving the stochastic landau-lifshitz-gilbert-slonczewski equation for monodomain nanomagnets: A survey and analysis of numerical techniques},}\ }\href@noop {} {\bibfield  {journal} {\bibinfo  {journal} {arXiv preprint arXiv:1607.04596}\ } (\bibinfo {year} {2016})}\BibitemShut {NoStop}%
\bibitem [{\citenamefont {Brown~Jr}(1963)}]{brown1963thermal}%
  \BibitemOpen
  \bibfield  {author} {\bibinfo {author} {\bibfnamefont {W.~F.}\ \bibnamefont {Brown~Jr}},\ }\bibfield  {title} {\enquote {\bibinfo {title} {Thermal fluctuations of a single-domain particle},}\ }\href@noop {} {\bibfield  {journal} {\bibinfo  {journal} {Physical review}\ }\textbf {\bibinfo {volume} {130}},\ \bibinfo {pages} {1677} (\bibinfo {year} {1963})}\BibitemShut {NoStop}%
\bibitem [{\citenamefont {Brown}(1979)}]{brown1979thermal}%
  \BibitemOpen
  \bibfield  {author} {\bibinfo {author} {\bibfnamefont {W.}~\bibnamefont {Brown}},\ }\bibfield  {title} {\enquote {\bibinfo {title} {Thermal fluctuation of fine ferromagnetic particles},}\ }\href@noop {} {\bibfield  {journal} {\bibinfo  {journal} {IEEE Transactions on Magnetics}\ }\textbf {\bibinfo {volume} {15}},\ \bibinfo {pages} {1196--1208} (\bibinfo {year} {1979})}\BibitemShut {NoStop}%
\bibitem [{\citenamefont {Wang}\ \emph {et~al.}(2022)\citenamefont {Wang}, \citenamefont {Zhang}, \citenamefont {Bheemarasetty}, \citenamefont {Zhou}, \citenamefont {Ying},\ and\ \citenamefont {Xiao}}]{wang2022single}%
  \BibitemOpen
  \bibfield  {author} {\bibinfo {author} {\bibfnamefont {K.}~\bibnamefont {Wang}}, \bibinfo {author} {\bibfnamefont {Y.}~\bibnamefont {Zhang}}, \bibinfo {author} {\bibfnamefont {V.}~\bibnamefont {Bheemarasetty}}, \bibinfo {author} {\bibfnamefont {S.}~\bibnamefont {Zhou}}, \bibinfo {author} {\bibfnamefont {S.-C.}\ \bibnamefont {Ying}}, \ and\ \bibinfo {author} {\bibfnamefont {G.}~\bibnamefont {Xiao}},\ }\bibfield  {title} {\enquote {\bibinfo {title} {Single skyrmion true random number generator using local dynamics and interaction between skyrmions},}\ }\href@noop {} {\bibfield  {journal} {\bibinfo  {journal} {Nature communications}\ }\textbf {\bibinfo {volume} {13}},\ \bibinfo {pages} {722} (\bibinfo {year} {2022})}\BibitemShut {NoStop}%
\bibitem [{\citenamefont {Chen}, \citenamefont {Zhang},\ and\ \citenamefont {Xiao}(2022)}]{chen2022magnetic}%
  \BibitemOpen
  \bibfield  {author} {\bibinfo {author} {\bibfnamefont {X.}~\bibnamefont {Chen}}, \bibinfo {author} {\bibfnamefont {J.}~\bibnamefont {Zhang}}, \ and\ \bibinfo {author} {\bibfnamefont {J.}~\bibnamefont {Xiao}},\ }\bibfield  {title} {\enquote {\bibinfo {title} {Magnetic-tunnel-junction-based true random-number generator with enhanced generation rate},}\ }\href@noop {} {\bibfield  {journal} {\bibinfo  {journal} {Physical Review Applied}\ }\textbf {\bibinfo {volume} {18}},\ \bibinfo {pages} {L021002} (\bibinfo {year} {2022})}\BibitemShut {NoStop}%
\bibitem [{\citenamefont {Suzuki}\ \emph {et~al.}(2017)\citenamefont {Suzuki}, \citenamefont {Koretsune}, \citenamefont {Ochi},\ and\ \citenamefont {Arita}}]{suzuki2017cluster}%
  \BibitemOpen
  \bibfield  {author} {\bibinfo {author} {\bibfnamefont {M.-T.}\ \bibnamefont {Suzuki}}, \bibinfo {author} {\bibfnamefont {T.}~\bibnamefont {Koretsune}}, \bibinfo {author} {\bibfnamefont {M.}~\bibnamefont {Ochi}}, \ and\ \bibinfo {author} {\bibfnamefont {R.}~\bibnamefont {Arita}},\ }\bibfield  {title} {\enquote {\bibinfo {title} {Cluster multipole theory for anomalous hall effect in antiferromagnets},}\ }\href@noop {} {\bibfield  {journal} {\bibinfo  {journal} {Physical Review B}\ }\textbf {\bibinfo {volume} {95}},\ \bibinfo {pages} {094406} (\bibinfo {year} {2017})}\BibitemShut {NoStop}%
\bibitem [{\citenamefont {Liu}\ and\ \citenamefont {Balents}(2017)}]{liu2017anomalous}%
  \BibitemOpen
  \bibfield  {author} {\bibinfo {author} {\bibfnamefont {J.}~\bibnamefont {Liu}}\ and\ \bibinfo {author} {\bibfnamefont {L.}~\bibnamefont {Balents}},\ }\bibfield  {title} {\enquote {\bibinfo {title} {Anomalous hall effect and topological defects in antiferromagnetic weyl semimetals: Mn 3 sn/ge},}\ }\href@noop {} {\bibfield  {journal} {\bibinfo  {journal} {Physical review letters}\ }\textbf {\bibinfo {volume} {119}},\ \bibinfo {pages} {087202} (\bibinfo {year} {2017})}\BibitemShut {NoStop}%
\bibitem [{\citenamefont {Kramers}(1940)}]{kramers1940brownian}%
  \BibitemOpen
  \bibfield  {author} {\bibinfo {author} {\bibfnamefont {H.~A.}\ \bibnamefont {Kramers}},\ }\bibfield  {title} {\enquote {\bibinfo {title} {Brownian motion in a field of force and the diffusion model of chemical reactions},}\ }\href@noop {} {\bibfield  {journal} {\bibinfo  {journal} {physica}\ }\textbf {\bibinfo {volume} {7}},\ \bibinfo {pages} {284--304} (\bibinfo {year} {1940})}\BibitemShut {NoStop}%
\bibitem [{\citenamefont {K{\"u}bler}\ and\ \citenamefont {Felser}(2014)}]{kubler2014non}%
  \BibitemOpen
  \bibfield  {author} {\bibinfo {author} {\bibfnamefont {J.}~\bibnamefont {K{\"u}bler}}\ and\ \bibinfo {author} {\bibfnamefont {C.}~\bibnamefont {Felser}},\ }\bibfield  {title} {\enquote {\bibinfo {title} {Non-collinear antiferromagnets and the anomalous hall effect},}\ }\href@noop {} {\bibfield  {journal} {\bibinfo  {journal} {Europhysics Letters}\ }\textbf {\bibinfo {volume} {108}},\ \bibinfo {pages} {67001} (\bibinfo {year} {2014})}\BibitemShut {NoStop}%
\bibitem [{\citenamefont {Markou}\ \emph {et~al.}(2018)\citenamefont {Markou}, \citenamefont {Taylor}, \citenamefont {Kalache}, \citenamefont {Werner}, \citenamefont {Parkin},\ and\ \citenamefont {Felser}}]{markou2018noncollinear}%
  \BibitemOpen
  \bibfield  {author} {\bibinfo {author} {\bibfnamefont {A.}~\bibnamefont {Markou}}, \bibinfo {author} {\bibfnamefont {J.}~\bibnamefont {Taylor}}, \bibinfo {author} {\bibfnamefont {A.}~\bibnamefont {Kalache}}, \bibinfo {author} {\bibfnamefont {P.}~\bibnamefont {Werner}}, \bibinfo {author} {\bibfnamefont {S.}~\bibnamefont {Parkin}}, \ and\ \bibinfo {author} {\bibfnamefont {C.}~\bibnamefont {Felser}},\ }\bibfield  {title} {\enquote {\bibinfo {title} {Noncollinear antiferromagnetic {Mn$_3$Sn} films},}\ }\href@noop {} {\bibfield  {journal} {\bibinfo  {journal} {Physical Review Materials}\ }\textbf {\bibinfo {volume} {2}},\ \bibinfo {pages} {051001} (\bibinfo {year} {2018})}\BibitemShut {NoStop}%
\end{thebibliography}%

\end{document}